\documentclass[11pt,a4paper]{article} 
\pdfoutput=1

\usepackage{amsmath,amsfonts,amsbsy,amssymb,array,accents}
\usepackage{enumerate,array,latexsym,graphicx,mathrsfs,verbatim,psfrag}
\usepackage{bm} 
\usepackage[normalem]{ulem}
\usepackage{booktabs} 
\usepackage[usenames]{color}

\usepackage[english]{babel}
\usepackage{fancybox}

\usepackage[nosort]{cite}
\usepackage{chngpage} 

\usepackage[utf8x]{inputenc}

\usepackage{enumitem}
\setlist{itemsep=0pt}

\usepackage[colorlinks=true,      linkcolor=darkblue,      urlcolor=darkblue,      
            filecolor=darkblue,      citecolor=darkblue,       pdfstartview=FitH,     
						pdfpagemode=UseNone,      bookmarksopen=true]{hyperref}  
\usepackage[all]{hypcap}     

\def\eq#1{(\ref{#1})}

\usepackage{mathtools}
\DeclarePairedDelimiter\bra{\langle}{\rvert}
\DeclarePairedDelimiter\ket{\lvert}{\rangle}
\DeclarePairedDelimiterX\braket[2]{\langle}{\rangle}{#1 \delimsize\vert #2}

\newcommand{\captionfonts}{\small}
\makeatletter  
\long\def\@makecaption#1#2{%
  \vskip\abovecaptionskip
  \sbox\@tempboxa{{\captionfonts #1: #2}}%
 \ifdim \wd\@tempboxa >\hsize
    {\captionfonts #1: #2\par}
  \else
    \hbox to\hsize{\hfil\box\@tempboxa\hfil}%
  \fi
  \vskip\belowcaptionskip}
\makeatother   

\DeclareMathSymbol{\medhatsym}{\mathord}{largesymbols}{"62} 

\DeclareMathSymbol{\medtildesym}{\mathord}{largesymbols}{"65}

\newcommand{\T}[3]{\ensuremath{ #1{}^{#2}_{\phantom{#2} \! #3}}}		

\mathchardef\mhyphen="2D

\def\({\left(}
\def\){\right)}
\def\[{\left[}
\def\]{\right]}

\def\barray{\begin{array}}
\def\earray{\end{array}}
\def\be{\begin{equation}}
\def\ee{\end{equation}}
\def\bea{\begin{eqnarray}}
\def\eea{\end{eqnarray}}
\def\bal{\begin{align}}
\def\eal{\end{align}}

\def\ba{\begin{aligned}}
\def\ea{\end{aligned}}

\numberwithin{equation}{section} %

\makeatletter
\g@addto@macro\bfseries{\boldmath}
\makeatother

\definecolor{cardinal}{rgb}{0.6,0,0}
\definecolor{darkgreen}{rgb}{0,0.4,0}
\definecolor{purple}{rgb}{0.5, 0, 0.5}
\definecolor{golden}{rgb}{0.92, 0.7, 0}
\definecolor{midnight}{rgb}{0, 0, 0.5}
\definecolor{darkblue}{rgb}{0, 0, 0.8}
\newcommand{\Red}{\color{red}}

\def\cA{{\cal A}}
\def\cB{{\cal B}}

\def\cM{{\cal M}}
\def\cN{{\cal N}}

\def\cP{{\cal P}}

\def\sst#1{\scriptscriptstyle{#1}}

\newcommand{\LL}{\ensuremath{\mathrm{\sst{L}}}}
\newcommand{\RR}{\ensuremath{\mathrm{\sst{R}}}}
\renewcommand{\k}{\mathsf{k}}
\newcommand{\m}{\mathsf{m}}
\newcommand{\n}{\mathsf{n}}
\newcommand{\q}{\mathsf{q}}
\newcommand{\h}{l}

\newcommand{\NS}{\ensuremath{\mathrm{\sst{NS}}}}

\def\k{\ensuremath{\mathsf{k}}}

\newcommand{\betab}{\ensuremath{\bm{\beta}}}

\newcommand{\su}{\text{su}}
\renewcommand{\sl}{\text{sl}}

\def\coeff#1#2{{\textstyle \frac{#1}{#2}}}

\begin{document}

\begin{flushright}
%
%
\end{flushright}

\vspace{19mm}

\begin{center}

{\huge \bf{Supercharged AdS$_3$ Holography}}

\vspace{22mm}

{\large
\textsc{Sami Rawash, ~David Turton}
}

\vspace{15mm}

Mathematical Sciences and STAG Research Centre, University of Southampton,\\
Highfield, Southampton SO17 1BJ, United Kingdom

\vspace{7mm}

{\footnotesize\upshape\ttfamily  s.rawash @ soton.ac.uk, ~d.j.turton @ soton.ac.uk } \\

\vspace{20mm}

\textsc{Abstract}
\vspace{7mm}
\begin{adjustwidth}{14mm}{14mm} 
\noindent
Given an asymptotically Anti-de Sitter supergravity solution, one can obtain a microscopic interpretation by
identifying the corresponding state in the holographically dual conformal field theory.
This is of particular importance for heavy pure states that are candidate black hole microstates. 
Expectation values of light operators in such heavy CFT states are encoded in the asymptotic expansion of the dual bulk configuration. 
In the D1-D5 system, large families of heavy pure CFT states have been proposed to be holographically dual to smooth horizonless supergravity solutions.
We derive the precision holographic dictionary in a new sector of light operators that are superdescendants of scalar chiral primaries of dimension (1,1). 
These operators involve the action of the supercharges of the chiral algebra, and they play a central role in the proposed holographic description of recently-constructed supergravity solutions known as ``supercharged superstrata''.
We resolve the mixing of single-trace and multi-trace operators in the CFT to identify the combinations that are dual to single-particle states in the bulk. 
We identify the corresponding gauge-invariant combinations of supergravity fields.
We use this expanded dictionary to probe the proposed holographic description of supercharged superstrata, finding precise agreement between gravity and CFT. 
\end{adjustwidth}

\end{center}

\thispagestyle{empty}

\newpage


%
%


\baselineskip=15pt
\parskip=3pt

\setcounter{tocdepth}{2}
\tableofcontents

\newpage

\section{Introduction}
\label{sec:intro}

Black holes are some of the most interesting objects in our universe. Despite many theoretical advances, we lack a detailed understanding of their internal structure at the quantum level, in particular the resolution of their singularities. 
String Theory is the leading framework within which to study the quantum properties of black holes and their microstates.

Holographic duality is a powerful tool in the study of black hole microstates.
Given an asymptotically AdS supergravity solution, by identifying  the corresponding state in the dual conformal field theory, we can gain valuable insight into its microscopic interpretation. Large families of pure CFT states are understood to be holographically dual to smooth horizonless supergravity solutions in the bulk, see e.g.~\cite{Lunin:2001jy,Lunin:2002iz,Lunin:2004uu,Giusto:2004id,Giusto:2004ip,Kanitscheider:2007wq,Mathur:2011gz,Lunin:2012gp,Giusto:2012yz}. We are primarily interested in pure CFT states that are `heavy' in the sense that their conformal dimension is proportional to the central charge of the CFT. When the mass and charges of such states agree with those of a black hole solution, such states are interpreted as microstates of the corresponding black hole.

In this paper we work in the D1-D5 system in Type IIB compactified on $\cM = T^4$ or K3, and consider heavy bound states  carrying  an additional momentum charge P along the spatial direction common to the D1 and D5 branes. The black holes we study are the BPS D1-D5-P black hole, either non-rotating~\cite{Strominger:1996sh} or rotating~\cite{Breckenridge:1996is}. On the bulk side, in this paper we will work in the supergravity approximation, though there has been recent progress on describing black hole microstates in worldsheet models~\cite{Martinec:2017ztd,Martinec:2018nco,Martinec:2019wzw,Martinec:2020gkv,Bufalini:2021ndn}.

The state-of-the-art constructions of supersymmetric microstate solutions in supergravity involve a breaking of the isometries that are preserved by the corresponding black hole, and are known as `superstrata'. For some early constructions and studies of superstrata, see e.g.~\cite{Bena:2015bea,Bena:2016agb,Bena:2016ypk,Bena:2017geu,Bena:2017xbt,Bena:2017upb}.
The construction of these solutions involves solving a set of layered BPS equations in sequence. A solution to the first two layers typically has a number of unfixed parameters. Solving the final layer of equations and imposing smoothness gives rise to algebraic relations on the parameters. This procedure is in general known as `coiffuring'~\cite{Mathur:2013nja,Bena:2013ora,Bena:2014rea}.

There is a proposal for the dual CFT description of superstrata, which has been developed hand-in-hand with (and indeed has informed) the supergravity constructions, see e.g.~\cite{Bena:2015bea,Bena:2016agb,Bena:2016ypk,Bena:2017geu,Shigemori:2020yuo}.
The parameters in the supergravity solutions have a specific holographic interpretation in the proposed dual CFT states. 
Precision holography, in particular protected three-point correlation functions that consist of the expectation value of a light operator in the heavy background, can be used to investigate such proposals~\cite{Kanitscheider:2007wq,Kanitscheider:2006zf,Taylor:2007hs,Giusto:2015dfa,Tormo:2019yus,Giusto:2019qig}.

The first families of superstrata that were constructed had limitations in that smooth solutions were not available for all choices of parameters. 
However recently, more general families of D1-D5-P superstrata have been constructed, that have resolved these limitations, as follows.

In the corresponding proposed dual CFT states, some of the momentum-carrying excitations are left-moving supercharges of the small (4,4) superconformal algebra. This is a novel feature with respect to the first families of superstrata that were constructed. As a result, these more recently constructed solutions are known as `supercharged superstrata'~\cite{Ceplak:2018pws,Heidmann:2019zws,Heidmann:2019xrd,Mayerson:2020tcl}. Supercharged superstrata and their proposed holographic description are a primary motivation for this work.

Precision holography relates expectation values of light operators in heavy CFT states to gauge-invariant combinations of supergravity fields in an asymptotic expansion at large radial distance in AdS. Coefficients of successively higher terms in the radial expansion in supergravity correspond to expectation values of successive higher dimension operators~\cite{Klebanov:1999tb,Skenderis:2006uy}.
Recently it was observed that in order to carry out precision holographic tests of superstrata, it is necessary to derive the explicit holographic dictionary to higher order than was previously known, and the precision holographic dictionary for a set of dimension two operators was derived~\cite{Giusto:2019qig}. Specifically, a set of chiral primary operators (and affine descendants) of dimension (1,1) was considered, and the mixing between single-trace operators and multi-trace operators was resolved. This enabled precision holographic tests of the first families of superstrata constructed in~\cite{Bena:2015bea,Bena:2016agb,Bena:2016ypk,Bena:2017geu} and their proposed dual CFT states. In addition, a simple preliminary test of the supercharged superstrata constructed in~\cite{Ceplak:2018pws} was computed. The results of~\cite{Giusto:2019qig} support the proposed dictionary in all examples, and also give a CFT interpretation of the coiffuring relations of non-supercharged superstrata.

In this paper we will derive the precision holographic dictionary in a novel sector, and use it to make precision holographic tests of supercharged superstrata.
This novel sector consists of superdescendants of the dimension (1,1) chiral primary operators studied in~\cite{Giusto:2019qig}.
Of particular interest to us is a new type of coiffuring relation, proposed in \cite{Heidmann:2019zws,Heidmann:2019xrd}, that resolves the limitations mentioned above. A main goal of this paper is to test this new coiffuring relation holographically. By doing so, we will also test the proposed holographic dictionary for the supercharged superstrata.

Although our primary motivations are in black hole physics and the fuzzball proposal~\cite{Mathur:2005zp,Skenderis:2008qn,Mathur:2012zp,Bena:2013dka}, our results are relevant to more general aspects of holography that have been discussed in recent (and less recent) literature, specifically the mixing of single and multi-trace operators in the CFT that are dual to supergravity states in the bulk.
As is well-known, supergravity fluctuations around the global AdS vacuum are holographically dual to
short multiplets of CFT operators whose top components are chiral primaries (or anti-chiral primaries). The remainder of the short multiplet is obtained by acting on the chiral primaries with the generators of the anomaly-free subalgebra of the superconformal algebra (see e.g.~the reviews~\cite{Aharony:1999ti,David:2002wn}).

However the precise form of this dictionary is not yet fully understood. In particular we will be interested in mixings between single-trace and multi-trace operators on the gauge theory side.
In the case of a bound state of $n_1$ D1 branes and $n_5$ D5 branes the holographic CFT is a permutation orbifold CFT, where the gauge group is the permutation group $S_N$, where $N=n_1n_5$. In this theory, traces are sums over the individual $N$ copies of the CFT.

In the large $N$ limit, for many purposes one can use the approximate dictionary that single-particle supergravity states correspond single-trace CFT operators. 
However for some time it has been understood that this approximate identification is not correct for all observables, for example extremal correlators, even at leading order in the large $N$ limit. (Extremal correlators are those for which the conformal dimension of one operator equals the sum of the conformal dimensions of the other operators in the correlator.)
The correct dictionary instead relates single-particle supergravity states to specific admixtures of single-trace and multi-trace operators. The coefficients of the multi-trace operators in the admixture involve powers of $1/N$, yet they can contribute to the value of extremal correlators at leading order in large $N$. %
The precise general form of this dictionary has until recently been elusive; for previous work, see e.g.~\cite{Arutyunov:1999en,DHoker:1999jke,Arutyunov:2000ima,DHoker:2000xhf,Corley:2001zk,Kanitscheider:2006zf,Taylor:2007hs,Uruchurtu:2011wh,Rastelli:2017udc}.

Aside from extremal correlators, certain (non-extremal) mixed heavy-light correlators require the correct dictionary to be used, even when working at leading order in the large $N$ limit.
It is these types of correlators that we consider in the present work.
We consider correlators in which the light CFT operator is dual to a single-particle supergravity state.
In the correlators we consider, the multi-trace admixtures in these light operators can contribute to the value of the correlator at leading order in large $N$ because they can have a Wick contraction with the heavy state that produces other factors of $N$. We will exhibit examples in detail.

Recently it has been proposed that single-particle supergravity fluctuations around global AdS$_5\times $S$^5$ are holographically dual to $\mathcal{N}=4$ SYM operators (in short multiplets) that are orthogonal to all multi-trace operators~\cite{Aprile:2018efk,Aprile:2020uxk}. 
This condition identifies specific admixtures of single-trace and multi-trace operators. For $\mathcal{N}=4$ SYM, this condition is sufficient to uniquely define this set of CFT operators up to normalization, and has passed recent checks~\cite{Aprile:2020uxk}.

In this paper we will emphasize that in AdS$_3 \times$S$^3 \times \cM$ the situation is more complicated than in AdS$_5\times $S$^5$. In the AdS$_3$ case, it is not sufficient to consider the set of orbifold CFT operators that are orthogonal to all multi-trace operators; one must resolve additional operator mixing. Such mixing has been studied previously in~\cite{Taylor:2007hs}. We will resolve this point fully in the sector in which we work, by combining and improving upon the results of~\cite{Giusto:2019qig} and~\cite{Taylor:2007hs}.

The method that we pursue is as follows. We first recast our recent work on AdS$_3$ holography~\cite{Giusto:2019qig} in the single-particle basis. This involves taking slightly different combinations of the single-trace and multi-trace operators of dimension (1,1) from those used in~\cite{Giusto:2019qig}. In doing so we resolve the additional operator mixing, combining~\cite{Giusto:2019qig}~and~\cite{Taylor:2007hs}. We then act with the supercharges of the small $\cN=4$ superconformal algebra to generate the superdescendants within this supermultiplet that are our primary interest in this paper. By construction, all the resulting CFT operators are in the single-particle half-BPS basis.

We then derive the gauge-invariant 
combinations of supergravity fields that describe the single-particle excitations of interest. We do so to the required order in the asymptotic expansion, and identify the terms in this expansion that encode the expectation values of the superdescendant operators that we study.

We determine the normalization coefficients of the holographic dictionary by taking a set of test CFT states and proposed dual supergravity superstrata solutions. For consistency these normalization coefficients must depend at most on charges and moduli of the theory, and not on any property of the microstates chosen for this calibration computation, and this is indeed the case. Furthermore one expects that these coefficients respect the symmetries of the supergravity theory, and our results indeed do so.

Having normalized the dictionary, we make two non-trivial holographic tests of superstrata: first we test the coiffuring proposal of general non-supercharged superstrata of~\cite{Heidmann:2019zws,Heidmann:2019xrd}. 
Second, we test a `hybrid' superstratum involving both supercharged and non-supercharged elements. These tests also represent further cross-checks on the holographic dictionary itself.

All our tests result in non-trivial and elegant agreement between supergravity and CFT. 
While the correlators we compute cannot prove that the proposed holographic description of superstrata is correct in all its details,
the agreement we find provides state-of-the-art evidence that supports the proposed holographic description, and gives a CFT interpretation to the supercharged coiffuring relation of~\cite{Heidmann:2019zws,Heidmann:2019xrd}.

The structure of this paper is as follows. In Section~\ref{sec:D1-D5-CFT} we review the D1-D5 CFT and perform a preliminary computation to be used later in the paper. Section~\ref{sec:new-sec-3} is a review of relevant aspects of the supergravity theory, superstrata, and precision holography. In Section \ref{sec:dict-1} we begin the construction of the dictionary in both gravity and CFT. In Section \ref{sec:normalizing-dictionary} we fix the normalization coefficients in the dictionary and perform tests of two distinct families of superstrata. In Section \ref{sec:disc} we discuss our results; technical details are recorded in four appendices.

\section{D1-D5 CFT}
\label{sec:D1-D5-CFT}

In this section we review the D1-D5 CFT, introduce the operators we shall consider in this work, and perform our first new calculation (in Section \ref{sec:norm}) for use later in the paper.

\subsection{D1-D5 CFT and structure of short multiplets}

We begin in Type IIB string theory compactified on $\mathcal{M}\times $S$^1$ where $\mathcal{M}$ is T$^4$ or K3. We take $\cM$ to be microscopic and the S$^1$ to be large. We coordinatize the S$^1$ by $y$, and we denote by 
$R_y$ 
the asymptotic proper radius of the S$^1$.
We consider bound states of $n_1$ D1-branes wrapped on the S$^1$ and $n_5$ D5-branes wrapped on  $\mathcal{M}\times $S$^1$, with $g_s n_1 \gg 1$ and $g_s n_5 \gg 1$. The AdS$_3$ decoupling limit can be formulated as a large $R_y$ scaling limit, see e.g.~\cite{Martinec:2018nco}. This decouples the original asymptotic region, resulting in an asymptotically AdS$_3 \times$S$^3 \times \cM$ bulk~\cite{Maldacena:1997re}, in which $y$ is the angular direction in AdS$_3$.

At low energies, the worldvolume theory on this D1-D5 system flows to a two-dimensional $\cN = (4, 4)$ SCFT with central charge $c = 6n_1n_5\equiv 6N$. There is considerable evidence that there is a locus in moduli space at which the theory is a sigma model with target space $\mathcal{M}^N/S_N$, where $S_N$ is the symmetric group; for some early work, see e.g.~\cite{Vafa:1995zh,Deger:1998nm,Larsen:1998xm,deBoer:1998ip,Larsen:1999uk,Strominger:1996sh}.\footnote{Recent work has explored a stringy version of this duality in the S-dual NS5-F1 system, for the case of a single NS5-brane; see~\cite{Eberhardt:2019ywk,Gaberdiel:2020ycd} and references within.}
We will work with $\cM\!=\:\!$T$^4$ for concreteness and ease of presentation, however our main results concern the universal sector of the compactification on $\cM$ and so all of our main results carry over appropriately to the K3 compactification.

The symmetric orbifold CFT for $\cM\!=\:\!$T$^4$ contains $N$ copies of the $c=6$ sigma model on T$^4$. This $c=6$ model has a free field realization in terms of four free bosons plus four left-moving and four right-moving free fermions. 

Let us introduce some notation that we will require. We label different copies of the $c=6$ CFT by the index $r = 1,2,\ldots,N$. The small $\cN=(4,4)$ superconformal algebra has left and right R-symmetry groups $SU(2)_L\times SU(2)_R$; we use indices $\alpha,\dot \alpha=\pm$ for the fundamental representation of each respectively. There is an additional organizational $SU(2)_C\times SU(2)_A\sim SO(4)_I$ which is inherited from the 
rotation group on the tangent space of $\mathcal{M}$; we use indices $A,\dot{A}=1,2$ for the fundamental representation of each respectively.

The four bosons and four left-moving and four right-moving free fermions on copy $r$ of the CFT are then denoted respectively by 
\be
X^{A\dot A}_{(r)} \,, \qquad \psi^{\alpha \dot A}_{(r)} \,, \qquad \bar \psi^{\dot \alpha \dot A}_{(r)} \,.
\ee

The orbifold theory also contains spin-twist operators, labelled by conjugacy classes of the permutation group, which modify the boundary conditions of the fields.
The orbifold nature of the target spaces decomposes the Hilbert space of the theory into twisted sectors corresponding to these operators.  
A generic element $g \in S_N$ involves various permutation cycles; let us denote their lengths by $\k_i$.
An orbifold CFT state involving twist operator excitations is often described as involving a collection of `strands' of length $\k_i$ occurring with degeneracy $N_i$, possibly carrying excitations and/or polarization indices, subject to the `strand budget' constraint that the total number of copies of the full CFT is $N$,
\begin{equation}
\sum_i \k_i N_i=N \,. 
\end{equation}

We now introduce the chiral primary operators (CPOs) of the theory, which are the top components of the short multiplets of the $SU(1, 1|2)_L\times SU(1, 1|2)_R$ symmetry. These operators, together with their descendants under the generators of the anomaly-free part of the small $\cN=(4,4)$ superconformal algebra, i.e.~$\{J^{-}_0,L_{-1},G^{-,A}_{-1/2}\}$ and $\{\bar J^{-}_0,\bar L_{-1},\bar G^{-,A}_{-1/2}\}$, play a central role in the construction of the holographic dictionary, because of their relation to single-particle excitations in supergravity~\cite{Deger:1998nm,Larsen:1998xm,deBoer:1998ip}.

\renewcommand\arraystretch{1.5}
\begin{table}[t]
\centering
\begin{tabular}{  c | c | c | c|c|c} 
State & $J^3$ & $L_0$ & $\bar J^3$ & $\bar L_0$ & $SU(2)_C$ \\ 
\hline
$\ket{CP}$ & $h$ & $h$ & $\bar{h}$ & $\bar{h}$ & $\mathbf{1}$  \\ 
$G^2\ket{CP}$ & $h-1$ & $h+1$ & $\bar{h}$ & $\bar{h}$ & $\mathbf{1}$  \\ 
$\bar{G}^2\ket{CP}$ & $h$ & $h$&$\bar{h}-1$&$\bar{h}+1$&$\mathbf{1}$  \\ 
$G\bar{G}\ket{CP}$ &$h-1/2$& $h+1/2$&$\bar{h}-1/2$&$\bar{h}+1/2$&$\mathbf{1}\oplus \mathbf{3}$  \\ 
$G^2\bar{G}^2\ket{CP}$ &$h-1$& $h+1$&$\bar{h}-1$&$\bar{h}+1$&$\mathbf{1}$  \\ 
\end{tabular}
\caption{Bosonic structure of the short multiplets.}
\label{table:1}
\end{table}

The bosonic structure of the $SU(1, 1|2)_L\times SU(1, 1|2)_R$ short multiplets is sketched in Table~\ref{table:1}. In Table~\ref{table:1}, $G^2$ is a short hand for the combination 
\begin{equation}
G_{-\frac12}^{+1}G_{-\frac12}^{+2} + \frac{1}{2h} J^+_0 L_{-1} \,,
\end{equation}
where $h$ is the eigenvalue of $L_0$ for the CPO we act upon; similarly for $\bar{G}^2$.
By working with this linear combination, one obtains a state that is orthogonal to the other descendants of the CPO, which is then dual to an independent supergravity fluctuation~\cite{Ceplak:2018pws,Shigemori_2019,Avery:2010qw}. Moreover, this combination gives a state that is an eigenstate of the Casimirs of $SU(2)_L\times  SU(2)_R$ and $SL(2,\mathbb{R})\times SL(2,\mathbb{R})$, as one can check  making use of the anomaly-free part of the chiral algebra in the NS-NS sector, composed of $L_0$, $L_{\pm 1}$, $J^a_0$, $G^{\alpha A}_{\pm 1/2}$. Setting temporarily $m,n=1,0,-1$ and $r,s=\pm \frac12$, the anomaly-free part of the chiral superconformal algebra is
\begin{equation}\label{anomaly free algebra}
\begin{aligned}
\big[L_m,L_n\big]\;\!&=\;\!(m-n)L_{m+n} \,, \qquad~ \big[J^{a}_0,J^{b}_0\big]\;\!=\;\!i\epsilon^{abc}J^c_0\,,\qquad \big[L_n,J^a_0\big]\;\!=\;\!0 \,,\\
\big[J^{a}_0,G^{\alpha A}_s\big]\;\!&=\;\!\frac{1}{2}G^{\beta A}_s
\big(\sigma^{a}\big)_{\beta}^{\alpha}\,, \qquad \big[L_m,G^{\alpha A}_s\big]\;\!=\;\!\Big(\frac{m}{2}-s\Big)G^{\alpha A}_{m+s}
\,,\\
\big\{G^{\alpha A}_r,G^{\beta B}_s\big\}\;\!&=\;\!\epsilon^{\alpha\beta}\epsilon^{AB}L_{r+s}+(r-s)\epsilon^{AB}\big(\sigma^{aT}\big)^{\alpha}_{\gamma}\epsilon^{\gamma \beta}J^{a}_{r+s}
 \,,
\end{aligned}
\end{equation}
where $a,b,c=\{\pm,3\}$ are indices in the adjoint of $SU(2)_L$. 

When discussing superstrata, it will be convenient for our conventions to work with anti-chiral primary operators (ACPOs): these are descendants of CPOs obtained acting the maximal number of times with the generators $J_0^{-},\bar J_0^{-}$ and are characterized by $h=-j$, $\bar{h}=-\bar j$. 
Denoting an anti-chiral primary by $O_{\sst{\mathrm{AC}}}$, we are interested in its bosonic descendants under the left-moving anomaly-free chiral algebra. These can be written in the form 
\bea \label{eq:cft-states}
\!\!  (J^+_0)^{\m-\q} L_{-1}^{\n-\q}  \left(G_{-\frac12}^{+1}G_{-\frac12}^{+2} + \frac{1}{2h} J^+_0 L_{-1}\right)^\q  |O_{\sst{\mathrm{AC}}}\rangle \,.
\eea
States with $\q=1$ are the building-blocks of the proposed holographic dual states of supercharged superstrata~\cite{Ceplak:2018pws}.

\subsection{Low-dimension operators}\label{Low dimensional operators}

We now introduce a particular set of chiral primaries and review the holographic dictionary for these operators, constructed in~\cite{Kanitscheider:2006zf,Kanitscheider:2007wq,Giusto:2015dfa,Giusto:2019qig}. It was shown in \cite{Baggio:2012rr} that expectation values of CPOs in 1/4 or 1/8-BPS states are independent of the moduli of the theory, so that their value at the free orbifold point can be reliably compared with the boundary expansion of the supergravity solution. In paricular, we will focus on light CPOs, with dimension $\Delta=h+\bar{h}\leq 2$.

In the symmetric product orbifold CFT, by definition operators must be invariant under permutations of the copies. One can construct gauge-invariant operators by tracing (i.e.~summing) over the $N$ copies of the basic fields and/or their products. 
We begin with single-trace CPOs, obtained taking a single sum over the $N$ copies.

For single-trace operators we will work with non-unit-normalized operators, for consistency with previous papers~\cite{Giusto:2015dfa}. This choice is particularly convenient for the $SU(2)$ currents, as it means that they satisfy the standard algebra (in the spin basis, i.e.~where $[J^+,J^-]=2J^3$). By contrast, for the multi-trace operators that we will introduce in the remainder of the paper, we will find it convenient to define them with unit normalization in the large $N$ limit.

Among the CPOs with $\Delta=1$, the only ones with non vanishing conformal spin $\mathsf{s}=h-\bar{h}$ are the R-symmetry currents (all sums over copy indices $r,s,\ldots$ run from $1$ to $N$ unless otherwise indicated):
\begin{equation}\label{op J}
J^{+}=\sum_r J^{+}_{(r)}=\sum_r \psi^{+1}_{(r)}\psi^{+2}_{(r)}~,\qquad \bar J^{+}=\sum_r \bar J^{+}_{(r)}=\sum_r \bar \psi^{+1}_{(r)}\bar \psi^{+2}_{(r)} ~.
\end{equation}
They are characterized respectively by ($h=j=1$, $\bar{h}=\bar{j}=0$) and ($h=j=0$, $\bar{h}=\bar{j}=1$).

Next, let us denote the dimension of the $(1,1)$ cohomology of $\mathcal{M}$ by $h^{1,1}(\mathcal{M})$. Then there are $n\equiv (h^{1,1}(\mathcal{M})+1)$ CPOs that have $\Delta=1$ and $\mathsf{s}=0$, i.e.~$(h,\bar{h})=\big(\frac12,\frac12\big)$ (For T$^4$, $n=5$, while for K3, $n=21$).
 First we have a twist-two operator. To write this operator we first define the `bare' twist-two operator $\sigma_{(rs)}$ to be the lowest-dimension (spin)-twist operator associated with the permutation $(rs)$.
We also define the spin fields $S^{+},\bar S^{+}$ that map NS to R boundary conditions on the fermions, and vice versa. The operator $\Sigma^{++}_2$ is given by 
\begin{equation}\label{op Sigma}
\Sigma^{++}_2 \,=\, \sum_{r<s}S^{+}\bar S^{+}\sigma_{(rs)} \,=\, \sum_{r<s} \sigma_{(rs)}^{++} \,.
\end{equation}

Our focus in this paper is on heavy pure microstates that are invariant on $\mathcal{M}$.
Of the $n-1$ remaining operators with $\Delta=1$ and $\mathsf{s}=0$, only the following one can have a non-vanishing expectation value on this class of microstates: 
\begin{equation}\label{op O}
O^{++}=\sum_r O^{++}_r=\sum_r\frac{\epsilon^{\dot{A}\dot B}}{\sqrt{2}}\psi^{+\dot{A}}_{(r)}\bar\psi^{+ \dot B}_{(r)} \,.
\end{equation}

Next, we discuss the set of CPOs of dimension $\Delta=2$ that are conformal scalars, and so have $h=j=\bar h=\bar j=1$.There are $n+1$ operators with these quantum numbers. First, in the untwisted sector,we have the single-trace product of one left and one right current:
\begin{equation}\label{op Omega}
\Omega^{++}=\sum_r J^{+}_{(r)}\bar J^{+}_{(r)}=\sum_r  \psi^{+1}_{(r)}\psi^{+2}_{(r)}  \bar \psi^{+1}_{(r)}\bar \psi^{+2}_{(r)} \,.
\end{equation}
Second, we have a twist-three operator which contains bare twist operators $\sigma_{(qrs)},\sigma_{(qsr)}$ associated with the inequivalent permutations $(qrs)$ and $(qsr)$ respectively, dressed with current modes that add the required charge:

\begin{equation}\label{op Sigma3}
\Sigma^{++}_3=\sum_{q<r<s} \bar{J}_{-1/3}^{+}J_{-1/3}^{+}\big(\sigma_{(qrs)}+\sigma_{(qsr)}\big)\,.
\end{equation}
Among the $n-1$ remaining operators, only the following one can have a non-vanishing expectation value on the class of $\mathcal{M}$-invariant microstates,
\begin{equation}\label{op O2}
O^{++}_2\,=\,\sum_{r<s}\big(O^{++}_{(r)}+O^{++}_{(s)}\big)\sigma^{++}_{(rs)}\,= \sum_{r<s}O^{++}_{(rs)} \,.
\end{equation}

We next consider double-trace operators, obtained taking the product of two single traces\footnote{\label{footnote definition multitrace}Our convention for the double-traces is different from that in Eq.~(4.5) of \cite{Giusto:2019qig}, where the double-traces were defined as the off-diagonal product of single-traces. The reason for this will be discussed in Section \ref{sec:spo basis}.}. We focus on the dimension $\Delta=2$ scalar double-trace operators, which are defined via
\begin{equation}\label{eq:doubletraces}
\begin{aligned}
&(\Sigma_2 \cdot \Sigma_2)^{++}\,=\, \frac{2}{N^2} \sum_{(r<s), (p<q)} \sigma_{(rs)}^{++} \sigma_{(pq)}^{++}\,,\qquad~~  (J \cdot \bar J)^{++}\,=\,\frac{1}{N} \sum_{r, s} J^{+}_{(r)} \bar J^{+}_{(s)}\,,\\
&~ (\Sigma_2 \cdot O)^{++}\,=\,\frac{\sqrt{2}}{N^{3/2}} \sum_{\substack{r<s,q}} \sigma^{++}_{(rs)} O^{++}_{(q)}\,,\qquad\qquad (O \cdot O)^{++}\,=\, \frac{1}{N} \sum_{r, s} O^{++}_{(r)} O^{++}_{(s)}\,.
\end{aligned}
\end{equation}
All these operators have unit norm in the large $N$ limit.
These operators are highest-weight with respect the $SU(2)_L\times SU(2)_R$ R-symmetry group; the rest of the R-symmetry multiplet can be constructed as usual by acting with the zero modes of $J^{-},\bar{J}^{-}$. 

For the scalar operators of dimension one, \eq{op Sigma}, \eq{op O}, we denote the members of the $(j,\bar{j})=\big(\frac12,\frac12\big)$ R-symmetry multiplet as $\Sigma_2^{\alpha,\dot{\alpha}}$ and $O^{\alpha,\dot{\alpha}}$ with $\alpha,\dot \alpha=\pm$. For the scalar operators of dimension two, \eq{op Omega}--\eq{eq:doubletraces}, we  label the $(j,\bar{j})=(1,1)$ R-symmetry multiplet with indices $a,\dot a=\pm,0$, and we choose to normalize the descendants such that they have the same norm as the highest-weight state: for instance, we define $\Sigma_3^{0,+}=\frac{1}{\sqrt{2}}[J^{-}_0,\Sigma_3^{++}]$. For the R-symmetry currents themselves, we denote the members of the multiplet by $J^a$, and we normalize the descendants such that the standard $SU(2)$ commutation relations $[J^{+},J^{-}]=2J^3$ hold (similarly for the right currents $\bar J^{\dot a}$).

\subsection{Holography for D1-D5(-P) black hole microstates}

We now review the holographic description of two-charge D1-D5 black hole microstates and three-charge D1-D5-P superstrata, including supercharged superstrata. For the two-charge states we follow in places the discussion in~\cite{Giusto:2015dfa}.

The relevant sector of the CFT to describe black hole microstates is the Ramond-Ramond (RR) sector. We begin with RR ground states. A strand of length $\k$ in a RR ground state is denoted by $\ket{s}_\k\equiv \ket{m,\bar m}_\k$ where\footnote{These are only 5 of the 16 RR ground states at twist $\k$: we restrict to these as we are interested in bosonic states which are invariant under rotations of the internal manifold $\mathcal{M}$.} $(m,\bar m)$ take values $(\pm,\pm)$ or $(0,0)$. The labels $(m,\bar m)$ denote the respective eigenvalues of the $SU(2)_L \times SU(2)_R$ currents $J_0^3, \bar J_0^3$ on the state $\ket{s}_\k$. 

These ground states are generated by acting on the strand $\ket{++}_\k$ with the zero modes of the untwisted chiral primaries: when $\k = 1$, for example, one has
\begin{equation}
J^-_0\ket{++}_1=\ket{-+}_1\, ,\quad~ \bar J^-_0\ket{++}_1=\ket{+-}_1\, ,\quad~ O^{--}_0\ket{++}_1=\ket{00}_1\, ,\quad~ \Omega_0^{--}\ket{++}_1=\ket{--}_1\;.
\end{equation}

A basis for the RR ground states is given by taking tensor products of $N_\k^{(s)}$ strands of type $\ket{s}_\k$, yielding the following eigenstates of the R-symmetry currents:
\begin{equation}\label{eq:RRstate}
\psi_{\{ N_\k^{(s)}\}} \equiv \prod_{\k,s} (\ket{s}_\k)^{N_\k^{(s)}}\,,
\end{equation}
subject to the ``strand budget'' constraint
\begin{equation}\label{eq:windingconstraint}
\sum_{\k,s} \k N_\k^{(s)} = N\,.
\end{equation}

It is convenient to work with non-normalized states. Due to this choice, the norm of the state is required when computing 3-pt functions. The norm of the state \eqref{eq:RRstate} is defined as the number of inequivalent elements that belong to the conjugacy class of $S_N$ defined by the partition \eqref{eq:windingconstraint}; it was derived in \cite{Giusto:2015dfa} and is given by
\begin{equation}\label{eq:norm}
 \left |\psi_{\{ N_\k^{(s)}\}}\right |^2 = \frac{N!}{\prod_{\k,s} N_\k^{(s)}!\, \k^{N_\k^{(s)}}}\;.
\end{equation}
The two-charge black hole microstates that are well-described in the supergravity limit are coherent states~\cite{Kanitscheider:2006zf} that are linear combinations of the $\psi_{\{ N_\k^{(s)}\}}$, weighted by coefficients $A_\k^{(s)}$
\begin{equation}\label{eq:coherent}
\psi(\{ A_\k^{(s)} \}) \equiv {\sum_{\{ N_\k^{(s)} \}}}^{\!\!\!\prime} \prod_{\k,s} \big(A_\k^{(s)} \ket{s}_\k\big)^{N_\k^{(s)}}\,,
\end{equation}
such that the sum is peaked for large values of $N_\k^{(s)}$; here the prime on the summation symbol indicates that the sum is subject to the constraint \eqref{eq:windingconstraint}.
The $A_\k^{(s)}$ can in general be complex, however for our practical applications in this paper we will take them to be real. The average number $\overline  N_\k^{(s)}$ of strands of type $\ket{s}_\k$ in the semi-classical limit is fixed by the coefficients $A_\k^{(s)}$. This relation is determined by calculating the norm of \eqref{eq:coherent} using \eqref{eq:norm}, and then taking its variation with respect to the $N_\k^{(s)}$: this leads to \cite{Giusto:2015dfa}
\begin{equation}\label{eq:average}
\k \,\overline{N}_\k^{(s)} \,=\, |A_\k^{(s)}|^2\,.
\end{equation}
Comparison with the strand budget constraint \eqref{eq:windingconstraint}  implies the relation
\begin{equation}\label{eq:windingconstraintbis}
\sum_{\k,s}|A_\k^{(s)}|^2\,=\,N\,.
\end{equation}

We now introduce the states that are proposed to be holographically dual to  superstrata.
To do so we will make use of the spectral flow transformation in the $\mathcal{N}=(4,4)$ superconformal algebra of the  D1-D5 CFT (see e.g.~\cite{Avery:2010qw,Giusto:2012yz, Chakrabarty:2015foa} for more details and~\cite{Shigemori:2021pir} for a very recent application), which connects the RR and the NS-NS sectors of the theory. In particular, it defines a map between RR ground states and anti-chiral primaries in the NS sector. Although the CFT states dual to superstrata are in the RR sector of the theory, it is convenient for ease of notation and of the computations to introduce their corresponding states in the NS-NS sector.

The momentum-carrying building blocks of the 1/8-BPS states dual to superstrata are the descendant states obtained by acting upon the anti-chiral primary $\ket{O^{--}_\k}_{\NS}$, which corresponds to the RR ground state $\ket{00}_\k$, with the holomorphic generators of the small $\mathcal{N}=4$ superconformal algebra $J^+_0$, $L_{-1}$ and $G^{+A}_{-\frac12}$. We will denote these by $\ket{\k,\m,\n,\q}$, and one has~\cite{Bena:2015bea,Bena:2016agb,Bena:2016ypk,Bena:2017xbt,Ceplak:2018pws,Heidmann:2019zws}
\bea \label{eq:cft-states-2}
\!\! |\k,\m,\n,\q\rangle &\!\!=\!\!& \frac{1}{(\m-\q)!(\n-\q)!}
  (J^+_0)^{\m-\q} L_{-1}^{\n-\q}  \left(G_{-\frac12}^{+1}G_{-\frac12}^{+2} + \frac{1}{\k} J^+_0 L_{-1}\right)^\q  |O^{--}_\k\rangle_{\NS} \,,
\eea
where the parameters can take values $\q=0$ or $1,\;$ $\n \ge \q,\;$ $\k >0,\;$ and $\q\le \m\le \k-\q$~\cite{Ceplak:2018pws} (our notation follows~\cite{Heidmann:2019zws}).
As discussed around Eq.\;\eqref{eq:cft-states}, we describe states with $\q=1$ as supercharged states.

We are interested in superstrata whose CFT dual state is made up of strands of type $\ket{\k,\m,\n,\q}$ and the NS-NS vacuum $\ket{0}_1$: a basis for these states is given by the following eigenstates of the $SU(2)_L\times SU(2)_R$ currents
\begin{equation}\label{basis CFT states}
\psi_{\{N_0, N_i \}}\equiv  \ket{0}_1^{N_0}\prod_i  \ket{\k_i,\m_i,\n_i,\q_i }^{N_i}
\end{equation}
subject to the constraint that the total number of copies must saturate the strand budget:
\begin{equation}\label{strand budget}
N_0+\sum_i \k_i N_i\;=\;N \,.
\end{equation}
Again, in order to construct states whose holographic duals are well described in supergravity, one must consider coherent states obtained superposing the $\psi_{\{N_0, N_i \}}$, weighted by coefficients $A, B_{i}$
\begin{equation}\label{coherent superposition superstrata}
\psi(\{A,B_{i}\}) \;\equiv\; {\sum_{\{N_0,N_i\}}}^{\!\!\!\!\!\!\prime} \,\,
\left[\;\prod_i \Big(A\ket{0}_1\Big)^{N_0}  \Big(B_i \ket{\k_i,\m_i,\n_i,\q_i }\Big)^{N_i} 
\right]\, .
\end{equation}
where the prime symbol denotes that the sum is subject to the constraint~\eqref{strand budget}. Again the coefficients $A, B_{i}$ can in principle be complex, however we shall take them to be real in all our examples.

In the remainder of the paper we will work with NS-NS sector states unless otherwise specified. As a result, for ease of notation in the following we will mostly suppress the subscript $\NS$ on the states.

\subsection{Norm of supercharged superstrata CFT states} \label{sec:norm}

In this section we compute the norm of the states~\eqref{coherent superposition superstrata}, for use later in the paper. This computation is new and 
generalizes the discussion in~\cite{Giusto:2015dfa} and~\cite{Bena:2017xbt} to the states~\eqref{coherent superposition superstrata}.

We do so by first computing the norm of the building-block states~\eqref{basis CFT states}. We then determine the average numbers $\{\bar{N}_0,\bar{N}_i\}$ over which the sum in the full coherent state~\eqref{coherent superposition superstrata} is peaked.
The norm of~\eqref{basis CFT states} is obtained combining the combinatorial contribution in \eqref{eq:norm} with the effects of the momentum carrying excitations on the norm of each strand.
In order to determine the norm of a single strand $\ket{\k,\m,\n,\q}$, we will make use of the algebra Eq.\;\eqref{anomaly free algebra}, together with the fact that an anti-chiral primary is annihilated by $G^{-A}_{-1/2}$ and $J^{-}_0$ (in addition to all positive modes of the anomaly-free subalgebra), plus the following relations on hermitian conjugation:
\begin{equation}
(G^{\alpha A}_n)^\dagger=-\epsilon_{\alpha\beta}\epsilon_{AB}G^{\beta B}_{-n}\,,\qquad (J^a_n)^{\dagger}=J^{-a}_{-n}\,,\qquad (L_{n})^{\dagger}=L_{-n} \,.
\end{equation}

We start by considering $\m=\n=\q=1$, which gives:
\begin{equation}\label{norm k111}
\braket{\k,1,1,1}{\k,1,1,1}\;=\;\k^2-1 \,.
\end{equation}
In order to compute the contribution to the norm coming from the contractions of the $J^+_0$ insertions, it is useful to consider the following state, using the shorthand $\hat{\m}=\m-\q$:
\begin{equation}
\begin{aligned}
J^-_0\big(J^+_0\big)^{\hat{\m}}&\Big(G_{-\frac12}^{+1}G_{-\frac12}^{+2} + \frac{1}{\k} J^+_0 L_{-1}\Big)^\q\ket{O^{--}_\k}\\ &=
\Big(-2(\hat \m-1)+\k-2q+J^+_0J^-_0\Big)\big(J^+_0\big)^{\hat \m-1}\Big(G_{-\frac12}^{+1}G_{-\frac12}^{+2} + \frac{1}{\k} J^+_0 L_{-1}\Big)^\q\ket{O^{--}_\k}
\\ &=\,
\hat \m\Big(\k-\hat \m+1-2\q\Big)\big(J^+_0\big)^{\hat{\m-1}}\Big(G_{-\frac12}^{+1}G_{-\frac12}^{+2} + \frac{1}{\k} J^+_0 L_{-1}\Big)^\q\ket{O^{--}_\k} \,.
\end{aligned}
\end{equation}
Here we have used the relations $J^3_0\ket{\k,\q,\q,\q}=(-\frac{\k}{2}+\q)\ket{\k,\q,\q,\q}$ and $J^-_0\ket{\k,\q,\q,\q}=0$.\\
Iterating this procedure and using \eqref{norm k111} one obtains
\begin{equation}\label{Norm k,m,q,q}
\braket{\k,\m,\q,\q}{\k,\m,\q,\q}\;=\;{\k-2\q\choose \m-\q}(\k^2-1)^\q \,.
\end{equation}

We proceed similarly for the Virasoro generators, using $\hat \n=\n-\q$:
\begin{equation}
\begin{aligned}
L_1\big(L_{-1}\big)^{\hat \n}&\Big(G_{-\frac12}^{+1}G_{-\frac12}^{+2} + \frac{1}{\k} J^+_0 L_{-1}\Big)^\q\ket{O^{--}_\k}\\ &=\,
\hat{\n} \big(\k+\hat \n-1+2\q\big)\Big(G_{-\frac12}^{+1}G_{-\frac12}^{+2} + \frac{1}{\k} J^+_0 L_{-1}\Big)^\q\ket{O^{--}_\k} \,.
\end{aligned}
\end{equation}
Here we have used the relations $L_0\ket{\k,\q,\q,\q}=(\frac{\k}{2}+\q)\ket{\k,\q,\q,\q}$ and $L_1\ket{\k,\q,\q,\q}=0$.  \\
Iterating this procedure and using \eqref{norm k111} one obtains
\begin{equation}\label{Norm k,q,n,q}
\braket{\k,\q,\n,\q}{\k,\q,\n,\q}\;=\;{\k+\n+\q-1\choose \n-\q}(\k^2-1)^\q \,.
\end{equation}
Since the Virasoro generators commute with $J_0^{\pm}$, we can directly combine the results in~\eqref{norm k111},~\eqref{Norm k,m,q,q} and~\eqref{Norm k,q,n,q} to obtain
\begin{equation}\label{Norm kmnq}
\braket{\k,\m,\n,\q}{\k,\m,\n,\q}={\k-2\q\choose \m-\q}{\k+\n+\q-1\choose \n-\q}(\k^2-1)^\q \,.
\end{equation}
This result, along with~\eqref{eq:norm}, gives the norm of the building-block state \eqref{basis CFT states},
\begin{equation}\label{Norm cft basis superstrata}
|\psi_{\{N_0, N_i \}}|^2=
\frac{N!}{N_0!}\prod_{i}\frac{1}{N_{i}!}\Big[\frac{\big(\k_i^2-1\big)^{\q_i}}{\k_i}{\k_i-2\q_i \choose \m_i-\q_i}{\k_i+\n_i+\q_i-1\choose \n_i-\q_i}
\Big]^{N_{i}}\,.
\end{equation}
This implies that the norm of the full coherent state \eqref{coherent superposition superstrata} is
\begin{equation}\label{coherent superposition superstrata 2}
|\psi(\{A,B_{i}\})|^2 \;\equiv\; {\sum_{\{N_0,N_i\}}}^{\!\!\!\!\!\!\prime} \,\,
\frac{N!}{N_0!}|A|^{2N_0}\prod_{i}\frac{1}{N_{i}!}\Big[|B_i^2|\frac{\big(\k_i^2-1\big)^{\q_i}}{\k_i}{\k_i-2\q_i \choose \m_i-\q_i}{\k_i+\n_i+\q_i-1\choose \n_i-\q_i}
\Big]^{N_{i}} \,.
\end{equation}

We now determine the average number of strands in the coherent state by requiring that the variation of the summand of~\eqref{coherent superposition superstrata 2} with respect to ${N_0,N_i}$ vanishes~\cite{Giusto:2015dfa}, obtaining
\begin{equation}\label{q=1 peak AB}
\bar N_0=|A|^2 \qquad \k_i\bar N_{i}=\big(\k_i^2-1\big)^{\q_i}{\k_i-2\q_i \choose \m_i-\q_i}{\k_i+\n_i+\q_i-1\choose \n_i-\q_i}|B_{i}|^2\,.
\end{equation}
This equation, combined with the strand budget constraint~\eqref{strand budget}, implies
\begin{equation}\label{strand budget A Bi}
 |A|^2+\sum_{i}\big(\k_i^2-1\big)^{\q_i}{\k_i-2\q_i \choose \m_i-\q_i}{\k_i+\n_i+\q_i-1\choose \n_i-\q_i}|B_{i}|^2=N \,.
 \end{equation}

\section{Supergravity and Superstrata}\label{sec:new-sec-3}

In this section we briefly review the supergravity theory in which we work, the superstratum solutions we study, and the Kaluza-Klein spectrum of the six-dimensional supergravity theory reduced on $S^3$.

\subsection{Six-dimensional supergravity fields}

The D1-D5 system admits an AdS$_3$ decoupling limit leading to configurations that have AdS$_3\times$ S$^3\times \mathcal{M}$ asymptotics~\cite{Maldacena:1997re}. In the limit in which the internal manifold $\mathcal{M}$ is microscopic, dimensional reduction of the 10-dimensional theory on $\cM$ gives a $D=6$ supergravity theory with $n$ tensor multiplets\footnote{Some details of the theory change depending on whether the internal manifold is $T^4$ or K3. In the first case the number of supersymmetries is $\mathcal{N}=(2,2)$, while in the other one has $\mathcal{N}=(2,0)$. Moreover, $n=5$ when the internal manifold is $T^4$ while $n=21$ when $\mathcal{M}=K3$.}, whose equations of motion were first derived by Romans in \cite{Romans:1986er}. The bosonic field content of the theory is as follows: a graviton $g_{MN}$ ($M,N=0,...,5$ are curved 6D indices), 5 2-forms whose field strengths $H^m$ are selfdual ($m=1,...,5$ is a vector index of $SO(5)$), $n$ 2-forms whose field strengths $H^r$ are anti-selfdual ($r=6,...,n+5$ is a vector index of $SO(n)$) and $5n$ scalars $\phi^{mr}$.
Dimensional reduction on the $T^4$ also gives rise to 16 vectors: their CFT duals, however, belong to the short multiplets of fermionic CPOs and we shall not consider them further in the present work.
The scalars live in the coset space $SO(5,n)/(SO(5)\times SO(n))$, which can be parametrized by vielbeins $(V^m_I,V^r_I)$ where $I=(m,r)$ is an $SO(5,n)$ vector index. We also introduce field strengths $G^I$ which are related to the selfdual and anti-selfdual field strengths through the vielbeins via $H^m=G^IV^m_I$ and $H^r=G^IV_I^r$. In order to support the global AdS$_3\times $S$^3$ vacuum, one must turn on one of the fluxes, which we will take to be  $H^{m=5}$.
We parametrize fluctuations of the six-dimensional supergravity fields around the AdS$_3\times $S$^3$ background as follows: 
\begin{equation}\label{linearization aroung background}
g_{MN}=g^{0}_{MN}+h_{MN}\,,\quad G^{(A)}=g^{0(A)}+g^{(A)}\,,\quad V^m_I=\delta_I^m+\phi^{(mr)}\delta^r_I\,,\quad V^r_I=\delta_I^r+\phi^{(mr)}\delta^m_I \,.
\end{equation} 
The explicit expression for the background fields $g^{0}_{MN}$ and $g^{0(A)}$ is given in Eq.~\eqref{background fields} below.

\subsection{Superstrata}
\label{sec:superstrata-review}

In this subsection we briefly review certain aspects of superstrata that will be relevant to the remainder of the paper.

Superstrata are supersymmetric supergravity solutions in which the isometries preserved by the corresponding black hole solution are broken by momentum-carrying waves, see e.g.~\cite{Bena:2015bea,Bena:2016agb,Bena:2016ypk,Bena:2017geu,Bena:2017xbt,Bena:2017upb,Ceplak:2018pws,Heidmann:2019zws,Heidmann:2019xrd,Mayerson:2020tcl,Shigemori:2020yuo}. These include the first families of smooth horizonless solutions with large BTZ-like AdS$_2$ throats, general angular momentum, and identified holographic duals in the AdS$_3$ limit~\cite{Bena:2016ypk,Bena:2017xbt}. In the D1-D5-P frame, these are typically constructed in the context of the general 1/8-BPS ansatz of Type IIB supergravity that carries D1, D5, P charges and is invariant on $\cM$. This was derived in~\cite{Giusto:2013rxa} and is reproduced in Appendix~\ref{app:sugra} for completeness.

Upon reduction to $6D$, this ansatz gives rise to minimal 6D supergravity coupled to $n=2$ tensor multiplets, which we will take to be labelled by $r=6,7$. The main interest of this work will be the holographic dictionary involving the field strength $G^6$ and $G^7$. Let us therefore discuss the relation between these fields and those given in Appendix \ref{app:sugra}. First, the 6D metric takes the form
\begin{equation}\label{metric superstrata}
ds_6^2=-\frac{2}{\sqrt{\mathcal{P}}}(dv+\betab)(du+\omega+\frac{\mathcal{F}}{2}(dv+\betab))+\sqrt{\mathcal{P}}ds^2_4 \,.
\end{equation}
We first define the following three-form field strengths (here and until the end of the subsection we use $a,b=1,2,4$)
\begin{equation}\label{3form G1G2G4}
G^{a}=d\Big[ -\frac{1}{2}\frac{\eta^{ab}Z_b}{P}(du+\omega)\wedge(dv+\betab)\Big]+\frac12 \eta^{ab}\star_4DZ_b+\frac{1}{2}(dv+\betab)\wedge\Theta^{a}
\end{equation}
where
\begin{equation}
\eta_{12}=\eta_{21}=-\eta_{44}=1 \,, \qquad P=Z_1Z_2-Z_4^2 \,.
\end{equation}
The field strengths $G^a$ respect the self-duality condition:
\begin{equation}
\star_6G^{a}=\T{M}{a}{b}G^b, \qquad M_{ab}=\frac{Z_aZ_b}{P}-\eta_{ab}.
\end{equation}
The $G^a$ arise from the dimensional reduction of the type IIB field strengths of $C_2,C_6$ and $B$, see App.~\ref{app:sugra}. 
The relation between these 3-forms and the field strengths $G^5, G^6, G^7$ introduced above was derived in~\cite{Kanitscheider:2007wq} and in our conventions is given by
\begin{equation}\label{relation G124 G567} 
G^5=\frac{Q_1 G^1+Q_5 G^2}{2Q_1 Q_5} \,,\qquad G^6=-\frac{Q_1 G^1-Q_5 G^2}{2Q_1 Q_5}\,,\qquad G^7=\frac{1}{\sqrt{Q_1Q_5}}G^4 \,.
\end{equation}
The scalar fields $\phi^{(56)}$ and $\phi^{(57)}$ arise from the dilaton, $C_0$, and the component of $C_4$ with all legs on $\cM$. These can be obtained from the vielbein matrix in \cite[Eqs.\;(3.33),\,(B.20)]{Kanitscheider:2007wq}, along with Eq.~\eqref{linearization aroung background}. In our conventions they take the form
\begin{equation}\label{phi56 phi57}
\phi^{(56)}=\frac{1}{2\sqrt{Q_1Q_5}}\left(\frac{Q_5Z_1-Q_1Z_2}{\sqrt{Z_1Z_2}}\right)\,,\qquad~~ \phi^{(57)}=\frac{Z_4}{\sqrt{Z_1Z_2}} \,.
\end{equation}

The general structure of superstratum solutions is as follows. The construction begins with a seed solution which is usually taken to be a circular supertube~\cite{Balasubramanian:2000rt,Maldacena:2000dr} with characteristic length-scale $a$. Momentum-carrying waves are added by a linear superposition of terms within the linear system of BPS equations, specifically at the level of the ``first layer'' recorded in Eq.\;\eqref{eq:firstlayer}. These momentum-carrying waves come with a set of  dimensionful Fourier coefficients $b_i$. This construction is designed to correspond to the structure of the CFT states in Eq.~\eqref{coherent superposition superstrata}. For further details, see e.g.~\cite{Bena:2017xbt,Heidmann:2019zws}.

Smoothness of the supergravity solutions imposes the relation~\cite[Eq.\;(4.13)--(4.14)]{Heidmann:2019zws}
\begin{equation}\label{regularity constraint a b}
\frac{Q_1Q_5}{R^2}=a^2+\sum_i\frac{ b^2_i}{2}\hat{x}_{i}\,, \qquad
\hat{x}_{i}=({k_i-2q_i\choose m_i-q_i}{k_i+n_i+q_i-1\choose n_i-q_i}(k_i^2-1))^{-1}
\end{equation}
which has the same form of the CFT strand budget constraints~\eqref{strand budget}, \eqref{strand budget A Bi}, and indeed this is no accident.
By comparing the two, the proposed superstratum holographic dictionary involves the following map between the CFT coefficients $A,B_{i}$ and the supergravity coefficients $a$ $b_i$:
\begin{equation}
\begin{aligned}
\label{q=1 AB ab}
\frac{A}{\sqrt{N}}&\;=\;R\sqrt{\frac{1}{Q_1Q_5}}a\;\equiv\;\mathsf{a}\,, \\ \frac{B_{i}}{\sqrt{N}}&\;=\; R\sqrt{\frac{1}{2Q_1Q_5}}\Big({k_i-2q_i\choose m_i-q_i}{k_i+n_i+q_i-1\choose n_i-q_i}(k^2-1)^q\Big)^{-1} b_{i}\\&\;=\;
\sqrt{\frac{1}{2}}\Big({k_i-2q_i\choose m_i-q_i}{k_i+n_i+q_i-1\choose n_i-q_i}(k^2-1)^q\Big)^{-1} \mathsf{b}_{i} \,,
\end{aligned}
\end{equation}
where we have also defined the quantities $\mathsf{a}$, $\mathsf{b}$ which will be used later in the paper.

\subsection{Kaluza-Klein spectrum}\label{KK spectrum and single-particle states}

In order to discuss the Kaluza-Klein spectrum of the 6D theory compactified on the $S^3$~\cite{Deger:1998nm,deBoer:1998ip} (see also e.g.~\cite{Kanitscheider:2006zf,Ceplak:2018pws}), we must expand the six-dimensional fluctuations in harmonics of $S^3$. Before doing this, however, it is convenient to perform the following rescalings:
\begin{align}\label{transformation standard ads}
r\rightarrow a_0\, \tilde{r}\,,\quad t\rightarrow R_y\, \tilde t\,,\quad y\rightarrow R_y\, \tilde y \,,\quad \beta\rightarrow R_y\, \tilde{\beta}\,,\quad \omega\rightarrow R_y\, \tilde{\omega}\,,\quad Z_i\rightarrow  \frac{\tilde{Z_i}}{a_0^2}\,,
\end{align}
with $a_0^2\equiv \frac{Q_1Q_5}{R_y^2}$. We then reabsorb the overall scale factor $\sqrt{Q_1Q_5}$ in the metric\footnote{This step allows us to use Eq.~\eqref{ads fields mass matrix diag} in the form given.} to obtain
\begin{equation}\label{background fields}
\begin{aligned}
g^{(0)}_{MN}&=\frac{d\tilde r^2}{\tilde r^2+1}-(\tilde r^2+1)d\tilde t^2+\tilde r^2d\tilde y^2+d\theta^2+\sin^2\theta d\phi^2+\cos^2\theta d\psi^2\,,\\
g^{0(A=5)}&= \cos\theta\sin\theta d\phi\wedge d\psi\wedge d\theta-\tilde r d\tilde r\wedge d\tilde t\wedge d\tilde y\,, \qquad g^{0(A\neq 5)}=0\,.
\end{aligned}
\end{equation}

We now introduce a multi-index $I$ for the $S^3$ harmonic degree $k$ and $(j_3,\bar{j}_3)$ quantum numbers, $I=(k,m,\bar{m})$. In some places we will write $k$ explicitly, and continue to use $I$ for the remaining quantum numbers $(m,\bar{m})$. Harmonics on S$^3$ and AdS$_3$ are reviewed in Appendix \ref{sec:harm S3 and AdS3}.
We split the 6D curved indices into AdS$_3$ indices $\mu,\nu=0,1,2$ and $S^3$ indices $a,b=1,2,3$. The subscript $(ab)$ denotes the symmetric traceless component of the field.  
Then expanding the six-dimensional fluctuations in harmonics of $S^3$, one obtains~\cite{Deger:1998nm,Kanitscheider:2006zf}
\begin{equation}\label{0611 geom harmonics expansion}
\begin{aligned}
h_{\mu\nu}&=\sum h^I_{\mu\nu}Y^I\\
h_{\mu a}&=\sum h^{I}_{(v)\mu}Y^I_a+h^{I}_{(s)\mu}D_aY^I\\
h_{(ab)}&=\sum \rho^IY^I_{(ab)}+\rho^{I}_{(v)}D_aY^I_b+\rho^{I}_{(s)}D_{(a}D_{b)}Y^I\\
h^a_{a}&=\sum \pi^IY^I\\
g^A_{\mu\nu\rho}&=\sum 3D_{[\mu}b^{(A)I}_{\nu\rho]}Y^I\\
g^A_{\mu\nu a}&=\sum b^{(A)I}_{\mu\nu}D_aY^I+2D_{[\mu}Z_{\nu]}^{(A)I}Y_a^{I}\\
g^A_{\mu a b}&=\sum D_\mu U^{(A)I}\epsilon_{abc}D^cY^I+2Z^{(A)I}_\mu D_{[b}Y^I_{a]}\\
g^A_{a b c}&=-\sum \epsilon_{abc} \Lambda^I U^{(A)I}Y^I\\
\phi^{mr}&=\sum \phi^{(mr)I}Y^I \,.
\end{aligned}
\end{equation}
In what follows, the main focus will be on $Z^{(6)}$ and $Z^{(7)}$.


\section{Constructing the supercharged holographic dictionary} \label{sec:dict-1}

In this section we review the single-particle basis and use it derive the correspondence between the operators of interest and their dual bulk fields. 
In Section \ref{sec:spo basis} we discuss how the single-particle basis can be used to determine most, but not all, of the mixing between single and multi-trace operators from CFT arguments alone.
In Section \ref{ssc:gauge invariance} we compute the gauge-invariant fluctuations of supergravity fields that are dual to the CFT operators of interest.
In Section \ref{duality fields operators} we derive the holographic map in the sector we study, in the single-particle basis. This involves resolving the operator mixing by combining and refining the results of~\cite{Giusto:2019qig} and \cite{Taylor:2007hs}. 
In Section \ref{ssc:superchaarged cft operators} we then generate the superdescendants within this supermultiplet that we use in the remainder of the paper. 
In Section~\ref{ssc:Holographic dictionary for scalar SPOs} we record the explicit holographic dictionary in the single-particle basis for convenient reference.

\subsection{Single-particle operator basis}\label{sec:spo basis}

AdS/CFT duality relates AdS fields and boundary CFT operators through a matching of the observables of the theories; we shall focus on protected correlation functions, which can be compared between supergravity and the orbifold CFT. 

On the bulk side, correlation functions can be computed by dimensionally reducing the 6D Lagrangian on S$^3$: the AdS$_3$ action takes the schematic form
\begin{equation}
S_{AdS_3}\sim \int_{AdS_3} \Big( \mathcal{L}_2+\mathcal{L}_3+\cdots \Big)
\end{equation}
where $\mathcal{L}_{n}$ contains the interactions between $n$ KK modes and is relevant for computing $n$-point functions and higher.

The first ingredient in the holographic dictionary is the identification of the quantum numbers of the fields and the dual operators.
With linear field redefinitions one can diagonalize the quadratic term $\mathcal{L}_2$, and thus identify the quantum numbers that the CFT operators dual to each supergravity field must carry. 

On the CFT side, however, there are degeneracies: in general there are single and multi-trace operators with the same quantum numbers. A further complication in the AdS$_3$ case (which  is absent, for example,  in the long-studied case of AdS$_5$) comes from the fact that there are degeneracies also between the single-traces: as discussed in Section \ref{Low dimensional operators}, the chiral primaries $\Sigma_{3}$ and $\Omega$ cannot be distinguished by their quantum numbers. 

In order to identify the mixing matrix, one must analyze the three-point functions on both sides of the duality \cite{Jevicki:1998bm,Lunin:2001pw,Mihailescu_2000,Arutyunov:2000by}. On the gravity side, these are generated by considering the cubic Lagrangian $\mathcal{L}_3$.
The cubic Lagrangian was derived in~\cite{Arutyunov:2000by} and a priori involves derivative couplings. It was shown, however, that the derivative couplings can be reabsorbed via a non-linear redefinition of the fields. 
While this transformation does not change non-extremal 3-point functions, it has been emphasized in~\cite{DHoker:1999jke,Arutyunov:2000ima,Rastelli:2017udc} that extremal 3-point functions require special attention. An important fact is that non-derivative extremal cubic couplings vanish. This is no coincidence: in the extremal case, the spacetime integral that occurs in the Witten diagram diverges, so the extremal coupling must vanish in order to avoid a divergence in the value of the correlator.
The extremal correlator can still gain contributions from the derivative couplings in the Lagrangian, since they give rise to boundary terms upon partial integration~\cite{DHoker:1999jke}. The field redefinition, however, removes the derivative terms and thus all extremal three-point functions vanish. 

The bulk field redefinition is interpreted on the CFT side as a change of basis~\cite{Arutyunov:2000ima}, since it amounts to forming an admixture between the operator dual to the original field and certain multi-trace operators. By AdS/CFT, in this basis all CFT extremal three-point functions vanish.
Let us denote by $\Phi^i$ the AdS field with respect to which the Lagrangian contains derivative terms, and let us denote its CFT dual by $\mathcal{O}^{\Delta_i}$. Then the operator $\tilde{\mathcal{O}}^{\Delta_i}$ dual to the redefined field $\tilde\Phi^i$ will take the form: $\tilde{\mathcal{O}}^{\Delta_i}=\mathcal{O}^{\Delta_i}+\frac{1}{\sqrt{N}}\sum_k c_{ik}\mathcal{O}^{\Delta_i-\Delta_k}\mathcal{O}^{\Delta_k}+...$ where the ellipses denote other double-traces or higher multi-trace operators, the only constraint being that they must have the same quantum numbers as the operator $\mathcal{O}^{\Delta_i}$. Note that the operators $\mathcal{O}^\Delta_i$ and $\tilde{\mathcal{O}}^\Delta_i$ coincide if $\Delta_i=1$, as at dimension 1 the spectrum of the theory consists only of single-trace operators.

In generic correlators, the contribution of the double-trace operators to the correlator is subleading in the $1/N$ expansion (see e.g.~the discussions in~\cite{DHoker:1999jke,Taylor:2007hs}). This is the CFT version of the bulk statement that the field redefinition $\Phi^i\rightarrow\tilde{\Phi}^i$ leaves non-extremal correlators unchanged.
However for certain correlators, the double-traces contribute at leading order in large $N$. This happens in extremal correlators and also in certain (non-extremal) mixed heavy-light correlators. In this paper we are interested in precisely such mixed heavy-light correlators.

In recent work it was proposed that single-particle supergravity excitations around global AdS$_5 \times$ S$^5$ are dual to CFT operators (in short multiplets) that are orthogonal to all multi-trace operators~\cite{Aprile:2018efk}. This was then extensively used in~\cite{Aprile:2020uxk} to discuss the properties of the single-particle operator basis in free $\mathcal{N}=4$ SYM.\footnote{See also~\cite{Yang:2021kot} for further discussion.}

We now argue that the redefinition that removes cubic couplings gives rise to precisely the same set of single-particle CFT operators defined as those that are orthogonal to all multi-trace operators. We follow in part a discussion in~\cite{DHoker:1999jke}. First of all, we recall that conformal symmetry implies that a two-point function can be non-zero only if the two operators have the same dimension. To show that an operator with dimension $k_1$ is a single-particle operator, we must therefore show that it has vanishing two point function with all multi-traces $\big (\mathcal{O}_{k_2}\mathcal{O}_{k_3}\big)(z)$ such that $k_1=k_2+k_3$.
The first non-singular term in the OPE $\mathcal{O}_{k_2}(z_2)\mathcal{O}_{k_3}(z_3)$, with coefficient one, is the multi-trace $\big (\mathcal{O}_{k_2}\mathcal{O}_{k_3}\big)(z_2)$.
Now consider the extremal ($k_1=k_2+k_3$) three-point function (the coefficient $c$ is zero in this basis however we keep it for convenience)
\begin{equation}\label{extremal 3pt general}
\langle \mathcal{O}_{k_1}(z_1)\mathcal{O}_{k_2}(z_2)\mathcal{O}_{k_3}(z_3)\rangle\,=\,\frac{c}{(z_1-z_2)^{2k_2}(z_1-z_3)^{2k_3}}\,.
\end{equation}
By taking the $z_2\rightarrow z_3$ limit (which is smooth at extremality) one obtains
\begin{equation}\label{2 pt from limit of 3 pt}
\langle \mathcal{O}_{k_1}(z_1)\big(\mathcal{O}_{k_2}\mathcal{O}_{k_3}\big)(z_2)\rangle\,=\,\frac{c}{(z_1-z_2)^{2k_1}}\,.
\end{equation}
This shows the equivalence of the single-particle CFT basis of~\cite{Aprile:2018efk,Aprile:2020uxk} and the CFT basis dual to supergravity fields with derivative cubic couplings removed.
The vanishing of the three-point coefficient $c$ in~\eqref{extremal 3pt general} implies the orthogonality between the operator $ \mathcal{O}_{k_1}$ and all multi-particle operators in Eq.~\eqref{2 pt from limit of 3 pt}, and vice versa. This discussion motivates the change in the definition of double-trace operators with respect to the convention in \cite{Giusto:2019qig} (see footnote \ref{footnote definition multitrace}).\\[-2mm]

Before proceeding to our analysis let us make a few comments on the derivation of the holographic dictionary for scalar operators of dimension two in~\cite{Giusto:2019qig}.
This work did not use the single-particle basis, however we shall use the single-particle basis in the present work, so let us describe the difference in the two approaches.

The method used in~\cite{Giusto:2019qig} was as follows.
First, the most general linear combination of single and double-trace operators allowed by the quantum numbers was worked out (higher multi-trace operators are trivially absent at dimension two). 
Then a set of different backgrounds were considered, for which there was already a well-established holographic description (the two-charge Lunin-Mathur solutions~\cite{Lunin:2001jy}).
CFT expectation values of these light fields in a selection of these heavy states were then matched to the expansion of the dual bulk fields identified in~\cite{Kanitscheider:2006zf} and~\cite{Kanitscheider:2007wq}.
By considering an exhaustive set of examples, 
the combinations of single and multi-trace operators in the CFT dual to certain supergravity fluctuations were fixed, as were the overall normalization coefficients of the holographic dictionary in this sector.

In the present paper we work in the single-particle basis. In this basis, the identification of the single-particle operators partially reduces to the identification of the operators that have the  property that all their extremal three point functions vanish\footnote{The authors thank Stefano Giusto and Rodolfo Russo for a discussion on this point.}. This is purely a  CFT computation, which works as an input in constructing the holographic dictionary. Importantly, this does not resolve all the mixing, as we shall discuss shortly.

\subsection{Gauge-invariant combinations of supergravity fields}\label{ssc:gauge invariance}

We now determine the gauge-invariant combinations of supergravity fluctuations that are dual to the operators we consider.

Not all the fluctuations in the KK harmonic expansion~\eqref{0611 geom harmonics expansion} are independent: some of them are connected to the background fields or to other fluctuations through coordinate transformations that tend to zero at infinity.
For instance, consider the vacuum state, which corresponds to empty global AdS$_3\times $S$^3$. If one performs such a change of coordinates, one can turn on some of the AdS$_3$ fields in the KK harmonic expansion~\eqref{0611 geom harmonics expansion}. Of course these are not physical excitations and there are no boundary operators that source them.

In the study of the KK spectrum in~\cite{Deger:1998nm}, the authors dealt with this redundancy by fixing the (de Donder) gauge. Here we follow instead the gauge-invariant KK reduction method developed in~\cite{Skenderis:2006uy}. The strategy is to organize the AdS$_3$ fields in combinations that have the correct transformation properties under a gauge transformation.

The coordinate transformation
\begin{equation}
x^M\rightarrow x'^M=x^M-\xi^M
\end{equation}
generates a perturbation of the metric and 3-form which, up to linear order in the gauge parameter, reads
\begin{equation}\label{Lie derivatives transformations}
\begin{aligned}
\delta h_{MN}&\;=\;D_M\xi_N+D_N\xi_M+D_M\xi^Rh_{RN}+D_N\xi^Rh_{RM}+\xi^RD_Rh_{MN}\,,\\
\delta g^A_{MNP}&\;=\;3D_{[M}\xi^R g^{0A}_{NP]R}+3D_{[M}\xi^R g^{A}_{NP]R}+\xi^RD_{R} g^{A}_{MNP}\,,
\end{aligned}
\end{equation}
where the background value $g^{0A}$ was introduced in  Eq.~\eqref{linearization aroung background}.
In order to deal with the non-linear terms, one must project onto the basis of $S^3$ harmonics.\\
As we will discuss in Section \ref{ssc:Holographic dictionary for scalar SPOs}, to study supercharged superstrata we will need to construct the holographic dictionary for the AdS$_3$ vector fields $Z^{(6)k=1}$ and $Z^{(7)k=1}$.
The operators dual to $Z^{(6)k=1}$ and $Z^{(7)k=1}$ have dimension 3: in principle, we would need the transformation up to second order in the gauge parameter~\cite{McFadden_2011,Bruni_1997}, namely the terms quadratic in  $\xi$ and linear in $g^{0(A)}$. However, since $g^{0(A)}=0$ for $A\neq 5$, these terms do not contribute in the analysis of the vector fields in the tensor multiplets and so the gauge-invariant combinations up to the order we are interested in can be obtained using Eq.~\eqref{Lie derivatives transformations}.
Using the KK spectrum \eqref{0611 geom harmonics expansion} together with the decomposition of $\xi^M$ in harmonics,
\begin{equation}
\xi_\mu=\sum_I \xi_\mu^I Y^I\,, \qquad \xi_a=\sum_{I,J} \xi^I_v Y^I_a+\xi^J_s D_aY^J\,,
\end{equation}
one obtains that under such a diffeomorphism, the second order transformation of $Z^{(6)k=1}$ and $Z^{(7)k=1}$ reads (here $A\neq 5$, so for us $A=6,7$):
\begin{equation}\label{gauge transf Zak1}
\begin{aligned}
\delta Z^{(A)K}_\mu\;=\;& D_\mu U^{(A)J}\big[\xi^I_s(n^{v}_{IJK}+c^{v}_{IJK})+\xi^I_v(p^{v}_{IJK}+g^{v}_{IJK})\big]\\&-\frac{E_{JIK}}{\lambda_k}\big[D_\mu \xi^{I\nu}D_\nu U^{(A)J}+\xi^{I\nu}D_\nu D_\mu U^{(A)J}\big]-\frac{\Lambda_k E_{KIJ}}{\lambda_k^2}\epsilon_{\mu\nu\rho}\xi^{I \nu}D_\rho U^{(A)J}\\&-\frac{\Lambda_j U^{(A)J}}{\lambda_k}\big[D_\mu \xi^{I}_v f_{IJK}+D_\mu \xi^{I}_s E_{IJK} \big]\,,
\end{aligned}
\end{equation}
where the degree $k$ associated with the multi-index $K$ must be equal to 1 in order for the equality to hold. The triple overlap coefficients  $n^{v}_{IJK},c^{v}_{IJK},p^{v}_{IJK},g^{v}_{IJK},E_{IJK}$ are defined in Appendix~\ref{sec: Def and identities projection harmonics}. 

The gauge-invariant combination associated with the ($A\neq 5$) field $Z^{(A),k=1}_\mu$ will take the form $\bm{Z}^{(A),k=1}_\mu=Z^{(A),k=1}_\mu+... \,$, where the ellipses represent fields and product of fields such that their transformation properties compensate those on the right-hand side of \eqref{gauge transf Zak1}, such that the redefined field has the correct transformation properties.
With this aim, we consider linear-order variations of 
\begin{equation}
\begin{aligned}
h_{(ab)}&\;=\;\sum_{I,J,K}\rho_t^IY^I_{ab}+\rho^J_v D_{(a}Y_{b)}^J+\rho^K_s D_{(a}D_{b)}Y^K\,,\\
h_{\mu a}&\;=\;\sum_{I,J}h^{v,I}_\mu Y^I_a+h^{s,J}_\mu D_aY^J\,,
\end{aligned}
\end{equation}
where $(ab)$ denotes symmetric traceless. The transformations read:
\begin{equation}
\begin{aligned}
\delta h_{(ab)}&\;=\;D_a\xi_b+D_b\xi_a=2\xi^I_vD_{(a}Y^I_{b)}+2\xi^J_sD_{(a}D_{b)}Y^J\,,
\\
 \delta h_{\mu a}&\;=\;D_\mu\xi_a+D_a\xi_\mu=D_\mu\xi^I_vY^I_a+(D_\mu\xi^J_s+\xi^J_\mu)D_aY^J\,.
\end{aligned}
\end{equation}
We thus obtain
\begin{equation}
\delta\rho^I_v=2\xi^I_v\,,\qquad \delta\rho^I_s=2\xi^I_s\,,\qquad \delta\hat{h}^{s,I}_\mu=\xi^I_\mu\,,\qquad \delta h_\mu^{v,I}=D_\mu \xi^I_v\,,
\end{equation}
where we have defined $\hat{h}^{s,I}_\mu=h^{s,I}_\mu-\frac{1}{2}D_\mu \rho^I_s$.
One can further check that the fields $U^{(A\neq 5)k=1}$ are gauge invariant, so one has (again for $A\neq 5$)
\begin{equation}
\begin{aligned}\label{ZA gauge invariant}
\bm{Z}^{(A)K}_\mu\;=\;&Z^{(A)K}_\mu+\frac{E^{JIK}}{\lambda_k}(\hat h^{s,I,\nu}D_\nu D_\mu U^{(A)J}+D_\mu \hat h^{s,I,\nu}D_\nu U^{(A)J})\\&
+\frac{E^{KIJ}}{\lambda_k^2}\epsilon_{\mu\nu\rho}\hat h^{s,I,\nu}D_\rho U^{(A)J}+
\frac{\Lambda_jU^{(A)J}}{\lambda_k}(\frac{1}{2}D_\mu\rho^I_vf^{IJK}+\frac{1}{2}D_\mu\rho^I_s E^{IJK})\\& -D_\mu U^{(A)J}(\frac{1}{2}\rho^I_sn^v_{IJK}+\frac{1}{2}\rho^I_v p^v_{IJK}+\frac{1}{2}\rho^I_s  c^v_{IJK}+\frac{1}{2}\rho^I_v g^v_{IJK})\,.
\end{aligned}
\end{equation}
In the following sections, we will also review the holographic dictionary for CPOs of dimension one and scalar chiral primaries of dimension two. Discussing this dictionary in an explicit gauge-invariant fashion requires studying the gauge-invariant combinations associated to other fields. In practise however, it is often convenient to partially fix the gauge and use the holographic dictionary in a preferred system of coordinates, and we will do this explicitly in Appendix \ref{sec:Gauge fixed dictionary}.

\subsection{Refining the existing holographic dictionary}\label{duality fields operators}

The action for the AdS$_3$ fields is obtained by substituting the KK harmonic expansion~\eqref{0611 geom harmonics expansion} into the six-dimensional Lagrangian. This procedure leads to a three-dimensional Lagrangian which has a non-diagonal mass matrix. The duality between 3D fields and operators prescribes that the dimension of the operator corresponds to the energy of the bulk excitation. Thus in order to identify the AdS fields dual to the operator of the D1-D5 CFT, one must perform the linear field redefinition that diagonalizes the mass matrix.

Moreover, as discussed in Section \ref{sec:spo basis}, by performing quadratic field redefinitions it possible to recast the cubic Lagrangian into a form with no derivative couplings: this corresponds to the basis of single-particle excitations.
For a general discussion we refer to \cite{Deger:1998nm,Arutyunov:2000by}; here we just discuss the AdS fields that will enter the holographic dictionary we are going to construct. Recall that in~\cite{Deger:1998nm,Arutyunov:2000by} the KK spectrum was been studied in de Donder gauge, while we are interested in a gauge independent discussion. It follows from~\cite{Skenderis:2006uy} that this can be obtained by simply replacing the fields with the corresponding gauge-invariant combination: in the following, unless explicitly stated, this replacement will be understood.

The field redefinitions that diagonalize the linearized field equations are ($r=6,7$)

 \begin{equation}
 \begin{aligned}\label{ads fields mass matrix diag}
 s^{(r)k}_I&\;=\;\frac{\sqrt{k}}{\sqrt{k+1}}\big(\phi^{(5r)k}_I+2(k+2)U^{(r)k}_I\big)\,,\\
 \sigma^{k}_I&\;=\;\frac{\sqrt{k(k-1)}}{3\sqrt{k+1}}\big(6(k+2)U^{(5)k}_I-\pi^{k}_I\big)\,,\\
 A^{(\pm)k}_{I\mu}&\;=\;\pm 2 Z^{(5)(\pm)k}_{I\mu}-h^{(\pm)k}_{I\mu}, \qquad Z^{(r)k}_{I\mu}\rightarrow 4\sqrt{k+1}Z^{(r)k}_{I\mu}\,,
 \end{aligned}
 \end{equation} 
where the superscripts $(\pm)$ are used to distinguish the fields that couple to left $(+)$ and right $(-)$ $SU(2)$ vector harmonics. The overall $k$-dependent factors are needed to canonically normalize the quadratic Lagrangian \cite{Arutyunov:2000by}\footnote{Note that  the canonically normalized fields that we have denoted by $s^k_I$ and $\sigma^k_I$ are denoted in~\cite[Eq.~(5.16)]{Kanitscheider:2006zf} by $S^k_I$ and $\Sigma^k_I$.}. These fields have masses:
\begin{equation}\label{mass ads fields}
m_{s^{(r)k}}^2=m_{\sigma^{k}}^2=k(k-2)\,, \qquad m_{A^{(\pm)k}}=k-1\,, \qquad m_{Z^{(r)k}}=k+1\,.
\end{equation}

\vspace{0.5mm}

\noindent
{\bf Restriction to fields with low $k$}

\vspace{0.5mm}

We focus on the low-order fields, in particular we shall restrict to $s^{(r)k}$ with $k=1,2\,$;\\
~$\sigma^{k}$ with $k=2\,$; ~ $A^{(\pm)k}$ with $k=1\,$;$\;$ and $\;Z^{(r)k}$ with $k=1$.$\;$ Among these fields, only $\sigma^{k=2}$ has a cubic coupling involving derivatives: the equation of motion reads\footnote{This is \cite[Eq. (5.8)]{Kanitscheider:2006zf} with implemented $SO(h^{1,1}(\mathcal{M})+1)$ invariance.}
\begin{equation}\label{quadratic eom sigma}
\Box \sigma^{(k=2)}_I\;=\;\frac{11}{12\sqrt{2}}\sum_{r=6,7}\big(s^{r(k=1)}_i s^{r(k=1)}_j -D_{\mu} s^{r(k=1)}_i D^{\mu} s^{r(k=1)}_j\big)a_{Iij}\,,
\end{equation}
where $a_{Iij}$ is defined as the following triple overlap in Eq.~\eqref{tripleoverlap}.
 
Following the discussion in Section \ref{sec:spo basis}, we wish to remove the derivative coupling, which can be done with the following field redefinition: 
\begin{equation}\label{field sigma quadratic redef}
\sigma^{(k=2)}_I\; \rightarrow\; \tilde{\sigma}^{(k=2)}_I=\sigma^{(k=2)}_I +\frac{11}{24 \sqrt{2}}\sum_{r=6,7}s^{r(k=1)}_i s^{r(k=1)}_j a_{Iij}\,.
\end{equation}

We now identify the single-particle operators of the D1-D5 CFT that are dual to the fields in~\eqref{ads fields mass matrix diag},~\eqref{field sigma quadratic redef}. At dimension one there are no multi-trace operators, nor there are degeneracies among the single traces, so the basis of single-trace  and single-particle operators coincide. The explicit  dictionary for these fields was derived in~\cite{Giusto:2015dfa} and is recorded in Table~\ref{table:2} below.

For dimension two operators the situation is more complicated, as follows. 
The spectrum of single-trace CPOs discussed in Section \ref{Low dimensional operators} splits into two subsectors, according to the quantum numbers.
The supergravity theory has an $SO(n)$ symmetry that acts on the tensor multiplets. However, in the full String Theory, only an   $SO(n-1)$ subgroup is preserved~\cite{Taylor:2007hs}. This means that the dimension two operator $O_2$ can only mix with the double trace $(\Sigma_2\cdot O)$. By contrast, since $\Sigma_3$ and $\Omega$ are scalars under this $SO(n-1)$, they mix with each other and with the multi-traces $(\Sigma_2 \cdot \Sigma_2)$, $(J \cdot \bar{J})$ and $(O\cdot O)$. This has been explicitly verified in \cite{Giusto:2019qig}.

The single-particle CPO in the first of these subsectors is~\cite{Giusto:2019qig}
\begin{equation}\label{single particle O2}
\tilde{O}_2^{++}=\Big(\frac{\sqrt{2}\,O_2^{++}}{N}-\frac{1}{\sqrt{N}}(\Sigma_2\cdot O)^{++}\Big).
\end{equation}
The coefficient in front of $O_2^{++}$ is chosen such that this first term on the right-hand side is unit-normalized in the large $N$ limit. Since $(\Sigma_2\cdot O)$ is unit-normalized at large $N$, the full operator $\tilde{O}_2^{++}$ also has unit norm at large $N$. We see that the mixing coefficient between the unit-normalized operators scales as $\:\! 1/{\sqrt{N}}$.
This is a general feature of all the examples we study.\footnote{The scaling of $\:\! 1/{\sqrt{N}}$ per additional trace for the mixing coefficients of  multi-trace operators has appeared before in discussions of extremal correlators~\cite{Taylor:2007hs}. Note that in the case of a bound state of $N_3$ D3 branes giving rise to $SU(N_3)$ $\cN=4$ SYM, the analogous scaling of such admixture coefficients is $\:\! 1/N_3$ per additional trace (see e.g.~\cite{Aprile:2020uxk} and references within).}
 As noted above, in generic correlators the multi-trace contribution is subleading at large $N$, however in extremal or certain heavy-light correlators it contributes at leading order in large $N$. Note that for this operator, all coefficients are fixed from CFT considerations. 

In the second subsector, $\Sigma_3$ and $\Omega$ mix among themselves, and also with the multi-traces $(\Sigma_2 \cdot \Sigma_2)$, $(J \cdot \bar{J})$ and $(O\cdot O)$. CFT considerations alone are not sufficient to
identify the two individual single-particle CPOs: if one imposes orthonormality and orthogonality with all multi-traces, one is left with a one-parameter family of possible pairs of candidate single-particle operators. We discuss this and the following steps in more detail in Appendix~\ref{sec:Extremal 3pt}.
To proceed, we fix the mixing among the single-traces $\Sigma_3$ and $\Omega$ using additional information from comparison with supergravity, using the mixing matrix derived in~\cite{Taylor:2007hs}. Once we incorporate this single-trace mixing, imposing orthonormality and orthogonality with all multi-traces determines all the remaining admixture coefficients, resulting in the single-particle operators: 
\begin{equation}\label{orthonormal single particle scalars}
\begin{aligned}
 \tilde{\Sigma}_3^{++}&\,\equiv\, \frac{3}{2} \left[\left(\frac{\Sigma_3^{++}}{N^{\frac32}}-\frac{\Omega^{++}}{3N^{\frac12}}\right)+\frac{1}{N^{\frac12}}\left(-\frac{2}{3}(\Sigma_2\cdot \Sigma_2)^{++}+\frac{1}{6}(O\cdot O)^{++}+\frac{1}{3}(J\cdot \bar{J})^{++}\right)\right]\,, \\
\tilde{\Omega}^{++}&\,\equiv\, \frac{\sqrt{3}}{2} \left[\left(  \frac{\Sigma_3^{++}}{N^{\frac32}}+\frac{\Omega^{++}}{N^{\frac12}}\right)+ \frac{1}{N^{\frac12}}\left(-(\Sigma_2\cdot \Sigma_2)^{++}-\frac12 (O\cdot O)^{++}-(J \cdot \bar{J})^{++}\right)\right] .
\end{aligned}
\end{equation}

Let us make a similar comment on the factors of $N$ and numerical coefficients in Eq.~\eqref{orthonormal single particle scalars}. 
The linear combinations of the single-particle operators inside $\Sigma_3^{++}$ and $\Omega^{++}$ ensure that these single-trace combinations (and thus the full single-particle operators) are orthonormal in the large $N$ limit, as we discuss in more detail in Appendix~\ref{sec:Extremal 3pt}. Again the admixture coefficients between the unit-normalized single-traces and multi-traces are of order $1/\sqrt{N}$; in generic correlators the contributions from the multi-traces are subleading; but in extremal or certain heavy-light correlators, the multi-traces contribute at leading order in large $N$.

We now make two observations. 
We note that in this paper, the coefficients of the multi-traces have been derived from a purely CFT calculation of orthogonality with all multi-traces. The only direct supergravity input here is the mixing between $\Omega$ and $\Sigma_3$ derived in~\cite{Taylor:2007hs}. This contrasts with the method of~\cite{Giusto:2019qig} which fixed the multi-trace coefficients holographically. The fact that these two methods agree is non-trivial, and is explored further in Appendix \ref{sec:Gauge fixed dictionary}.

Secondly, let us emphasize that the non-trivial mixing between $\Omega$ and $\Sigma_3$ demonstrates that there is no one-to-one correspondence between $\k$-cycles in the CFT and single-particle supergravity excitations, even at the level of single traces: single-particle states in the bulk are dual to a linear combination of single cycles of the symmetric group (c.f.~\cite{Taylor:2007hs,Ceplak:2018pws}).

{\renewcommand\arraystretch{1.5}
\begin{table}[t]
\centering
 \begin{tabular}{||c c c c||} 
 \hline
 AdS$_3$ field & Dual operator & $(j_{\sl},{\bar j}_{\sl})$ & $(j_{\su},{\bar j}_{\su})$ \\  
 \hline 
 $s^{(6)k=1}$ & $\Sigma_2$ & $(\frac{1}{2},\frac{1}{2})$ & $(\frac{1}{2},\frac{1}{2})$ \\ 
 \hline
 $s^{(6)k=2}$ & $\tilde{\Sigma}_3$ & $(1,1)$ & $(1,1)$ \\ 
 \hline
 $s^{(7)k=1}$ & $O$ & $(\frac{1}{2},\frac{1}{2})$ & $(\frac{1}{2},\frac{1}{2})$ \\ 
 \hline
 $s^{(7)k=2}$ & $\tilde{O}_2$ & $(1,1)$ & $(1,1)$ \\ 
 \hline
$\tilde{\sigma}^{k=2}$ & $\tilde{\Omega}$ & $(1,1)$ & $(1,1)$ \\ 
 \hline
$A^{(+)k=1}_{\mu}$ & $J$ & $(1,0)$ & $(1,0)$ \\ 
 \hline
 $A^{(-)k=1}_{\mu}$ & $\bar{J}$ & $(0,1)$ & $(0,1)$ \\ 
 \hline
 $Z^{6(-)k=1}_{\mu}$ & $GG\tilde{\Sigma}_3$ & $(2,1)$ & $(0,1)$ \\ 
 \hline
$Z^{7(-)k=1}_{\mu}$ & $GG\tilde{O}_2$ & $(2,1)$ & $(0,1)$ \\ 
 \hline
\end{tabular}
\caption{This table shows the duality between AdS$_3$ fields and the CFT operator. We denote with $(j_{\sl},{\bar j}_{\sl})$ the quantum numbers associated with the Casimirs of the two copies of $SL(2,\mathbb{R)}$ and with $(j_{\su},{\bar j}_{\su})$ those associated to the Casimir of the two copies of $SU(2)$.}
\label{table:2}
\end{table}}


\subsection{Supercharged CFT operators}\label{ssc:superchaarged cft operators}

In the following, we will construct the holographic dictionary for the bosonic $1/8$-BPS supercharged descendants of the single-particle operators $\tilde{O}_2$ and $\tilde{\Sigma}_3$. We denote these by $GG\tilde{O}_2$ and $GG\tilde{\Sigma}_3$ respectively, and we obtain:
\begin{align}\label{Normalized supercharged operators}
\big(G&G\tilde{O}_2\big)^{(0,a)}\equiv \frac{1}{\sqrt{3}}\big(G_{-\frac12}^{+1}G_{-\frac12}^{+2}+ \frac12  J^+_0 L_{-1}\big)\tilde{O}_2^{-,a}\cr
& \qquad\qquad=\frac{1}{\sqrt{3}}\big(G_{-\frac12}^{+1}G_{-\frac12}^{+2}+ \frac12  J^+_0 L_{-1}\big)\bigg[
\frac{\sqrt{2}\,O_2}{N}-\frac{1}{N^{1/2}}(\Sigma_2\cdot O)
\bigg]^{-,a},\\
\big(G&G\tilde{\Sigma}_3\big)^{(0,a)} \equiv  \frac{1}{\sqrt{3}}\big(G_{-\frac12}^{+1}G_{-\frac12}^{+2}+ \frac12  J^+_0 L_{-1}\big)\tilde{\Sigma}_3^{-,a} \cr
&= \frac{\sqrt{3}}{2}\big(G_{-\frac12}^{+1}G_{-\frac12}^{+2}\!\!\:+\!\!\: \frac12  J^+_0 L_{-1}\big) \!
\left[\!\!\:\left(
\frac{\Sigma_3}{N^{\frac32}}\!\!\:-\!\!\:\frac{\Omega}{3N^{\frac12}}
\!\right)\!+\!\frac{1}{N^{\frac12}}\!\left(\!-\frac{2}{3}(\Sigma_2\cdot \Sigma_2)\!\!\:+\!\!\:\frac{1}{6}(O\cdot O)\!\!\:+\!\!\:\frac{1}{3}(J\cdot \bar{J})\right)\!\!\:\right]^{-,a} \!\!\!. \nonumber
\end{align}
The overall numerical coefficients follow from Eqs.~\eqref{norm k111},~\eqref{single particle O2} and~\eqref{orthonormal single particle scalars} and are required to normalize the operators to one at large $N$. Being descendants of single-particle operators, they are orthogonal to all multi-trace operators.
They carry quantum numbers $(j_{\sl},\bar j_{\sl})=(2,1)$ associated with the Casimirs of $SL(2,\mathbb{R})$ and quantum numbers $(j_{\su},{\bar j}_{\su})=(0,1)$ associated to the Casimirs of $SU(2)_L\times SU(2)_R$ (we refer the reader to Appendix \ref{sec:harm S3 and AdS3} for  explicit definitions).

Using the relation between the mass $m$ of a field in AdS$_{d+1}$ and the dimension of the dual operator (see for example \cite{Aharony:1999ti}) one can identify the map between single-particle operators and AdS$_3$ fields. The residual degeneracy between $s^{(6)k=2}$ and $\tilde{\sigma}^{k=2}$ can be fixed by comparing non-extremal three-point functions \cite{Taylor:2007hs}: $s^{(6)k=2}$ is dual to $\tilde{\Sigma}_3$ and $\tilde{\sigma}^{k=2}$ is dual to $\tilde{\Omega}$. These results are recorded in Table \ref{table:2}.

\subsection{Refined holographic dictionary at dimension one and two}\label{ssc:Holographic dictionary for scalar SPOs}

For convenient reference we now record the  holographic dictionary for CPOs of dimension one and scalar CPOs of dimension two derived in \cite{Kanitscheider:2006zf,Kanitscheider:2007wq,Giusto:2015dfa,Giusto:2019qig}, after having recasted it in the single-particle basis. The dictionary relates the asymptotic expansion of the AdS$_3$ fields in a non-trivial background with the expectation value of the dual operators $\mathcal{O}$ in the dual heavy CFT state $\ket{H}$,
\begin{equation}
\langle \mathcal{O}\rangle\;\equiv\; \bra{H}\mathcal{O}(\tilde t,\tilde y)\ket{H}\,,
\end{equation}
where $(\tilde t,\tilde y)$ is a generic insertion point on the CFT cylinder.

For scalars, the holographic prescription relates the expectation value of a scalar operator of dimension $\Delta$ with the coefficient of $\tilde{r}^{-\Delta}$ of the large $\tilde{r}$ expansion of the dual scalar field. The mass of the scalar fields $s^{(r)k}$ and $\sigma^k$ in Eq.~\eqref{mass ads fields} implies that their dual operators have dimension $\Delta=k$. This motivates introducing the following asymptotic expansion of scalar fields: we denote with  $\Big[\Phi_k\Big]$ the first non-vanishing term of the expansion of the scalar field $\Phi_k$,
\begin{equation}
\Phi_k\;=\;\frac{\Big[\Phi_k\Big]}{\tilde{r}^k}+O(\tilde{r}^{-(k+1)})\,.
\end{equation}
Similarly, for the one-forms $A^{a(\pm)}_{k=1}$ we expand as \cite{Hansen:2006wu,Kanitscheider:2006zf}
\begin{equation}
A^{a(\pm)}_{k=1}\;=\;\Big[A^{a(\pm)}_{k=1} \Big](d\tilde t\pm d\tilde y)+O(\tilde r^{-1})\,.
\end{equation}
We then have the dictionary
\begin{align}\label{gauge invariant old dict}
\frac{1}{\sqrt{N}}\langle J^{\pm}\rangle&=-\frac{\sqrt{N}}{\sqrt{2}} \Big[A^{\mp(+)}_{k=1}\Big]\,,
\hspace{6.8 em}
\frac{1}{\sqrt{N}}\langle \bar J^{\pm}\rangle=-\frac{\sqrt{N}}{\sqrt{2}} \Big[A^{\mp(-)}_{k=1}\Big]\,,
\cr
\frac{1}{\sqrt{N}}\langle J^{3}\rangle&=-\frac{\sqrt{N}}{2} \Big[A^{0(+)}_{k=1}\Big]\,,\hspace{7.3 em} \frac{1}{\sqrt{N}}\langle \bar J^{3}\rangle=-\frac{\sqrt{N}}{2}\Big[ A^{0(-)}_{k=1}\Big]\,,
\cr
\frac{\sqrt{2}}{N}\big\langle\Sigma_2^{\alpha\dot \alpha}\big\rangle&=(-1)^{\alpha\dot\alpha}\frac{\sqrt{N}}{\sqrt{2}} \Big[s^{(6)(-\alpha,-\dot\alpha)}_{k=1}\Big]\,,
\qquad  
\frac{1}{\sqrt{N}}\big\langle O^{\alpha\dot\alpha}\big\rangle=(-1)^{\alpha\dot\alpha}\frac{\sqrt{N}}{\sqrt{2}} \Big[s^{(7)(-\alpha,-\dot\alpha)}_{k=1}\Big]\,,
\cr
\big\langle\tilde{\Omega}^{a,\dot a}\big\rangle&=-(-1)^{a+\dot a}\frac{\sqrt{N}}{\sqrt{2}}\Big[\tilde{\sigma}^{(-a,-\dot a)}_{k=2}\Big]\,,
\\
 \big\langle\tilde{\Sigma}_3^{a,\dot a}\big\rangle&=(-1)^{a+\dot a}\frac{\sqrt{N}}{\sqrt{2}}\Big[s^{(6)(-a,-\dot a)}_{k=2}\Big]\,,\hspace{3.8em}
\big\langle\tilde{O}_2^{a,\dot a}\big\rangle=(-1)^{a+\dot a}\frac{\sqrt{N}}{\sqrt{2}}\Big[s^{(7)(-a,-\dot a)}_{k=2}\Big]\,.
\nonumber
\end{align}
The numerical coefficients and factors of $N$ on the left hand side of each equality are such that the operators are unit-normalized at large $N$. With this choice, we note that the coefficients in the third and fifth lines of the dictionary respect the $SO(n)$ symmetry between the $n$ tensor multiplets of the supergravity theory.


\newpage

\section{Supercharged holographic dictionary}
\label{sec:normalizing-dictionary}

In this section we construct the holographic dictionary for the single-particle operators $GG\tilde{O}_2^{(0, a)}$ and $GG\tilde{\Sigma}_3^{(0,a)}$ defined in \eqref{Normalized supercharged operators}. We saw in Table \ref{table:2} that the expectation value of these operators corresponds to the bulk asymptotic expansion of the vector fields $Z^{7(-)k=1}_{\mu}$ and $Z^{6(-)k=1}_{\mu}$. 
The linearized equation of motion for these fields is \cite{Arutyunov:2000by}
\begin{equation}
\star dZ^{(A)(-)}_k\;=\;-(k+1) Z^{(A)(-)}_k\,,
\end{equation}
for $A=6,7$. We note that this is the equation obeyed by the left AdS$_3$ harmonics $B_{\LL,\h}^{(\pm)l,\bar l}$ in Eq.~\eqref{vector harmonic eigenstate}, with the identification $l=k+3$.
Left vector harmonics on AdS$_3$ are discussed in Appendix \ref{AdS vector harmonics}: their large $\tilde r$ expansion reads
\begin{equation}
B_{\LL,\h}^{(\pm)l,\bar l}\;\sim\; \frac{d\tilde t+d\tilde y}{\tilde r^{(\h-2)}}+O\Big(\frac{1}{r^{\h-1}}\Big)\,.
\end{equation}
This motivates, for $k=1$, the following asymptotic expansion of the bulk fields, where we use the same square bracket notation introduced in the previous subsection for the leading term:
\begin{equation}\label{asympt expansion Z6 Z7}
\ba
Z^{7(a,-)}_{k=1}= \Big[Z^{7(a,-)}_{k=1}\Big]\frac{d\tilde t+d\tilde y}{\tilde r^2}+O\Big(\frac{1}{\tilde r^3}\Big)\,,\qquad Z^{6(a,-)}_{k=1}=  \Big[Z^{6(a,-)}_{k=1}\Big]\frac{d\tilde t+d\tilde y}{\tilde r^2}+O\Big( \frac{1}{\tilde r^3}\Big )\,.
\ea
\end{equation}

We consider the following ansatz for the dictionary involving supercharged operators dual to vector fields in the tensor multiplet:
\begin{equation}\label{ansatz supercharged dictionary}
\begin{aligned}
\big\langle GG\tilde{O}_2^{(0, a)}\big\rangle&\;=\;\alpha  \Big[Z^{7(a,-)}_{k=1}\Big]\,,\\
\big\langle GG\tilde{\Sigma}_3^{(0,a)}\big\rangle&\;=\;\beta  \Big[Z^{6(a,-)}_{k=1}\Big]\,.
\end{aligned}
\end{equation}
where $\alpha$ and $\beta$ are unknown coefficients that we will determine by evaluating the ansatz \eqref{ansatz supercharged dictionary} on some reference heavy states. Consistency of the dictionary requires that $\alpha$ and $\beta$ depend neither on the $SU(2)_R$ quantum number $a$ nor on the heavy state considered. 

Moreover, based on the $SO(n)$ symmetry between the tensor multiplets of the supergravity theory, one expects to find $\alpha=\beta$, and we shall verify explicitly that this is the case.

\subsection{Normalizing the supercharged holographic dictionary}\label{ssc:Normalizing the supercharged holographic dictionary}

In this section we fix the coefficient $\alpha$ in the supercharged holographic dictionary in Eq.~\eqref{ansatz supercharged dictionary} by looking at one of the simplest examples of supercharged superstrata: the one sourced by the mode $(\k,\m,\n,\q)=(2,1,\n,1)$. This supergravity solution was constructed in \cite{Ceplak:2018pws}. 
Since we work in the conventions of~\cite{Heidmann:2019zws}, a slightly more convenient reference for the explicit form of the supergravity quantities we need in the following\footnote{In what follows we will label with $b$ the supergravity coefficient that is denoted by $c_4$ in \cite{Heidmann:2019zws}.} is~\cite[Eqs.\;(2.4),\,(4.1),\,(4.4),\,(4.13),\,(4.14)]{Heidmann:2019zws}.

The CFT state that is proposed to be dual to this bulk solution is
\begin{equation}\label{21n1 NS}
\sum_p \Big(A\ket{0}_1\Big)^{N-2p}\Big(B\frac{L_{-1}^{\n-1}}{(\n-1)!}\big(G_{-\frac12}^{+1}G_{-\frac12}^{+2} + \frac{1}{2} J^+_0 L_{-1}\big)\ket{O^{--}_2}\Big)^{p}\,.
\end{equation}

The single-particle operator $GG\tilde{O}_2$ has a non-vanishing expectation value on this state, sourced by its single-trace constituent. In order to compute the correlator, we first consider the basic process described by the correlator
\begin{align}\label{21n1 process ty to zbarz}
\bra{O^{++}_2}&\Big(-G_{\frac12}^{-1}G_{\frac12}^{-2} + \frac{1}{2} J^-_0 L_{+1}\Big)\frac{L_{1}^{\n-1}}{(\n-1)!}\Big(G_{-\frac12}^{+1}G_{-\frac12}^{+2} + \frac{1}{2} J^+_0 L_{-1}\Big)O_2^{--}(\tilde t,\tilde y)\ket{0}^{\otimes 2}_1\\
&=z^2\bar z\bra{O^{++}_2}\Big(-G_{\frac12}^{-1}G_{\frac12}^{-2} + \frac{1}{2} J^-_0 L_{+1}\Big)\frac{L_{1}^{\n-1}}{(\n-1)!}\Big(G_{-\frac12}^{+1}G_{-\frac12}^{+2} + \frac{1}{2} J^+_0 L_{-1}\Big)O_2^{--}(z,\bar z)\ket{0}^{\otimes 2}_1\,,
\nonumber
\end{align}
where we have mapped the one-point function from the cylinder to the plane and inserted the appropriate conformal factor.

Upon expanding, we obtain four terms. The one with four supercharge modes evaluates to
\begin{align}
\!\!\!\!\!
-\bra{O^{++}_2}&\big(G^{-1}_{\frac12}G^{-2}_{\frac12}\big)\frac{L_1^{\n-1}}{(\n-1)!}\big(G^{+1}_{-\frac12}G^{+2}_{-\frac12}\big)O^{--}_2(z,\bar z)\ket{0}^{\otimes 2}_1\cr
&=-\frac{1}{(\n-1)!}
\bra{O^{++}_2}\big(G^{-1}_{\frac12}G^{-2}_{\frac12}\big)\bigg[(\n^2-3\n)G^{+2}_{\frac12}G^{+1}_{\frac12}L_1^{\n-3}+(\n-1)G^{+2}_{\frac12}G^{+1}_{-\frac12}L_1^{\n-2}\cr
&\qquad\qquad\qquad\qquad\qquad \;\;\;\;+(\n-1) G^{+2}_{-\frac12}G^{+1}_{\frac12}L_1^{\n-2}+G^{+1}_{-\frac12}G^{+2}_{-\frac12}L_1^{\n-1}\bigg]O^{--}_2(z,\bar z)\ket{0}^{\otimes 2}_1\cr
&= \frac{(\n^2+2\n+1)}{(\n-1)!}\bra{O^{++}_2}L_1^{\n-1}O^{--}_2(z,\bar z)\ket{0}^{\otimes 2}_1\,,
\end{align}
where we have used the anomaly-free algebra \eqref{anomaly free algebra}.
By similar standard manipulations the other three terms evaluate to
\begin{equation}
-\frac{1}{2(\n-1)!}\bra{O^{++}_2}L_1^{\n}L_{-1}O^{--}_2(z,\bar z)\ket{0}\;=\;\frac{\n(\n+1)}{(\n-1)!}\bra{O^{++}_2}L_1^{\n-1}O^{--}_2(z,\bar z)\ket{0}^{\otimes 2}_1\,.
\end{equation}
We note that the commutation relation between $L_1^\n$ and a primary with left dimension $h$ is
\begin{equation}\label{L1 descendant primary}
[L_1^\n,O_h]\;=\;\sum_{m=0}^{\n}\frac{\n!}{(\n-m)!m!}w^{\n+m}\frac{(2h+\n-1)!}{(2h+m-1)!}\partial^m O_h\,.
\end{equation}
Collecting all terms and using \eqref{21n1 process ty to zbarz} and \eqref{L1 descendant primary} we obtain
\begin{equation}\label{21n1 on 21n1 process}
\begin{aligned}
\bra{O^{++}_2}&\big(-G_{\frac12}^{-1}G_{\frac12}^{-2} + \frac{1}{2} J^-_0 L_{+1}\big)\frac{L_{1}^{\n-1}}{\n-1!}\big(G_{-\frac12}^{+1}G_{-\frac12}^{+2} + \frac{1}{2} J^+_0 L_{-1}\big)O_2^{--}(\tilde t,\tilde y)\ket{0}^{\otimes 2}_1\\
&=\frac{\n (\n+1)(\n+2)}{2}e^{i ((\n+2)\tilde t+\n\tilde y)}\,.
\end{aligned}
\end{equation}

This basic process describes the contribution to the expectation value of the single-particle operator when it acts on two copies of the CFT vacuum. 
We now use this to compute the effect of the single-particle operator acting on the full state \eqref{21n1 NS}.
To do so, we must compute a combinatorial factor, as we shall describe momentarily.
Combining the amplitude in \eqref{21n1 on 21n1 process} with this combinatorial factor, the relevant contribution is represented by
\begin{align}\label{process coherent state 1}
GG\tilde{O}_2^{(0,-)}&(\tilde t,\tilde y)\Big[\Big(\ket{0}_1^{N-2p}\Big)\Big(\frac{L_{-1}^{\n-1}}{(\n-1)!}\big(G_{-\frac12}^{+1}G_{-\frac12}^{+2} + \frac{1}{2} J^+_0 L_{-1}\big)\ket{O^{--}_2}\Big)^{p}\Big]\\
&=
\frac{\sqrt{2}(p+1)}{\sqrt{3}N}e^{i ((\n+2)\tilde t+\n\tilde y)}\Big[\Big(\ket{0}_1^{N-2p-2}\Big)\Big(\frac{L_{-1}^{\n-1}}{(\n-1)!}\big(G_{-\frac12}^{+1}G_{-\frac12}^{+2} + \frac{1}{2} J^+_0 L_{-1}\big)\ket{O^{--}_2}\Big)^{p+1}\Big]\,.
\nonumber
\end{align}
The factors in~\eqref{process coherent state 1} arise as follows (c.f.~\cite{Giusto:2015dfa}). The norm of the states on each side of the equality must match.
The factor of ${\sqrt{2}}/({\sqrt{3}N})$ comes from the normalization of $\tilde{O}_2$,~\eqref{Normalized supercharged operators}. The factor of $(p+1)$ is a combination of a combinatorial factor and the $\n$-dependent factor in~\eqref{21n1 on 21n1 process}. The norm of the state on the left-hand side (LHS) of the equation is given by the norm of the state in square brackets (on the LHS) multiplied by the number of ways in which the single-particle operator can act on any two of the $N-2p$ vacua, namely ${N-2p\choose 2}$. The norm of the states in square brackets on both the LHS and RHS are given in Eq.~\eqref{Norm cft basis superstrata}. These factors all combine with the $\n$-dependent prefactor in~\eqref{21n1 on 21n1 process} to give~\eqref{process coherent state 1}.

When we compute the amplitude with the full coherent state~\eqref{21n1 NS}, we obtain an additional factor of $A^2/B$ due to the fact that the process annihilates two $A$-type strands and creates one $B$-type strand (c.f.~\cite[Eq.\;(4.19)]{Giusto:2015dfa}).
Furthermore, we work at large $N$ and with coherent states in which the average $\bar{p}$ is of order $ N$, so we approximate $(\bar{p}+1) \simeq \bar{p}\:\! $.
We then obtain the amplitude
\begin{equation}\label{2111 on 2111 VeV pre}
\begin{aligned}
\big\langle GG\tilde{O}_2^{(0,-)} \big\rangle\;=\;\frac{\sqrt{2}\;\! \bar p }{\sqrt{3}N} \;\! \frac{A^2}{B} \:\! e^{i ((\n+2)\tilde t+\n\tilde y)} \,.
\end{aligned}
\end{equation}
From Eq.~\eqref{q=1 peak AB}, $\bar{p}$ is of order $B^2$. Using Eqs.~\eqref{q=1 peak AB} and~\eqref{q=1 AB ab}, we obtain the final result for the correlator,
\begin{equation}\label{2111 on 2111 VeV}
\begin{aligned}
\big\langle GG\tilde{O}_2^{(0,-)} \big\rangle\;=\;\sqrt{N}\frac{\mathsf{a}^2 \mathsf{b}}{2\sqrt{3}} e^{i ((\n+2)\tilde t+\n\tilde y)} \,.
\end{aligned}
\end{equation}
Since $GG\tilde{O}_2^{(0,+)}=\big(GG\tilde{O}_2^{(0,-)}\big)^{\dagger}$, the expectation value of the operator $GG\tilde{O}_2^{(0,+)}$ must also be non-vanishing. We thus obtain
\begin{equation}\label{2111 on 2111 VeV dagger}
\begin{aligned}
\big\langle GG\tilde{O}_2^{(0,+)} \big\rangle=\big\langle GG\tilde{O}_2^{(0,-)} \big\rangle^*=\sqrt{N}\frac{\mathsf a^2\mathsf b}{2\sqrt{3}} e^{-i ((\n+2)\tilde t+\n\tilde y)}\,. 
\end{aligned}
\end{equation}

In order to evaluate the coefficient $\alpha$ in Eq.~\eqref{ansatz supercharged dictionary}, we must perform an asymptotic expansion of the gauge-invariant combination $\bm{Z}^7_{k=1}$, given in Eq.~\eqref{ZA gauge invariant}. 
The supergravity solution with modes $(\k,\m,\n,\q)=(2,1,\n,1)$ is characterized by $U^7_{k=1}=0\:\!$, which implies that the field  $Z^7_{k=1}$ that appears in the Kaluza-Klein reduction \eqref{0611 geom harmonics expansion} coincides with the gauge-invariant combination $\bm{Z}^7_{k=1}$. We thus obtain (recall $\n\ge 1$)
\begin{equation}\label{Z7 for 21n1}
\ba
\bm Z^{7(--)}_{k=1}\;=\;Z^{7(--)}_{k=1}&\;=\;-\mathsf a^2 \mathsf b\, e^{-i ((\n+2)\tilde t+\n\tilde y)}\, \frac{\tilde{r}^{\n-1}}{(\tilde r^2+1)^{\n/2+2}} \Big(\!\!\:+i  d\tilde r+\tilde r(\tilde r^2+1)(d\tilde t+d\tilde y)\Big) \,, \\
\bm Z^{7(+-)}_{k=1}\;=\;Z^{7(+-)}_{k=1}&\;=\;-\mathsf{a}^2\mathsf{b}\,\, e^{i ((\n+2)\tilde t+\n\tilde y)}\, \frac{\tilde r^{\n-1}}{(\tilde r^2+1)^{\n/2+2}}\Big(\!\!\: -i d\tilde r+\tilde r(\tilde r^2+1)(d\tilde t+d\tilde y)\Big)\,,
\ea
\end{equation}
where we recall that the rescaling in Eq.\;\eqref{ads fields mass matrix diag} has been performed, and where the superscript $(\pm-)$ indicates  that the field couples to the $Y^{(\pm-)}$ harmonic respectively.
The large $\tilde r$ expansion of \eqref{Z7 for 21n1} gives
\begin{equation}\label{f7 for 21n1}
\Big[Z^{7(--)}_{k=1}\Big]\;=\;-\mathsf{a}^2\mathsf{b}\,e^{-i ((\n+2)\tilde t+\n\tilde y)}\,,\qquad \Big[Z^{7(+-)}_{k=1}\Big]\;=\;-\mathsf{a}^2\mathsf{b}\,e^{i ((\n+2)\tilde t+\n\tilde y)}\,.
\end{equation}

Using the ansatz for the holographic dictionary in Eq. \eqref{ansatz supercharged dictionary}, along with Eqs. \eqref{2111 on 2111 VeV}, \eqref{2111 on 2111 VeV dagger} and  \eqref{f7 for 21n1}, we obtain
\begin{equation}
\alpha=-\frac{ N^{1/2}}{2\sqrt{3}} \,.
\end{equation}
We observe that the value of $\alpha$ is independent of the quantum numbers that specify the state, as required.

\subsection{Holographic test of general non-supercharged superstrata} \label{sec:beta-calc}

We now compute the coefficient $\beta$ defined in Eq.~\eqref{ansatz supercharged dictionary} and find that $\alpha=\beta$, verifying the expectation discussed below Eq.~\eqref{ansatz supercharged dictionary}.
In doing so, we will also test the coiffuring proposal for general multi-mode superstrata developed in~\cite{Heidmann:2019zws,Heidmann:2019xrd}. In the present subsection we test the non-supercharged part of this proposal, and in the following subsection we shall test the hybrid supercharged plus non-supercharged part.

As described in the Introduction, ``coiffuring'' refers to imposing a set of algebraic relations on the parameters in the supergravity solution, required for smoothness~\cite{Mathur:2013nja,Bena:2013ora,Bena:2014rea}. 
A significant achievement of~\cite{Heidmann:2019zws} was a proposal for particular families of multi-mode superstrata that had proven impossible to construct with previous methods, as we describe in more detail below.
This came as part of a more general proposal for the coiffuring of multi-mode superstrata, where the modes could be either of supercharged or non-supercharged type. This was expressed in a more general holomorphic formalism in~\cite{Heidmann:2019xrd}.

From the supergravity point of view, coiffuring relations are often not easy to give an interpretation to, beyond being a consequence of requiring solutions to be smooth.
By contrast, holographic calculations can give a microscopic interpretation to coiffuring relations, as has been done in~\cite{Giusto:2015dfa,Giusto:2019qig}.
We will now test the new type of coiffuring relation proposed in \cite{Heidmann:2019zws,Heidmann:2019xrd}, and we will find perfect agreement.

We focus on the family of $(1,\m,\n)$ multimode non-supercharged superstrata constructed in \cite[App,~D]{Mayerson:2020tcl}. This family of solutions is given in terms of two holomorphic functions $F_0$, $F_1$ of a complex variable $\xi$ which is related to the standard six-dimensional coordinates in Section~\ref{KK spectrum and single-particle states} via 
\be\label{variable xi}
\xi\;=\;\frac{\tilde r}{\sqrt{\tilde r+1}}e^{i(\tilde t+\tilde y)} \,.
\ee
We consider the two-mode solution in which both $F_0$ and $F_1$ consist of a single mode,
\be
\label{eq:F0-F1}
F_0(\xi) \;=\; b \;\! \xi^{\n_b} \,, \qquad
F_1(\xi) \;=\; d \;\! \xi^{\n_d} \,.
\ee
The explicit supergravity solution is given in full detail in \cite[Eqs.~(D.15)--(D.29)]{Mayerson:2020tcl} and we shall not reproduce it here.

The proposed family of dual CFT states is
\begin{equation}\label{state 110nb1nd}
\sum_{p,q}\Big(A\ket{0}_1\Big)^{N-p-q}\Big(B\frac{1}{\n_b!}L_{-1}^{\n_b}\ket{O^{--}}\Big)^p\Big(D \frac{1}{\n_d!}J^+_{0}L_{-1}^{\n_d}\ket{O^{--}}\Big)^q\,.
\end{equation}
The supergravity mode parameters $(b,d)$, the CFT parameters $(B,D)$, and the convenient parameters $(\mathsf{b},\mathsf{d})$ are related as before by Eq.~\eqref{q=1 AB ab}, which in this specific case takes the form
\be
\label{BD bd coiffuring}
\frac{B}{\sqrt{N}} \;=\;\frac{R}{\sqrt{2\,Q_1Q_5}}b\;=\;\frac{\mathsf{b}}{\sqrt{2}} \,, \qquad \frac{D}{{\sqrt{N}}} \;=\;  \frac{R}{\sqrt{2\,Q_1Q_5}}d\;=\;\frac{\mathsf{d}}{\sqrt{2}}    \,.
\ee

When $\n_b\neq \n_d$, this is a class of multimode superstrata where $(\k_b\m_d-\k_d\m_b)(\k_b \n_d-\k_d \n_b)\neq 0$. This is the class of superstrata that had evaded construction since~\cite{Bena:2017xbt} until the proposal of~\cite{Heidmann:2019zws}.
We have taken $k_b=k_d=1$, since higher values of $k_b$ and/or $k_d$ would require extending the holographic dictionary even further beyond the sector of conformal dimensions that we consider in this work. 
With $k_b=k_d=1$, this family of states essentially contains the full class of states in which $(\k_b\m_d-\k_d\m_b)(\k_b \n_d-\k_d \n_b)\neq 0$: since $\m \leq \k$, we must have one excited strand with $\m=0$ and one with $\m=1$. 
Furthermore, once one has control over the general two-mode family \eqref{eq:F0-F1}--\eqref{state 110nb1nd}, adding further modes of the same type is a straightforward generalization in both supergravity and CFT.

We start from the CFT side. We shall see that when $\n_b\neq \n_d$, the operator $GG\tilde{\Sigma}_3^{(0,a)}$ has a non-vanishing expectation value in this state. Expanding the definition of $GG\tilde{\Sigma}_3^{(0,a)}$, we have

\begin{align}\label{def sigma3 two}
\big\langle & GG\tilde{\Sigma}_3^{(0,a)}\big\rangle \equiv \Big\langle \frac{1}{\sqrt{3}}\big(G_{-\frac12}^{+1}G_{-\frac12}^{+2}+ \frac12  J^+_0 L_{-1}\big)\tilde{\Sigma}_3^{-,a}\Big\rangle =\\&
\Big\langle \frac{\sqrt{3}}{2}\big(G_{-\frac12}^{+1}G_{-\frac12}^{+2}+ \frac12  J^+_0 L_{-1}\big)
\left[\!\!\:\left(
\frac{\Sigma_3}{N^{\frac32}}\!\!\:-\!\!\:\frac{\Omega}{3N^{\frac12}}
\!\right)\!+\!\frac{1}{N^{\frac12}}\!\left(\!-\frac{2}{3}(\Sigma_2\cdot \Sigma_2)\!\!\:+\!\!\:\frac{1}{6}(O\cdot O)\!\!\:+\!\!\:\frac{1}{3}(J\cdot \bar{J})\right)\!\!\:\right]^{-,a}\Big\rangle \,.\nonumber
\end{align}

This expectation value is sourced only by the double-trace term (we suppress overall factors and restore them at the end)
\begin{equation}
\begin{aligned}\label{splitting of GGOO-pre}
GG(O\cdot O) \;=\; 
\frac{1}{N} \Big(&G_{-\frac12}^{+1}G_{-\frac12}^{+2} + \frac{1}{2} J^+_0 L_{-1}\Big)\sum_{r,s}O_{r}^{--} O_{s}^{--} \,.
\end{aligned}
\end{equation}
Expanding this product, the $GG$ combination of modes can act either on strand $r$ or strand $s$, giving rise to eight terms. Two of these terms give rise to fermionic strands, and so do not contribute to the correlator. The remaining terms can be written as 
\begin{equation}\label{relevant terms for 110nb1nd}
\frac{1}{N}\sum_{r,s}\left[ -\frac12 O^{--}_{r} \big( J^+_0 L_{-1}O^{--}_{s}\big)+\frac12  \big(L_{-1}O^{--}_{r}\big)\big( J^+_0 O^{--}_{s}\big)\right],
\end{equation}
where we have used the relation $\big(G_{-\frac12}^{+1}G_{-\frac12}^{+2} +  J^+_0 L_{-1}\big)O^{--}=0$, which holds because $O^{--}$ is a scalar operator of dimension one.

The operator in Eq.~\eqref{relevant terms for 110nb1nd} transforms two copies of the vacuum into one strand of type $\ket{1,0,\n_b,0}$ and another of type $\ket{1,1,\n_d,0}$. We now compute the contribution of this fundamental process. After doing so, we will again dress it with the appropriate combinatorial factor.
The initial state is given by two copies of the vacuum $\ket{0}_{r=1}\ket{0}_{r=2}$, which can be transformed into the state $\,\ket{1,0,\n_b,0}_{r=1}\ket{1,1,\n_d,0}_{r=2}\,$ or $\,\ket{1,0,\n_d,0}_{r=1}\ket{1,1,\n_b,0}_{r=2}\:\!$: the two processes contribute with equal amplitudes, so we compute only one of them, and multiply the result by two. For ease of notation (and to avoid confusion with the twist-two operator $O_2$), we shall abbreviate the subscripts $r=1,2$ to $(1), (2)$.

We proceed to compute
\begin{align}\label{process GGOO 110nb1nd}
{}_{(1)}\bra{O^{++}}&\frac{L_1^{\n_b}}{\n_b!}\,{}_{(2)}\bra{O^{++}}\frac{L_1^{\n_d}}{\n_d!}J^{-}_0\Big[\big(G_{-\frac12}^{+1}G_{-\frac12}^{+2} + \frac{1}{2} J^+_0 L_{-1}\big)O_{(1)}^{--} O_{(2)}^{--}\Big](\tilde t,\tilde y)\,\ket{0}_{(1)}\ket{0}_{(2)}\\&=z^2\bar z
\Big[-\frac{1}{2}\,{}_{(1)}\bra{O^{++}}\frac{L_1^{\n_b}}{\n_b!}O^{--}_{(1)}(z,\bar{z})\ket{0}_{(1)}\,\,{}_{(2)}\bra{O^{++}}\frac{L_1^{\n_d}}{\n_d!}J^{-}_0 \big(J^+_0L_{-1}O^{--}_{(2)}\big)(z,\bar{z})\ket{0}_{(2)}\nonumber\\&
\qquad+\frac{1}{2}\,{}_{(1)}\bra{O^{++}}\frac{L_1^{\n_b}}{\n_b!}\big(L_{-1}O^{--}_{(1)}\big)(z,\bar{z})\ket{0}_{(1)}\,\,{}_{(2)}\bra{O^{++}}\frac{L_1^{\n_d}}{\n_d!}J^{-}_0 \big(J^+_0O^{--}_{(2)}\big)(z,\bar{z})\ket{0}_{(2)}\Big]\,.\nonumber
\end{align}

Using standard manipulations, together with the commutation relation in Eq.~\eqref{L1 descendant primary} and the anomaly-free algebra \eqref{anomaly free algebra}, one can rewrite this amplitude as
\begin{equation}\label{process GGOO 110nb1nd 2}
\begin{aligned}
{}_{(1)}\bra{O^{++}}&\frac{L_1^{\n_b}}{\n_b!}\,{}_{(2)}\bra{O^{++}}\frac{L_1^{\n_d}}{\n_d!}J^{-}_0\Big[\big(G_{-\frac12}^{+1}G_{-\frac12}^{+2} + \frac{1}{2} J^+_0 L_{-1}\big)O_{(1)}^{--} O_{(2)}^{--}\Big](\tilde t,\tilde y)\,\ket{0}_{(1)}\ket{0}_{(2)}\\&
=\frac{1}{2}\Big[ - (\n_d+1)+ (\n_b+1)\Big]e^{i(\n_b+\n_d+2)\tilde t+i(\n_b+\n_d)\tilde y}\\&
=\frac{1}{2}(\n_b-\n_d)e^{i(\n_b+\n_d+2)\tilde t+i(\n_b+\n_d)\tilde y}\,.
\end{aligned}
\end{equation}
Already we see that the expectation value is non-zero only when $\n_b\neq \n_d$.

We now dress this with the combinatorial factor as before. The relevant contribution can be represented as
\begin{equation}
\begin{aligned}\label{amplitude 2nd example}
 \big(GG (O\cdot O)\big)^{0-} &(\tilde t,\tilde y)\:\!  \Big[\Big(\ket{0}_1^{N-p-q}\Big)\Big(\frac{1}{\n_b!}L_{-1}^{\n_b}\ket{O^{--}}\Big)^p\Big( \frac{1}{\n_d!}J^+_{0}L_{-1}^{\n_d}\ket{O^{--}}\Big)^q\Big]\\
=\;&\Big((\n_b-\n_d)e^{i(\n_b+\n_d+2)\tilde t+i(\n_b+\n_d)\tilde y}\Big)\frac{(p+1)(q+1)}{N}\\&
\times\Big[\Big(\ket{0}_1^{N-p-q-2}\Big)\Big(\frac{1}{\n_b!}L_{-1}^{\n_b}\ket{O^{--}}\Big)^{p+1}\Big( \frac{1}{\n_d!}J^+_{0}L_{-1}^{\n_d}\ket{O^{--}}\Big)^{q+1}\Big]\,.
\end{aligned}
\end{equation}
The first term on the RHS is the contribution of the fundamental process, given by twice the result in Eq.~\eqref{process GGOO 110nb1nd 2}. The second term comes requiring that the 
normalization of the two sides of the equality are the same.
The expectation value of the single-particle operator is then obtained by combining  Eq.~\eqref{amplitude 2nd example} with the normalization factors we suppressed from  Eq.~\eqref{def sigma3 two}, along with the relation between the CFT and the supergravity coefficients in Eq.~\eqref{BD bd coiffuring}. This gives
\begin{equation}\label{VeV GGOO 110nb1nd}
\begin{aligned}
\big\langle GG\tilde\Sigma_3^{0-}\big\rangle&\,=\,\frac{1}{8\sqrt{3}\;\! N^{1/2}}\big\langle \big(GG( O\cdot O)\big)^{0-}\big\rangle\\&
\,=\,\frac{1}{4\sqrt{3}}\frac{\bar p\,\bar q}{N^{3/2}} \;\! \frac{A^2}{BD}\;\!(\n_b-\n_d)\;\! e^{i(\n_b+\n_d+2)\tilde t+i(\n_b+\n_d)\tilde y} \\&
\,=\,\frac{\sqrt{N}}{8\sqrt{3}}\;\!\mathsf{a}^2\,\mathsf{b}\;\!\mathsf{d}\;\!(\n_b-\n_d)\;\!e^{i(\n_b+\n_d+2)\tilde t+i(\n_b+\n_d)\tilde y}\,,\\
\big\langle GG\tilde\Sigma_3^{0+}\big\rangle&\,=\,\big( \big\langle GG\tilde\Sigma_3^{0-}\big\rangle\big)^*\\
&\,=\,\frac{\sqrt{N}}{8\sqrt{3}}\;\! \mathsf{a}^2\,\mathsf{b}\;\!\mathsf{d}\;\!(\n_b-\n_d)\;\!e^{-i(\n_b+\n_d+2)\tilde t-i(\n_b+\n_d)\tilde y}\,.
\end{aligned}
\end{equation}
These CFT expectation values are holographically encoded in the expansion of the gauge-invariant vector field $\bm Z^6_{k=1}$. 
The supergravity solution is obtained combining \cite[Eqs.~(D.15)-(D.29)]{Mayerson:2020tcl}
with the holomorphic functions $F_0$ and $F_1$ in Eq.~\eqref{eq:F0-F1}.
One can check that, as in the example studied in Section \ref{ssc:Normalizing the supercharged holographic dictionary}, the AdS$_3$ scalar field $U^6_{k=1}$ vanishes on this background. Eq.~\eqref{ZA gauge invariant} then implies that $\bm Z^6_{k=1}=Z^6_{k=1}$. The vector fields are given by
\begin{equation}\label{Z6 110nb1nd}
\ba
Z^{6(+-)}_{k=1}&\,=\,-\frac{\mathsf{a}^2\, \mathsf{b}\, \mathsf{d}}{4}\:\!(\n_b-\n_d)\;\!e^{i((\n_b+\n_d+2)\tilde t+(\n_b+\n_d)\tilde y)}\frac{\tilde r^{\n_b+\n_d-1}}{(1+\tilde r^2)^{\frac{(\n_b+\n_d+4)}{2}}}\Big(-id\tilde r+\tilde r(\tilde r^2+1)(d\tilde t+d\tilde y)\Big)\,, \\
Z^{6(--)}_{k=1}&\,=\,-\frac{\mathsf{a}^2\, \mathsf{b}\, \mathsf{d}}{4}\:\!(\n_b-\n_d)\;\!e^{-i((\n_b+\n_d+2)\tilde t+(\n_b+\n_d)\tilde y)}\frac{\tilde r^{\n_b+\n_d-1}}{(1+\tilde r^2)^{\frac{(\n_b+\n_d+4)}{2}}}\Big(id\tilde r+\tilde r(\tilde r^2+1)(d\tilde t+d\tilde y)\Big)\,,
\ea
\end{equation}
where we have used the normalization in Eq.~\eqref{ads fields mass matrix diag} and the relation between the CFT and the supergravity modes in Eq.~\eqref{BD bd coiffuring}.

The coefficient $\beta$ in Eq.~\eqref{ansatz supercharged dictionary} is obtained performing the asymptotic expansion~\eqref{asympt expansion Z6 Z7} of the three-dimensional vectors  \eqref{Z6 110nb1nd} and comparing it with the CFT result in Eq.~\eqref{VeV GGOO 110nb1nd}.
We find that
\begin{equation}
\beta=-\frac{N^{1/2}}{2\sqrt{3}} \,,
\end{equation}
thus explicitly verifying that $\alpha=\beta$. 

We emphasize again that this amplitude is non-vanishing only when $n_b\neq n_d$. This can be interpreted as the CFT telling us that for the set of states~\eqref{state 110nb1nd}, when $n_b\neq n_d$ the bulk solution must involve an extra field that is not turned on when $n_b = n_d$. This is precisely the field that was introduced in the more general coiffuring proposal of~\cite{Heidmann:2019zws}. 
So we have seen that a very non-trivial  holographic test of this more general coiffuring is passed.

\subsection{Holographic test of hybrid supercharged superstrata}\label{ssc:cross check GGsigma3}

At this point we have fixed all coefficients of the holographic dictionary in the sector in which we work. We now use the dictionary to make a precision holographic test of a ``hybrid'' superstratum solution that combines non-supercharged and supercharged elements. Our computation tests both the proposed holographic dictionary for hybrid superstrata and the supergravity coiffuring procedure for combining non-supercharged and supercharged modes of~\cite{Heidmann:2019zws,Heidmann:2019xrd}.

This test also serves as an additional non-trivial cross-check of the operator mixing in the dictionary. Note that this operator mixing has already passed thorough holographic tests~\cite{Giusto:2019qig}. The tests in~\cite{Giusto:2019qig}  were performed in a different basis to the single-particle basis, however our results are equivalent, as demonstrated in Appendix~\ref{sec:Gauge fixed dictionary}.
The test that follows involves a non-trivial and delicate cancellation between a set of terms. 

We consider the multi-mode hybrid superstratum composed of modes $(\k_1,\m_1,\n_1,\q_1) = (2,1,0,0)$, and $(\k_2,\m_2,\n_2,\q_2) = (2,1,1,1)$, constructed in~\cite[App.~(B.1)]{Heidmann:2019xrd}.  There, the solution was given in terms of two holomorphic functions $F$ and $S$ of the complex variable $\xi$ defined in Eq.~\eqref{variable xi}. In our conventions, we have
\be
\label{eq:F-S}
F(\xi) \;=\; b \,, \qquad
S(\xi) \;=\; d \, \frac{\xi}{6} \,.
\ee
The supergravity solution is given explicitly in \cite[Eqs.\;(6.8),\,(6.9),\,(B.1)--(B.12)]{Heidmann:2019xrd} and so we shall not reproduce it here.
Performing the Kaluza-Klein reduction of this background, one obtains that the gauge-invariant field  $\bm Z^6_{k=1}$ in Eq.~\eqref{ZA gauge invariant} vanishes. The holographic dictionary in Eq.~\eqref{ansatz supercharged dictionary} then predicts that the expectation values of the single-particle operator $(GG\tilde\Sigma_3)^{0a}$ on the dual CFT state must vanish. We will now explicitly check that this is indeed the case.

The proposed dual CFT state is
\begin{equation}\label{CFT dual example 3}
\sum_{p,q} \Big(A\ket{0}_1\Big)^{N-2p-2q} \Big(B J^+_0\ket{O^{--}_2}\Big)^p\Big(D\big(G_{-\frac12}^{+1}G_{-\frac12}^{+2}+ \frac12  J^+_0 L_{-1}\big)\ket{O^{--}_2}\Big)^{q}\,.
\end{equation}
The CFT coefficients $(B,D)$ and the supergravity Fourier coefficients $(b,d)$ in~\eqref{eq:F-S} are again related via Eq.~\eqref{q=1 AB ab}, which in the present example becomes 
\be
\label{BD bd multimode}
\frac{B}{\sqrt{N}} \;=\;\frac{R}{2\sqrt{2\,Q_1Q_5}}\,b \,, \qquad \frac{D}{\sqrt{N}} \;=\;  \frac{R}{3\sqrt{2\,Q_1Q_5}}\,d\,.
\ee

 The only $SU(2)_L\times SU(2)_R$ component of the single-particle operator $(GG\tilde\Sigma_3)^{0a}$ that can have a non-vanishing expectation value is $(GG\tilde\Sigma_3)^{00}$.  Indeed, this single-particle operator contains three operators that have non-zero expectation values in the state~\eq{CFT dual example 3}: the single-trace operators $(GG\Sigma_3)^{00}$, $(GG\Omega)^{00}$ and the double-trace operator $(GG J\bar J)^{00}$.
 
First, we compute the expectation value of $(GG\tilde\Sigma_3)^{00}$, which arises from the basic process in which one $B$-type strand plus one $A$-type vacuum strand are converted into one $D$-type strand plus one $A$-type vacuum strand:
\begin{equation}\label{GGSigma 210 2111 process}
\begin{aligned}
\bigg({}_1\bra{0}\bra{O^{++}_2}&\big(-G_{\frac12}^{-1}G_{\frac12}^{-2} + \frac{1}{2} J^-_0 L_{+1}\big)\bigg)\bigg(\big(G_{-\frac12}^{+1}G_{-\frac12}^{+2}+ \frac12  J^+_0 L_{-1}\big)\Sigma^{-0}_3\bigg) \bigg(J^{+}_0\ket{O^{--}_2}\ket{0}_1\bigg)\\
&=\;3\,{}_1\bra{0}\bra{O^{++}_2}\Sigma^{-0}_3 J^{+}_0\ket{O^{--}_2}\ket{0}_1\\
&=\;-3\Big(\sqrt{2}\,{}_1\bra{0}\bra{O^{++}_2}\Sigma^{00}_3\ket{O^{--}_2}\ket{0}-{}_1\bra{0}\bra{O^{++}_2}J^+_0 \Sigma^{-,0}_3\ket{O^{--}_2}\ket{0}_1\Big)\\
&=\;-3\sqrt{2}\,{}_1\bra{0}\bra{O^{++}_2}\Sigma^{00}_3\ket{O^{--}_2}\ket{0}_1\\&=-\frac{1}{\sqrt{2}}\,.
\end{aligned}
\end{equation}
In the equation above, following the prescription after Eq. \eqref{eq:doubletraces}), the descendant is normalized so that it has the same norm as the highest weight state, that is $\Sigma_3^{00}=\frac{1}{\sqrt{2}}[J^+_0,\Sigma_3^{-0}]$. 
This three-point function is independent of the insertion point of the light operator on the cylinder. 
The last equality follows from
\begin{equation}
{}_1\bra{0}\bra{O^{++}_2}\Sigma^{00}_3 \ket{O^{--}_2}\ket{0}_1\;=\;\frac{1}{6}\,,
\end{equation}
which can be derived using the covering-space method of Lunin-Mathur~\cite{Lunin:2000yv,Lunin:2001pw} analogously to the computation in \cite[App.~(B.2)]{Giusto:2019qig}.

We now use the basic amplitude in Eq.~\eqref{GGSigma 210 2111 process} to compute the expectation value of the single-trace operator $(GG\Sigma_3)^{00}$ on the full coherent state~\eqref{CFT dual example 3}. The relevant contribution can be represented by
\begin{equation}
\begin{aligned}
\big(GG\Sigma_3\big)^{00}\Big[&\big(\ket{0}_1^{N-2p-2q}\big) \Big( J^+_0\ket{O^{--}_2}\Big)^p\Big(\big(G_{-\frac12}^{+1}G_{-\frac12}^{+2}+ \frac12  J^+_0 L_{-1}\big)\ket{O^{--}_2}\Big)^{q}\Big]\\
&=-\frac{2\sqrt{2}}{3}\big(N-2(p+q)\big)(q+1)\\
&\qquad\times\Big[\Big(\ket{0}_1^{N-2(p+q)}\Big) \Big( J^+_0\ket{O^{--}_2}\Big)^{p-1}\Big(\big(G_{-\frac12}^{+1}G_{-\frac12}^{+2}+ \frac12  J^+_0 L_{-1}\big)\ket{O^{--}_2}\Big)^{q+1}\Big]\,.
\end{aligned}
\end{equation}
The overall factor on the RHS is obtained as before by requiring that the norms on the two sides of the equality sign are the same. In doing so, we combined Eq.~\eqref{GGSigma 210 2111 process} with a combinatorial factor that represents the number of ways that the operator $(GG\Sigma_3)^{00}$ can act on the state on the LHS. In particular, it can act on any of the $(N-2p-2q)p\;\!$ pairs of strands $\ket{0}_1\big(J^+_0\ket{O^{--}}_2\big)$ and can cut-and-join these in two inequivalent ways.

We now use Eq.~\eqref{q=1 peak AB} to express the average number of strands in a coherent state with the CFT coefficients $A,B,D$. We obtain that, in the large $N$-limit,
\begin{equation}\label{GGSigma 210 2111 VeV}
\big\langle \big(GG\Sigma_3\big)^{00}\big\rangle\;=\;-\frac{2\sqrt{2}}{3}(N-2\bar p-2\bar q)\,\bar q\,\frac{B}{D}\;=\;-\sqrt{2}A^2BD \,.
\end{equation}

Second, we compute the expectation value of the single-trace operator $GG\Omega$, which arises from the basic process in which a $B$-type strand is converted into a $D$-type strand:
\begin{equation}\label{GGOmega 210 2111 process}
\begin{aligned}
\bigg(\bra{O^{++}_2}&\big(-G_{\frac12}^{-1}G_{\frac12}^{-2} + \frac{1}{2} J^-_0 L_{+1}\big)\bigg)\bigg(\big(G_{-\frac12}^{+1}G_{-\frac12}^{+2}+ \frac12  J^+_0 L_{-1}\big)\Omega^{-0}\bigg) \bigg(J^{+}_0\ket{O^{--}_2}\bigg)\\
&=\;3\,\bra{O^{++}_2}\Omega^{-0}_3 J^{+}_0\ket{O^{--}_2}\\
&=\;-3\,\sqrt{2}\bra{O^{++}_2}\Omega^{00}\ket{O^{--}_2}\\&
=-3\sqrt{2}\,.
\end{aligned}
\end{equation}
The last equality follows from \cite[Eq.~(5.40)]{Giusto:2019qig}.
To compute the expectation value of the operator on the full coherent state we again combine the above result with a combinatorial factor. Doing so, we obtain
\begin{align}\label{GGOmega 210 2111 combinatorics}
\big(GG\Omega\big)^{00}&\Big[\Big(\ket{0}_1^{N-2p-2q}\Big) \Big( J^+_0\ket{O^{--}_2}\Big)^p\Big(\big(G_{-\frac12}^{+1}G_{-\frac12}^{+2}+ \frac12  J^+_0 L_{-1}\big)\ket{O^{--}_2}\Big)^{q}\Big]\\
&=-2\sqrt{2}\big(q+1\big)\Big[\Big(\ket{0}_1^{N-2p-2q}\Big) \Big( J^+_0\ket{O^{--}_2}\Big)^{p-1}\Big(\big(G_{-\frac12}^{+1}G_{-\frac12}^{+2}+ \frac12  J^+_0 L_{-1}\big)\ket{O^{--}_2}\Big)^{q+1}\Big]\,.\nonumber
\end{align}
Here we have used the fact that the operator $(GG\Omega)$ can act on any of the $p$ strands of type $J^+_0\ket{O^{--}}_2$, and we have matched the norms on the two sides of the equation.
The expectation value on the full coherent state then follows from Eq.~\eqref{q=1 peak AB}. We obtain
\begin{equation}\label{GGOmega 210 2111 VeV}
\big\langle \big(GG\Omega\big)^{00}\big\rangle\;=\;-\frac{2\sqrt{2}\,\bar q\, B}{D}\;=\;-\frac{3\sqrt{2}}{N}BD\big(A^2+3D^2+2B^2\big)\,.
\end{equation}
For later convenience, in the last equality we have used the strand budget constraint~\eqref{strand budget A Bi}.

The final operator that contributes to the expectation value of $(GG\tilde\Sigma_3)^{00}$ is
\begin{equation}\label{GGJbarJ example 3}
    (GGJ\bar J)^{00} \,=\,\sqrt{2}\Big(\big(G_{-\frac12}^{+1}G_{-\frac12}^{+2}+ \frac12  J^+_0 L_{-1} \big) J^- \Big)\bar J^3 \,,
\end{equation}
where the factor of $\sqrt{2}$ follows from the normalization of the $SU(2)_R$ descendant.

This operator contributes through the basic process in which a $B$-type strand is converted into a $D$-type strand.  This process is mediated only by the holomorphic part of the operator $(GGJ\bar J)^{00}$. Both $B$-type and $D$-type strands are eigenstates of $\bar J^3_0$, so we treat this contribution separately below. Focusing for now on the holomorphic part, we have the amplitude
\begin{equation}\label{GGJJ 210 2111 process}
\begin{aligned}
\bigg(&\bra{O^{++}_2}\big(-G_{\frac12}^{-1}G_{\frac12}^{-2} + \frac{1}{2} J^-_0 L_{+1}\big)\bigg)\bigg(\big(G_{-\frac12}^{+1}G_{-\frac12}^{+2}+ \frac12  J^+_0 L_{-1}\big)J^- \bigg) \bigg(J^{+}_0\ket{O^{--}_2}\bigg)\\
&=-6\,\bra{O^{++}_2}J^3\ket{O^{--}_2}\\
&=6\,.
\end{aligned}
\end{equation}
We now compute the expectation value of the multi-trace operator in the full coherent state~\eqref{CFT dual example 3}. We include the antiholomorphic part at this point. The relevant  contribution can be represented by
\begin{align}
&\big(GGJ\bar J\big)^{00}\Big[\Big(\ket{0}_1^{N-2p-2q}\Big) \Big( J^+_0\ket{O^{--}_2}\Big)^p\Big(\big(G_{-\frac12}^{+1}G_{-\frac12}^{+2}+ \frac12  J^+_0 L_{-1}\big)\ket{O^{--}_2}\Big)^{q}\Big]\\
&=-4\sqrt{2}\,\big(p+q\big)\big(q+1\big)\Big[\Big(\ket{0}_1^{N-2p-2q}\Big) \Big( J^+_0\ket{O^{--}_2}\Big)^{p-1}\Big(\big(G_{-\frac12}^{+1}G_{-\frac12}^{+2}+ \frac12  J^+_0 L_{-1}\big)\ket{O^{--}_2}\Big)^{q+1}\Big]\,.\nonumber
\end{align}
The prefactor on the RHS is a combination of the basic amplitude \eq{GGJJ 210 2111 process}, the action of $\bar J$, and a combinatorical factor. The combinatorics has two parts: first, the operator $\bar J$ can act either on one of the $p$ strands of type $J^+_0\ket{O^{--}_2}$ or on one of the $q$ strands of type $GG\ket{O^{--}_2}$. Second, the operator $GGJ$ can only act upon strands of type $GG\ket{O^{--}_2}$. 
Using Eqs.~\eqref{q=1 peak AB} and~\eqref{strand budget A Bi} we obtain
\begin{equation}\label{GGJJ 210 2111 VeV}
\big\langle \big(GG J\bar J\big)^{00}\big\rangle\;=\;-\frac{4\sqrt{2}\bar q (\bar p+\bar q)B}{D}\;=\;-6\sqrt{2}BD\Big(\frac{3}{2}D^2+B^2\Big)\,.
\end{equation}
Finally, we combine the three contributions in Eqs.~\eqref{GGSigma 210 2111 VeV},~ \eqref{GGOmega 210 2111 VeV} and~\eqref{GGJJ 210 2111 VeV} using the definition of the single-particle operator $GG\tilde\Sigma$ in Eq.~\eqref{Normalized supercharged operators} to obtain the anticipated cancellation:
\begin{equation}
\big\langle GG\tilde{\Sigma}_3^{00}\big\rangle=0 \,.
\end{equation}
This result agrees with the vanishing of the dual AdS$_3$ field $\bm Z^6_{k=1}$ in the proposed dual supergravity solution. 
Thus we see that the proposed holographic dictionary for hybrid superstrata has passed a non-trivial test. This computation also represents a non-trivial cross-check of the operator mixing involved in the single-particle operator dual to the AdS$_3$ vector field $\bm Z^6_{k=1}$.

\subsection{Summary of precision holographic dictionary}

For convenient reference we record here a summary of the precision holographic dictionary for single-particle scalar operators of dimension one and two, together with our new entries for the superdescendant operators $GG\tilde{\Sigma}_3$, $GG\tilde{O}_2$.
\begin{align}\label{Summary}
\frac{1}{\sqrt{N}}\langle J^{\pm}\rangle&=-\frac{\sqrt{N}}{\sqrt{2}} \Big[A^{\mp(+)}_{k=1}\Big]\,,
\hspace{6.8 em}
\frac{1}{\sqrt{N}}\langle \bar J^{\pm}\rangle=-\frac{\sqrt{N}}{\sqrt{2}} \Big[A^{\mp(-)}_{k=1}\Big]\,,
\cr
\frac{1}{\sqrt{N}}\langle J^{3}\rangle&=-\frac{\sqrt{N}}{2} \Big[A^{0(+)}_{k=1}\Big]\,,\hspace{7.3 em} \frac{1}{\sqrt{N}}\langle \bar J^{3}\rangle=-\frac{\sqrt{N}}{2}\Big[ A^{0(-)}_{k=1}\Big]\,,
\cr
\frac{\sqrt{2}}{N}\big\langle\Sigma_2^{\alpha\dot \alpha}\big\rangle&=(-1)^{\alpha\dot\alpha}\frac{\sqrt{N}}{\sqrt{2}} \Big[s^{(6)(-\alpha,-\dot\alpha)}_{k=1}\Big]\,,
\qquad  
\frac{1}{\sqrt{N}}\big\langle O^{\alpha\dot\alpha}\big\rangle=(-1)^{\alpha\dot\alpha}\frac{\sqrt{N}}{\sqrt{2}} \Big[s^{(7)(-\alpha,-\dot\alpha)}_{k=1}\Big]\,,
\cr
\big\langle\tilde{\Omega}^{a,\dot a}\big\rangle&=-(-1)^{a+\dot a}\frac{\sqrt{N}}{\sqrt{2}}\Big[\tilde{\sigma}^{(-a,-\dot a)}_{k=2}\Big]\,,
\\
 \big\langle\tilde{\Sigma}_3^{a,\dot a}\big\rangle&=(-1)^{a+\dot a}\frac{\sqrt{N}}{\sqrt{2}}\Big[s^{(6)(-a,-\dot a)}_{k=2}\Big]\,,\hspace{3.8em}
\big\langle\tilde{O}_2^{a,\dot a}\big\rangle=(-1)^{a+\dot a}\frac{\sqrt{N}}{\sqrt{2}}\Big[s^{(7)(-a,-\dot a)}_{k=2}\Big]\,,
\cr
\big\langle GG\tilde{\Sigma}_3^{(0,a)}\big\rangle&=-\frac{\sqrt{N}}{2\sqrt{3}}\Big[ Z^{6(-a,-)}_{k=1}\Big]\,, \hspace{5em} \langle GG\tilde{O}_2^{(0, a)}\big\rangle =-\frac{\sqrt{N}}{2\sqrt{3}}\Big[ Z^{7(-a,-)}_{k=1}\Big]\,.
\nonumber
\end{align}

\section{Discussion}
\label{sec:disc}

In this paper we have derived the precision holographic dictionary in a new sector of superdescendants of scalar operators of dimension two. In doing so we also expressed the existing precision dictionary in the single-particle basis. Our results for the dictionary in these sectors are summarized in Eq.~\eqref{Summary}.

We considered the recent proposal that single-particle supergravity fluctuations around global AdS are holographically dual to half-BPS operators that are orthogonal to all multi-trace operators~\cite{Aprile:2018efk,Aprile:2020uxk}. We emphasized that this proposal is not sufficient to determine the single-particle basis in the D1-D5 CFT. We thus combined this proposal with the mixing between single-trace operators worked out in~\cite{Taylor:2007hs} to obtain the refined dictionary for operators of dimension one, and scalar operators of dimension two summarized in the first five lines of Eq.~\eqref{Summary}. We then derived the new part of the dictionary in the last line of Eq.~\eqref{Summary}.

We note that the dictionary in Eq.~\eqref{Summary} has a lot of structure: (i) the normalization coefficients in the various sectors respect the $SO(n)$ symmetry of the supergravity theory; (ii) single-particle states are orthogonal to all multi-particle states; (iii) the mixing between single-trace operators of dimension two is fixed as described above. Furthermore, the overall normalization of each subsector of the dictionary has been calibrated successively on the well-established holographic description of two-charge microstates with $\mathbb{R}^4$ base polarizations, and non-supercharged superstrata, in~\cite{Giusto:2019qig}. Our computation of the coefficient $\beta$ in Section~\ref{sec:beta-calc} can also be regarded as a calibration on a non-supercharged superstratum, and is consistent with the value of $\alpha$ that we obtained by comparison with a supercharged superstratum.

Having derived the new sector of the dictionary for superdescendant operators, we used this to perform tests of a set of non-supercharged superstrata of particular interest, and a set of `hybrid' superstrata that involve both supercharged and non-supercharged elements. We found precise agreement between gravity and CFT.

The agreement we find in this work does not prove that the proposal for the dual CFT states of supercharged superstrata is precisely correct. The reason is simply the usual limitation of precision holographic studies: for a given superstratum solution in supergravity, and a given precision involving a finite set of expectation values of light operators, there can be other CFT states that have the same values of those correlators.
However, our results provide evidence in support of the proposed holographic description of both non-supercharged and supercharged superstrata, and demonstrate that the proposal for the dual CFT states passes all available state-of-the-art tests.

Our results open up various possibilities for future work. The most obvious ones are to use our dictionary~\eqref{Summary} to perform precision tests of other families of solutions, and to generalize the dictionary  both to primaries of higher dimension and to other superdescendants. Of course, the higher one goes in dimension, the more complex the task of constructing the explicit precision dictionary.
Our present extension of the dictionary has allowed us to test the main features of the  supercharged and hybrid superstrata that have been constructed to date. Future constructions of more general classes of superstrata, or the desire for input into the construction of such new solutions, may provide particular motivation to expand the dictionary into further new sectors.

Looking beyond three-point functions, there has been much progress on holographic four-point functions in the D1-D5 system in recent years. Heavy-heavy-light-light (HHLL) four-point functions were computed in~\cite{Galliani_2016,Galliani:2017jlg,Bombini:2017sge}, and generalized to correlators in which the heavy operators are simple non-supercharged and supercharged superstrata~\cite{Bombini:2019vnc}.
This enabled new results on LLLL four-point functions to be derived~\cite{Giusto:2018ovt,Rastelli_2019,Giusto:2019pxc,Giusto:2020neo,Aprile:2021mvq}.
More recently, HHLL correlators in the Regge limit have been computed~\cite{Giusto:2020mup,Ceplak:2021wak}, including those in which the light operators are multi-trace operators~\cite{Ceplak:2021wzz}. Our results on resolving the operator mixing among light operators should enable new classes of four-point functions to be studied.

Holography has been an invaluable tool in the development of the fuzzball description of black hole microstates, and in this work we have further enlarged the precision holographic dictionary. We expect that our results should prove useful to further our understanding of the heavy bound states that comprise the quantum description of black holes in String Theory.

\vspace{5mm}
\section*{Acknowledgements}

We thank Davide Bufalini, Stefano Giusto, Bin Guo, Emil Martinec, Samir Mathur, Rodolfo Russo, Michele Santagata, Kostas Skenderis and Marika Taylor for fruitful discussions.
The work of SR was supported by a Royal Society URF Enhancement Award.
The work of DT was supported by a Royal Society Tata University Research Fellowship.

\vspace{7mm}

\appendix

\section{Harmonics on S${}^3$ and AdS${}_3$}
\label{sec:harm S3 and AdS3}
\subsection{Harmonics on S${}_3$}
\subsubsection{Spherical harmonics}
The spherical harmonics on S$^3$ are a representation of the isometry group of the three-sphere $SO(4)\simeq SU(2)_L\times SU(2)_R$. 
We will use spherical coordinates in the $\mathbb{R}^4$ base space that are related to the  Cartesian coordinates via 
\begin{equation}\label{cartesian to spherical coordinates} \begin{split}
x^1&=r\sin\theta\cos\phi \;,\qquad x^2=r\sin\theta\sin\phi\,,\\
x^3&=r\cos\theta\cos\psi \;,\qquad x^4=r\cos\theta\sin\psi\,,
\end{split}
\end{equation}
where $\theta\in[0,\frac{\pi}{2}]$ and $\psi,\phi\in[0,2\pi)$. With this coordinate choice, the S$^3$ line element $ds_3^2$ is given by $ds_3^2=d\theta^2+\sin^2\theta d\phi^2+\cos^2\theta d\psi^2 $. We use conventions in which  $\epsilon_{\theta\phi\psi}$=1.
The generators of the isometry group of S$^3$, written in terms of the standard $SU(2)$ generators, are
\begin{equation}\label{generators isometry S3} \begin{split}
	J^{\pm}&=\frac{1}{2}  e^{\pm i(\phi+\psi)}  \big(\pm \partial_\theta +i \cot\theta  \partial_\phi  - i \tan\theta  \partial_\psi\big)\, , \quad 
	J^3=-\frac{i}{2}\big(\partial_{\phi}+\partial_{\psi}\big)\,,
	\\
	\bar{J}^{\pm}&=\frac{1}{2}  e^{\pm i(\phi-\psi)} \big(\mp \partial_\theta -i \cot\theta  \partial_\phi- i \tan\theta \partial_\psi\big)
	\, , \quad 
	\bar{J}^3=-\frac{i}{2}\big(\partial_{\phi}-\partial_{\psi}\big)\,,
\end{split}
\end{equation}
which satisfy the $SU(2)_L\times SU(2)_R$ algebra:
\begin{equation}
\begin{aligned}\label{SU2timesSU2 algebra}
&\big[J^+,J^-\big]=2J^3\,, \qquad \big[J^3,J^+\big]=J^+\,,  \qquad \big[J^3,J^-\big]=-J^-\,,\\
&\big[\bar J^+,\bar J^-\big]=2\bar J^3\,, \qquad \big[\bar J^3,\bar J^+\big]=\bar J^+\,,  \qquad \big[\bar J^3,\bar J^-\big]=-\bar J^- \,.
\end{aligned}
\end{equation}
The left quadratic Casimir operator is $J^2=\frac{1}{2}\big(J^+J^-+J^-J^+\big)+\big(J^3\big)^2$, and likewise for $\bar J^2$. A state in a representation with principal quantum number $j_\su$ has $J^2$ eigenvalue $j_\su(j_\su+1)$.

Degree $k$ scalar harmonics live in the $(j_\su,\bar{j}_\su)=(k/2,k/2)$ representation of $SU(2)_L\times SU(2)_R$. We denote these by 
 $Y_k^{m,\bar m}$, and $(m,\bar{m})$ are the spin charges under $(J^3,\bar{J}^3)$. They satisfy the following Laplace equation:
\begin{equation}\label{laplace eq sph harm}
\Box_{S_3}Y^{m_1,\bar m_2}_k=-k(k+2) Y^{m_1,\bar m_2}_k \,.
\end{equation}

Denoting the volume of S$^3$ by $\Omega_3=2\pi^2$, we use normalized spherical harmonics
\begin{equation}\label{normalization sph harmonic} \begin{split}
\int Y_{k_1}^{*m_1,\bar{m}_1}Y_{k_2}^{m_2,\bar{m}_2}\;=\;\Omega_3 \,\delta_{k_1,k_2}\delta^{m_1,m_1}\delta^{\bar{m}_1,\bar{m}_2}\,.
\end{split}
\end{equation}

One can generate the degree $k$ scalar spherical harmonic wavefunctions acting with the lowering operators in \eqref{generators isometry S3} on the highest-weight wavefunctions, which are 
\begin{equation}\label{highest sph harmonic} \begin{split}
Y_k^{\pm\frac{k}{2},\pm\frac{k}{2}}=\sqrt{k+1}\sin^k\theta e^{\pm i k \phi} \, .
\end{split}
\end{equation}
The degree $k=1,2$ normalized scalar spherical harmonics are given by:
\begin{equation}	\label{sph har k=1} \begin{split}
Y_1^{+\frac{1}{2},+\frac{1}{2}}= \sqrt{2} \sin\theta \,e^{i \phi}\quad &, \quad 
Y_1^{+\frac{1}{2},-\frac{1}{2}}=  \sqrt{2} \cos\theta \,e^{i \psi} \,,\\
Y_1^{-\frac{1}{2},+\frac{1}{2}}= - \sqrt{2} \cos\theta \,e^{-i \psi}\quad &, \quad 
Y_1^{-\frac{1}{2},-\frac{1}{2}}=  \sqrt{2} \sin\theta \,e^{-i \phi}\,;
\end{split}
\end{equation}

\begin{equation}\label{sph har k=2}	 \begin{split}
Y_2^{+1,+1}&=\sqrt{3} \sin^2\theta\, e^{2 i \phi}\quad , \quad 
Y_2^{+1,0}= \sqrt{6} \sin\theta \cos\theta \,e^{ i (\phi+\psi)} \quad , \quad  Y_2^{+1,-1}=\sqrt{3} \cos^2\theta e^{2 i \psi} \,,\\
Y_2^{0,+1}&=-\sqrt{6} \sin\theta \cos\theta \,e^{ i (\phi-\psi)} \quad , \quad  Y_2^{0,0}=-\sqrt{3} \cos2\theta\,\quad , \quad  Y_2^{0,-1}=\sqrt{6} \sin\theta \cos\theta \,e^{ -i (\phi-\psi)} \\
Y_2^{-1,+1}&=\sqrt{3} \cos^2\theta \,e^{-2 i \psi} \quad , \quad Y_2^{-1,0}=-\sqrt{6} \sin\theta \cos\theta \,e^{ -i (\phi+\psi)} \quad , \quad 
Y_2^{-1,-1}=\sqrt{3} \sin^2\theta \,e^{-2 i \phi} \,.
\end{split}
\end{equation}
We also define the following triple overlap,
\begin{equation}\label{tripleoverlap}
\int Y^{M_k,\bar M_k}_k Y^{m_1,\bar m_1}_{k_1} \Big(Y^{m_2,\bar m_2}_{k_2}\Big)^*=\Omega_3 \,\,a^{M_k,\bar M_k}_{(m_1,\bar m_1)(m_2,\bar m_2)} \,.
\end{equation}

\subsubsection{Vector harmonics}

Degree $k$ left vector harmonics live in the
$(j_\su,\bar{j}_\su)=(\frac{k+1}{2},\frac{k-1}{2})$ representation of $SU(2)_L\times SU(2)_R$. We denote these by $Y^{m,\bar m}_{\LL,k}$ where $\LL$ stands for left; we shall suppress the label $\LL$ when we write explicit $S^3$ vector indices (which will be denoted by $a,b$).
Similarly, degree $k$ right vector harmonics
have $(j_\su,\bar{j}_\su)=(\frac{k-1}{2},\frac{k+1}{2})$ and we denote them by $Y^{m,\bar m}_{\RR,k}$. We shall use $Y^{m,\bar m}_{v,k}$ to denote a vector harmonic which can be either right or left. Vector harmonics satisfy
\begin{equation}\label{laplace eq vector harm}
\nabla^2_{S^3}Y^{m,\bar m}_{v,k}=-(k^2+2k-1)Y^{m,\bar m}_{v,k} \,, \qquad D^a(Y^{m,\bar m}_{v,k})_a=0   \,,
\end{equation}
where $\nabla^2_{S^3}=g^{ab}D_aD_b$ and where $D_a$ is the covariant derivative with Levi-Civita connection.

One could generate the $(1,0)$ and $(0,1)$ vector harmonics by dualizing~\eqref{generators isometry S3}, i.e.~through $V_a=g_{ab}V^b$.
We choose a different normalization for these harmonics by imposing
\begin{equation}\label{normalization vec harmonic} 
\begin{split}
\int (Y_1^{\hat{a} A})_a^* (Y_1^{\hat{b} B})^{a} \;=\;\Omega_3\, \delta^{\hat{a},\hat{b}} \delta^{A,B}\,,
\end{split}
\end{equation}
where $\hat{a},\hat{b}=\pm,0$ denote the range of $m$ and $A,B=\pm$ denotes left/right vector harmonics.
With this choice, the degree 1 vector spherical harmonics expressed as one-forms are
\begin{equation}\label{vector harmonics}
\begin{split}
Y_1^{++}&=\frac{1}{\sqrt{2}}e^{i(\phi+\psi)}\,\left[-i \,d\theta+\sin\theta\cos\theta\,d(\phi-\psi)\right]\,,\\
Y_1^{-+}&=\frac{1}{\sqrt{2}}e^{-i(\phi+\psi)}\,\left[ i \,d\theta+\sin\theta\cos\theta\,d(\phi-\psi)\right]\,,\\
Y_1^{0+}&=-\cos^2\theta\,d\psi-\sin^2\theta\,d\phi\,,\\
Y_1^{+-}&=\frac{1}{\sqrt{2}}e^{i(\phi-\psi)}\left[i \,d\theta-\sin\theta\cos\theta\, d(\phi+\psi)\right]\,,\\
Y_1^{--}&=-\frac{1}{\sqrt{2}}e^{-i(\phi-\psi)}\left[i \,d\theta+\sin\theta\cos\theta\, d(\phi+\psi)\right]\,,\\
Y_1^{0-}&=\cos^2\theta\,d\psi-\sin^2\theta\,d\phi\,.\\
\end{split}
\end{equation}

One can generate a degree $k$ vector harmonic using the $SU(2)$ tensor product decomposition
\begin{equation}
(\frac{k}{2},\frac{k}{2})\otimes (1,0)=(\frac{k}{2}+1,\frac{k}{2})\oplus (\frac{k}{2},\frac{k}{2})\oplus (\frac{k}{2}-1,\frac{k}{2})\,.
\end{equation}
The highest weight state of $(\frac{k}{2}+1,\frac{k}{2})$ is obtained by simply multiplying the highest weight states of $(\frac{k}{2},\frac{k}{2})$ and that of $ (1,0)$ (the Clebsch-Gordan coefficient in this case is always one). One can then generate all the descendants by acting with the lowering operator $J^-$.
In order to satisfy the generalized normalization condition
\begin{equation}\label{normalization vec harmonic high k} 
\begin{split}
\int (Y^{m_1,\bar m_1}_{v,k_1})_a^* (Y^{m_2,\bar m_2}_{v,k_2})^{a} \;=\;\Omega_3\, \delta^{k_1,k_2} \delta^{m_1,m_2}\, \delta^{\bar m_1,\bar m_2}\,,
\end{split}
\end{equation}
one can use the $SU(2)$ algebra to write
\begin{equation}
\label{eq:su2-normalization}
Y^{\frac{k+1}{2}-m,\frac{k-1}{2}-\bar m}_{\LL,k}=\sqrt{\frac{(k+1-m)!}{(k+1)!m!}}\sqrt{\frac{(k-1-\bar m)!}{(k-1)!\bar m!}}\Big[(J^-)^m (\bar J^-)^{\bar{m}},Y^{\frac{k+1}{2},\frac{k-1}{2}}_{\LL,k}\Big] \,.
\end{equation}
All the previous discussion proceeds analogously for right vector harmonics.
We define the following triple integral:
\begin{equation}\label{triple overlap} \begin{split}
\int Y_k^{(m_k,\bar{m_k})}\big(Y_1^{a-}\big)_\mu \big(Y_1^{b+}\big)^\mu\;=\;\Omega_3\:\!  f^{(k)}_{(m_k,\bar{m_k}) a b}\,.
\end{split}
\end{equation}
The explicit value of the components of $f^{(k)}_{(m_k,\bar{m_k}) ab}$, defined in~\eqref{triple overlap}, that have been used in this paper are
\begin{equation}\label{f overlap} \begin{split}
f^{(2)}_{(0,0)00}\,=\,\frac{1}{\sqrt{3}} \,,\hspace{2em}
f^{(2)}_{(1,1)--}\,=\,\frac{1}{\sqrt{3}} \,,\hspace{2em}
f^{(2)}_{(\pm1,\pm1)00}\,=\,0\,.
\end{split}
\end{equation}
One also has $\,\epsilon_{abc}D^b(Y^{m,\bar m}_{\LL,k})^c=(k-1)(Y^{m,\bar m}_{\LL,k})_a\,$ and $\,\epsilon_{abc}D^b(Y^{m,\bar m}_{\RR,k})^c=-(k-1)(Y^{m,\bar m}_{\RR,k})_a\,$.

\subsubsection{Useful definitions}\label{sec: Def and identities projection harmonics}

Due to the non-linearity of the gauge-invariant combinations at higher order, one must project products of harmonics into harmonics of higher order. The following definitions are useful:
\begin{enumerate}[label=(\Roman*)]

\item $(Y^{I})^\mu Y^J=\frac{E^{IJK}}{\Lambda_K}D^\mu Y^K+f^{IJK}(Y^{K})^\mu$

\item $\epsilon_{\mu\nu\rho}D^\sigma Y^ID_\sigma D_\rho  Y^J=c_{IJK}^s\epsilon_{\mu\nu\rho}D^\rho Y^K+c_{IJK}^vD_{[\nu}Y^K_{\mu]}$

\item $Y^{\sigma,I}\epsilon_{\mu\nu\rho}D_\sigma D^\rho Y^J=g_{IJK}^s\epsilon_{\mu\nu\rho}D^\rho Y^K+g_{IJK}^vD_{[\nu}Y^K_{\mu]}$

\item $2D_{[\mu}D^\rho Y^I\epsilon_{\nu]\rho\sigma}D^\sigma Y^J=n_{IJK}^s\epsilon_{\mu\nu\rho}D^\rho Y^K+n_{IJK}^vD_{[\nu}Y^K_{\mu]}$

\item $2D_{[\mu}(Y^{I})^{\rho}\epsilon_{\nu]\rho\sigma}D^\sigma Y^J=p_{IJK}^s\epsilon_{\mu\nu\rho}D^\rho Y^K+p_{IJK}^vD_{[\nu}Y^K_{\mu]}$

\end{enumerate}
where $I,J,K$ are the multi-indices defined below Eq.~\eqref{background fields}, and where on the right-hand side the index $K$ is summed over.

\subsection{Harmonics on AdS$_3$}\label{sec:harmonics ads3}
\subsubsection{Scalar harmonics}
The spherical harmonics are a representation of the AdS$_3$ isometry group  $SO(2,2)\simeq SL(2,\mathbb{R})\times SL(2,\mathbb{R})$. We use coordinates where the line element reads:
\begin{equation}\label{AdS3 metric 2}
ds^2_{AdS_3}=-(\tilde r^2+1)d\tilde t^2+\frac{d\tilde r^2}{\tilde r^2+1}+\tilde r^2d\tilde y^2 \,.
\end{equation}
In  our conventions  $\epsilon_{\tilde t\tilde r\tilde y}$=1.
The generators of the isometry group are given by:
\begin{equation}\label{generators sl2R spacetime}
\begin{aligned}
L_{\pm 1}&=i e^{\pm i(\tilde t+\tilde y)}\Big({}-\frac{1}{2}\frac{\tilde r}{\sqrt{\tilde r^2+1}}\partial_{\tilde t}-\frac{1}{2}\frac{\sqrt{\tilde r^2+1}}{r}\partial_{\tilde y}\pm\frac{i}{2}\sqrt{r^2+1} \, \partial_{\tilde r} \Big)\,,\qquad
L_0=\frac{i }{2}\big(\partial_{\tilde t}+\partial_{\tilde y} \big)\,,\\
\bar L_{\pm 1}&=i e^{\pm i(\tilde t-\tilde y)}\Big(-\frac{1}{2}\frac{\tilde r}{\sqrt{\tilde r^2+1}}\partial_{\tilde t}+\frac{1}{2}\frac{\sqrt{\tilde r^2+1}}{\tilde r}\partial_{\tilde y}\pm\frac{i}{2}\sqrt{\tilde r^2+1} \, \partial_{\tilde r}\Big)\,,\qquad
\bar L_0=\frac{i }{2}(\partial_{\tilde t}-\partial_{\tilde y}) \,
\end{aligned}
\end{equation}
and they respect the algebra
\begin{equation}
\big[L_0,L_\pm\big]=\mp L_\pm \hspace{2em} \big[L_1,L_{-1}\big]=2L_0 \hspace{2em}
\big[\bar L_0,\bar L_\pm\big]=\mp \bar L_\pm \hspace{2em} \big[\bar L_1,\bar L_{-1}\big]=2\bar L_0 \,.
\end{equation}
The quadratic Casimirs are 
$L^2=\frac{1}{2}\big(L_{1}L_{-1}+L_{-1}L_{1}\big)-\big(L_0\big)^2$ and the corresponding antiholomorphic operator $\bar{L}^2$.
For a state of $L^2$ quantum number $j_\text{sl}$, we have $L^2\ket{j_\text{sl}}=-j_\text{sl}(j_\text{sl}-1)\ket{j_\text{sl}}.$

Scalar harmonics have $j_\text{sl}=\bar{j}_\text{sl}$.
We introduce for convenience $\h=2j_\text{sl}$.
We introduce scalar harmonics $B_{\h}^{(\pm)}$, where the superscript $\pm$ denotes the positive and negative frequency modes. $B^{(+)00}_{\h}$ is the lowest-weight state in the discrete series representation $D^+$, and $B^{(-)00}_{\h}$ is the highest-weight state in the discrete series representation $D^-$.  For ease of language we refer to these both as highest-weight states. These harmonics solve the following Laplace equation:
\begin{equation}
\Box_{AdS_3} B_{\h}^{(\pm)}=\h(\h-2)B_{\h}^{(\pm)}
\end{equation}
We then have
\begin{equation}
B^{(\pm)}_{\h}=\frac{e^{\mp i \h \tilde t}}{\sqrt{\tilde r^2+1}^\h} \,, \qquad
L_0B^{(\pm)}_{\h}=\bar L_0 B^{(\pm)}_{\h}=\pm \frac{\h}{2}B^{(\pm)}_{\h} \,.
\end{equation}
The fact that these are highest-weight states can be seen as follows:
\begin{equation}
\big[L_1, B^{(+)}_{\h}\big]=\big[\bar L_1, B^{(+)}_{\h}\big]=0\qquad \big[L_{-1}, B^{(-)}_{\h}\big]=\big[\bar L_{-1}, B^{(-)}_{\h}\big]=0 \,.
\end{equation}

\subsubsection{Vector Harmonics}\label{AdS vector harmonics}
We now introduce vector harmonics on AdS$_3$. Those that live in the representation labeled by $L^2$ quantum numbers  $(j_\text{sl},\bar{j}_\text{sl})=(\frac{\h-2}{2},\frac{\h}{2})$  will be denoted by $B_{\RR,\h}^{(\pm)}$, where $\RR$ stands for right. Analogously, there is also the left representation with quantum numbers $(j_\text{sl},\bar{j}_\text{sl})=(\frac{\h}{2},\frac{\h-2}{2})$, and we will denote it by $B_{\LL,\h}^{(\pm)}$.  The vector harmonics satisfy the following Laplace equation 
\begin{equation}
(d\delta+\delta d)B_{v,\h}^{(\pm)}=(\h-2)^2B_{v,\h}^{(\pm)} \,.
\end{equation}
Here the subscript $v$ stands for vector, meaning that the formula applies to both left and right vector harmonics.
The $\h=0$ vector harmonics are the Killing one forms, which can be obtained can be obtained dualizing Eq.~\eqref{generators sl2R spacetime},
\begin{equation}\label{killing vectors ads3}
\begin{aligned}
L_0&=-\frac{i}{2}\big((\tilde r^2+1)d\tilde t-\tilde r^2d\tilde y\big) \,,\\
L_\pm&= e^{\pm i (\tilde t+\tilde y)}\frac{\tilde r}{2\sqrt{\tilde r^2+1}}\big(\mp \frac{d\tilde r}{\tilde r}+i(\tilde r^2+1)(d\tilde t-d\tilde y)\big)\,,\\
\bar L_0&=-\frac{i}{2}\big((\tilde r^2+1)d\tilde t+\tilde r^2d\tilde y\big)\,,\\
\bar L_\pm&= e^{\pm i (\tilde t-\tilde y)}\frac{\tilde r}{2\sqrt{\tilde r^2+1}}\big(\mp \frac{d\tilde r}{\tilde r}+i(\tilde r^2+1)(d\tilde t+d\tilde y)\big)\,.
\end{aligned}
\end{equation}
One can generate degree $\h$ vector harmonics, as in the $S^3$ case, by multiplying scalar harmonics with the one-forms in Eq.~\eqref{killing vectors ads3} and exploiting the $SL(2,\mathbb{R})$ tensor product decomposition. For concreteness, let us consider right vector harmonics, which live in the representation in the first term of the following direct sum:
\begin{equation}\label{sl2r tensor prod decomp}
 \begin{aligned}
\Big(\frac{\h}{2},\frac{\h}{2}\Big)\otimes \big(-1,0\big)=\Big(\frac{\h-2}{2}, \frac{\h}{2} \Big)\oplus \Big(\frac{\h}{2}, \frac{\h}{2} \Big)\oplus \Big(\frac{\h+2}{2}, \frac{\h}{2}\Big)\,.
\end{aligned}
\end{equation}
Let us focus on the $B^{(+)}$ modes. It is important to note that the scalar harmonic $B^{(+)}_{\h}$ is a lowest weight state; in order to obtain the lowest weight of the vectorial representation we need to take the product with $L_1$, and one has:
\begin{align}
\begin{aligned}
& B^{(+)}_{\RR,\h}=B^{(+)}_{\h}\otimes L_1\,, \qquad \big[L_0,B^{(+)}_{\RR,\h}\big]=\frac{\h-2}{2}B^{(+)}_{\RR,\h}\,,\\ 
& \big[\bar L_0,B^{(+)}_{\RR,\h}\big]=\frac{\h}{2}B^{(+)}_{\RR,\h}\,, \qquad
 \big[L_1,B^{(+)}_{\RR,\h}\big]=\big[\bar L_1,B^{(+)}_{\RR,\h}\big]=0\,.
\end{aligned}
\end{align}
Analogous relations hold for the left vector harmonics.
 Vector harmonics on AdS$_3$ are also eigenstates of the following operator, where $\star$ is the Hodge star on global AdS$_3$, \eqref{AdS3 metric 2}:
\begin{equation}\label{vector harmonic eigenstate}
\star d B^{(\pm)}_{\RR,\h}=(\h-2)B^{(\pm)}_{\RR,\h}\,,\qquad \star d B^{(\pm)}_{\LL,\h}=-(\h-2)B^{(\pm)}_{\LL,\h}\,.
\end{equation}

\section{Extremal 3-point functions}
\label{sec:Extremal 3pt}

In this Appendix we will derive the scalar chiral primary operators of dimension two in the single-particle basis. The vanishing of extremal three-point functions built with the operator $\tilde O_2$ in~\eqref{single particle O2} was discussed in Section (4.1) of~\cite{Giusto:2019qig}; here we shall focus on the operators $\tilde\Omega$ and $\tilde\Sigma_3$ defined in Eq.~\eqref{orthonormal single particle scalars}.
These operators involve a mixing between the single-trace operators
$\Sigma_3$, $\Omega$ and the double-trace operators $(\Sigma_2\cdot \Sigma_2)$, $(J \cdot \bar{J})$ and $(O\cdot O)$.
Let us first discuss the mixing matrix between the single-traces~\cite{Taylor:2007hs}. The single-particle operators take the form
\begin{equation}\label{mixing matrix single trace}
    \tilde\Omega\;\sim\; a_1\frac{\Sigma_3}{N^{\frac32}}+a_2\frac{\Omega}{N^{\frac12}}\,,\qquad \tilde\Sigma_3\;\sim\; b_1\frac{\Sigma_3}{N^{\frac32}}+b_2\frac{\Omega}{N^{\frac12}}\,, 
\end{equation}
where $\sim$ means that we are retaining only the single-trace contributions, and where we have included factors of $N$ such that each term contributes at order $N^0$ to the norm of the respective single-particle operator at large $N$. As we shall discuss below, at large $N$, the multi-trace contribution to the norm of the single-particle operator is subleading. Thus, the orthonormality conditions
\begin{equation} 
\langle\tilde{\Sigma}_3^{++}\tilde{\Sigma}_3^{--}\rangle\;=\;\langle\tilde{\Omega}^{++}\tilde{\Omega}^{--}\rangle\;=\;1\,,\qquad \langle\tilde{\Sigma}_3^{++}\tilde{\Omega}^{--}\rangle\;=\;0\,,
\end{equation}
give three constraints on the four coefficients $a_1$, $a_2$, $b_1$ and $b_2$. 
In order to completely fix the mixing matrix in Eq.~\eqref{mixing matrix single trace} we need a fourth constraint, which was derived in \cite{Taylor:2007hs} from comparison with non-extremal supergravity correlators.
The result is
\begin{equation}\label{mixing matrix single trace2}
\tilde{\Omega}\;\sim\; \frac{\sqrt{3}}{2} \left(  \frac{\Sigma_3}{N^{\frac32}}+\frac{\Omega}{N^{\frac12}}\right)\,, \qquad
     \tilde{\Sigma}_3\;\sim\;\frac{3}{2} \left(\frac{\Sigma_3}{N^{\frac32}}-\frac{\Omega}{3N^{\frac12}}\right)\,.
\end{equation}

We now turn to the mixing of multi-trace operators in the single-particle basis. As we shall see, having fixed the single-trace mixing in \eqref{mixing matrix single trace2}, the multi-trace admixtures can then be completely fixed by CFT computations.
By contrast, if we had not fixed the single-trace mixing in \eqref{mixing matrix single trace2} and performed the following steps with the general admixture~\eqref{mixing matrix single trace}, we would not have enough constraints to determine the mixing.

Let us consider the most general linear combination allowed by the quantum numbers that can give rise to the single-particle operators:
\begin{equation}\label{extremal 3point ansatz}
\alpha\frac{\Sigma_3}{N^{3/2}}+\beta \frac{\Omega}{N^{1/2}}+\gamma (\Sigma_2\cdot \Sigma_2)+\delta (J \cdot \bar{J})+\epsilon (O\cdot O)^{++}\,.
\end{equation}
The two sets of values for the coefficients $\alpha$ and $\beta$ are given by Eq.~\eqref{mixing matrix single trace2}. We now fix the two sets of coefficients $\gamma$, $\delta$ and $\epsilon$ by imposing that all extremal three-point functions containing the operator \eq{extremal 3point ansatz} and operators of lower dimension vanish. Equivalently, this imposes that the operator \eqref{extremal 3point ansatz} is orthogonal to all multi-trace operators, as discussed around Eq.~\eq{2 pt from limit of 3 pt}.

There are three non-trivial such correlators, corresponding respectively to the last three terms in \eqref{extremal 3point ansatz}. We work at leading order in large $N$.
For ease of notation we will suppress the standard space-time dependence in the following correlators. 
\begin{itemize}
\item The first constraint follows from imposing that the three-point function containing the operator~\eqref{extremal 3point ansatz} and two $O^{--}$ operators vanishes.  
Since $O^{--}$ belongs to the untwisted sector of the theory, the contributions coming from the twist operators in~\eqref{extremal 3point ansatz} are trivially zero; moreover the contribution of the multi-trace $(J \bar J)$ is subleading at large $N$.  We obtain
\begin{equation}\label{extremal 3point ansatz OO}
\langle \Big(\beta \frac{\Omega^{++}}{N^{1/2}}+\epsilon (O\cdot O)^{++}\Big) O^{--}O^{--}\rangle\;=\;\beta N^{1/2}+ 2\epsilon N\;=\;0\,,
\end{equation}
where we have used the definitions of $O$, $\Omega$ and $(O\cdot O)$ in Eqs.~\eqref{op O},~\eqref{op Omega} and~\eqref{eq:doubletraces}.
\item Let us now consider the current insertions: again, since they carry no twist, the twist operators in Eq.~\eqref{extremal 3point ansatz} do not contribute to the correlator; the contribution of the double-trace operator $(O\cdot O)^{++}$ is subleading at large $N$. The computation leads to
\begin{equation}\label{extremal 3point ansatz JbarJ}
\langle \Big(\beta \frac{\Omega^{++}}{N^{1/2}}+\delta (J\cdot \bar J)^{++}\Big)J^{-}\bar J^{-}\rangle\;=\;\beta N^{1/2}+\delta N\;=\; 0\,,
\end{equation}
where we have used the definitions of $J$, $\bar J$, $\Omega$ and $(J\cdot \bar J)$ in Eqs.~\eqref{op J},~\eqref{op Omega} and~\eqref{eq:doubletraces}.
\item The final constraint comes from inserting two twist operators $\Sigma_2^{--}$. There are three leading-order contributions to the extremal three-point function, from the operators $\Sigma_3^{++}$, $(\Sigma_2\cdot \Sigma_2)^{++}$ and $\Omega^{++}$ in~\eqref{extremal 3point ansatz}. 
First, the contraction with $\Sigma_3^{++}$ gives at large $N$
\begin{equation}\label{extremal sigma3sigma2sigma2}
\Big\langle \sum_{r<s<t} \big(\sigma^{++}_{(rst)}+ \sigma^{++}_{(rts)}\big) \sum_{a<b} \sigma_{(ab)}^{--}\sum_{c<d} \sigma_{(cd)}^{--} \,\Big\rangle\;=\;\frac{3 N^{3/2}}{4}\,.
\end{equation}

The combinatorics works as follows. The choice of the copies $r,s,t$ gives $N\choose 3$ and, for each choice, there are two inequivalent cycles. Thus the twist-three operator can glue three strands out of $N$ in $2 {N\choose 3}$ ways. For concreteness let us suppose that the orientation of the 3 copies $r,s,t$ is chosen to be $(123)$. Then there remains the freedom to take the first $\sigma_2$ to be the 2-cycle $(12)$, $(23)$ or $(13)$ (this fixes the second $\sigma_2$ to be $(23)$, $(13)$ and $(12)$ respectively): this gives an additional factor of three. Eq.~\eqref{extremal sigma3sigma2sigma2} follows from combining this combinatorial factor with the building block 
\begin{equation}
\big\langle \sigma^{++}_{(123)}\sigma_{(12)}^{--} \sigma_{(23)}^{--} \big\rangle\;=\; \frac34 \,,
\end{equation}
derived in \cite[Eq.\;(5.25)]{Lunin:2001pw}.

Second, the contribution from the double-trace $(\Sigma_2\cdot \Sigma_2)^{++}$ in \eqref{extremal 3point ansatz} is
\begin{equation}\label{extremal sigma2sigma2sigma2sigma2}
\Big\langle \frac{2}{N^2} \sum_{(r<s)\neq (p<q)} \sigma_{(rs)}^{++} \sigma_{(pq)}^{++} \sum_{a<b} \sigma_{(ab)}^{--}\sum_{c<d} \sigma_{(cd)}^{--}\Big \rangle\;=\;N^2 \,.
\end{equation}
 
The computation goes as follows. The choice of the copies $r,s$ and $p,q$ gives the combinatorial factors $N(N-1)/2$ and $(N-2)(N-3)/2$ respectively; moreover, one can perform two inequivalent Wick contractions, which give an extra factor of two. Taking into account the normalization of the multi-trace operator, one obtains Eq.~\eqref{extremal sigma2sigma2sigma2sigma2}.

The third contribution comes from the operator $\Omega^{++}$ in \eqref{extremal 3point ansatz}. One could evaluate this amplitude either by using the techniques developed in~\cite{Lunin:2000yv,Lunin:2001pw}, or by exploiting Ward identities that relate different $n$-point functions in the same R-symmetry multiplet together with some results obtained in~\cite{Giusto:2019qig}; we shall use the latter method. 

We first use the permutation property of the correlator (see e.g. \cite[Eq.\;(2.2.48)]{Ribault:2014hia}) to reorder the operators as
\begin{equation}
\begin{aligned}\label{extremal omegasigma2sigma2-pre}
\langle \Omega^{++}\,\Sigma_2^{--}\,\Sigma_2^{--} \rangle&=\langle \Sigma_2^{--}\, \Omega^{++}\, \Sigma_2^{--} \rangle\,.
\end{aligned}
\end{equation}
We next write the operator $\Sigma_2^{--}$ as $[J^-_0 \bar J^-_0 ,\Sigma_2^{++}]$ and move the current modes onto the other operators, to obtain
\begin{equation}
\begin{aligned}\label{extremal omegasigma2sigma2}
\langle \Sigma_2^{--}\, \Omega^{++}\, \Sigma_2^{--} \rangle
&= \big\langle \Sigma_2^{++}\, \big(2\Omega^{00}\big)\, \Sigma_2^{--} \big\rangle=\frac{N^2}{4}\,.
\end{aligned}
\end{equation}
The last equality is obtained by spectral flowing the correlator to the RR sector. On doing so we notice that $\Omega$ carries no twist so that the twist operators need to act on the same copies, which introduces a combinatorial factor of ${N \choose 2}$. We then use~\cite[Eq.\;(5.40)]{Giusto:2019qig}.
\end{itemize}

Combining Eqs.~\eqref{extremal 3point ansatz OO}--\eqref{extremal sigma3sigma2sigma2},~\eqref{extremal sigma2sigma2sigma2sigma2}, and \eqref{extremal omegasigma2sigma2}, we find that the required vanishing of the three-point functions considered imposes the three constraints 
\begin{equation}\label{extremal constraints} \gamma\,=\,-\frac{3\alpha}{4 N^{1/2}}-\frac{\beta}{4N^{1/2}}\,, \qquad \delta\,=\,-\frac{\beta}{N^{1/2}} \,, \qquad \epsilon\,=\,-\frac{\beta}{2N^{1/2}} \,.
\end{equation}
Combining these constraints with the two sets of values of $\alpha$ and $\beta$ in Eq.~\eqref{mixing matrix single trace2}, we obtain
\begin{equation}
\begin{aligned}
 \tilde{\Sigma}_3^{++}&\,=\, \frac{3}{2} \left[\left(\frac{\Sigma_3^{++}}{N^{\frac32}}-\frac{\Omega^{++}}{3N^{\frac12}}\right)+\frac{1}{N^{\frac12}}\left(-\frac{2}{3}(\Sigma_2\cdot \Sigma_2)^{++}+\frac{1}{6}(O\cdot O)^{++}+\frac{1}{3}(J\cdot \bar{J})^{++}\right)\right]\,, \\
\tilde{\Omega}^{++}&\,=\, \frac{\sqrt{3}}{2} \left[\left(  \frac{\Sigma_3^{++}}{N^{\frac32}}+\frac{\Omega^{++}}{N^{\frac12}}\right)+ \frac{1}{N^{\frac12}}\left(-(\Sigma_2\cdot \Sigma_2)^{++}-\frac12 (O\cdot O)^{++}-(J \cdot \bar{J})^{++}\right)\right] ,
\end{aligned}
\end{equation}
as written in the main text in Eq.~\eqref{orthonormal single particle scalars}. Recall that the factors of $N$ multiplying the single-trace operators are such that these terms all contribute at leading order (order $N^0$) to the norm of the operators. By contrast, since the double-trace operators are already unit normalized, the admixture factors of $1/\sqrt{N}$ imply for instance that the double-trace contribution to the norm of the single-particle operators is subleading at large $N$.

\newpage

\section{Type IIB supergravity ansatz and BPS equations}\label{app:sugra}

In this appendix we record for completeness the  general solution to Type IIB supergravity compactified on T$^4$ that is 1/8-BPS, carries D1-D5-P charges, and is invariant on T$^4$~\cite[(E.7)]{Giusto:2013rxa}:
\begin{align}\label{ansatzSummary}
d s^2_{10} &~=~ \sqrt{\alpha} \,ds^2_6 +\sqrt{\frac{Z_1}{Z_2}}\,d \hat{s}^2_{4}\, ,\nonumber\\
d s^2_{6} &~=~-\frac{2}{\sqrt{\cP}}\,(d v+\betab)\,\Big[d u+\omega + \frac{\mathcal{F}}{2}(d v+\betab)\Big]+\sqrt{\cP}\,d s^2_4\,,\nonumber\\
e^{2\Phi}&~=~\frac{Z_1^2}{\cP}\, ,\nonumber\\
B&~=~ -\frac{Z_4}{\cP}\,(d u+\omega) \wedge(d v+\betab)+ a_4 \wedge  (d v+\betab) + \gamma_4\,,\nonumber\\ 
C_0&~=~\frac{Z_4}{Z_1}\, ,\\
C_2 &~=~ -\frac{Z_2}{\cP}\,(d u+\omega) \wedge(d v+\betab)+ a^1 \wedge  (d v+\betab) + \gamma_2\,,\nonumber\\ 
C_4 &~=~ \frac{Z_4}{Z_2}\, \widehat{\mathrm{vol}}_{4} - \frac{Z_4}{\cP}\,\gamma_2\wedge (d u+\omega) \wedge(d v+\betab)+x_3\wedge(d v + \betab) \,, \nonumber\\
C_6 &~=~\widehat{\mathrm{vol}}_{4} \wedge \left[ -\frac{Z_1}{\cP}\,(d u+\omega) \wedge(d v+\betab)+ a^2 \wedge  (d v+\betab) + \gamma_1\right]\nonumber , 
%
\end{align}

%
where
\begin{equation}
\alpha \;=\; \frac{Z_1 Z_2}{Z_1 Z_2 - Z_4^2}~,\qquad~~
\cP   \;=\;     Z_1  Z_2  -  Z_4^2 \,.
\label{Psimp}
\end{equation}
Here $d \hat{s}^2_4$ denotes the flat metric on $T^4$, and $\widehat{\mathrm{vol}}_{4}$ denotes the corresponding volume form. This ansatz contains all fields that are known to arise from world-sheet calculations of the backreaction of D1-D5-P bound states invariant on $\cM$~\cite{Black:2010uq,Giusto:2011fy}.

The BPS equations for this ansatz have the following structure. The base metric, $ds^2_4$, and the one-form $\betab$ satisfy non-linear equations known as the ``zeroth layer''. Having solved these initial equations, the remaining BPS equations are organized into two further layers of linear equations~\cite{Giusto:2013rxa,Bena:2011dd}.

We denote the exterior differential on the spatial base $\cB$ by $\tilde d$, and following~\cite{Gutowski:2003rg} we introduce
\begin{equation}
\mathcal{D} \;\equiv\; \tilde d - \betab\wedge \frac{\partial}{\partial v}\,.
\end{equation}

In the current work we consider only solutions in which the four-dimensional base space is flat $\mathbb{R}^4$, and in which $\betab$ is independent of $v$.
The BPS equation for $\betab$ is then
 \begin{equation}\label{eqbeta}
d \betab \;=\; *_4 d\betab\,,
 \end{equation}
where $*_4$ stands for the flat $\mathbb{R}^4$ Hodge dual.

To write the remaining layers of BPS equations in a covariant form, we make the rescaling $(Z_4,a_4,\gamma_4) \to (Z_4,a_4,\gamma_4)/\sqrt{2}$ for the remainder of this appendix (and only here).  We introduce the $SO(1,2)$ Minkowski metric $\eta_{ab}$ ($a=1,2,4$) in the null form
\be \label{eq:etaab}
\eta_{12} ~=~ \eta_{21} ~=~ 1\,, \qquad \eta_{44} = -1 \,.
\ee
This metric is used to raise and lower $a,b$ indices. 
We introduce the two-forms $\Theta^1$, $\Theta^2$, $\Theta^4$ as follows:\footnote{The relation to the notation in~\cite{Bena:2017xbt} is that $\Theta^1_{\mathrm{here}}=\Theta_1^{\mathrm{there}}$, $\Theta^2_{\mathrm{here}}=\Theta_2^{\mathrm{there}}$, $(1/\sqrt{2})\Theta_4^{\mathrm{here}}=\Theta_4^{\mathrm{there}}$.}
\begin{equation}
\label{Thetadefs}
\Theta^b ~\equiv~ \mathcal{D} a^b + \eta^{bc} \:\! \dot{\gamma}_c  \;.
\end{equation}
We then have
\be
\cP ~\equiv~ \coeff{1}{2} \eta^{ab} Z_a Z_b ~=~ Z_1 Z_2 - \coeff12 Z_4^2 \,.
\ee
The ``first layer'' of the BPS equations is then 
\bea \label{eq:firstlayer}  
 *_4 D\dot{Z}_a ~=~ & \eta_{ab} D\Theta^{b}\,,\qquad D*_4 DZ_a ~=~  - \eta_{ab} \Theta^{b} \! \wedge d\betab\,,
\qquad \Theta^{a} ~=~ *_4 \Theta^{a} \,,
\eea
while the ``second layer'' is given by
\begin{equation}
 \begin{aligned}
D \omega + *_4 D\omega + \mathcal{F} \,d\betab 
~=~ & Z_a \Theta^{a}\,,  \\ 
 *_4D*_4\!\Bigl(\dot{\omega} -\coeff{1}{2}\,D\mathcal{F}\Bigr) 
~=~& \ddot \cP  -\coeff{1}{2} \eta^{ab} \dot{Z}_a \dot{Z}_b 
-\coeff{1}{4} \eta_{ab} *_4\! \Theta^{a}\wedge \Theta^{b} \,.
\end{aligned}
\label{eqFomega-app}
\end{equation} 

\newpage

\section{Gauge-fixed holographic dictionary}\label{sec:Gauge fixed dictionary}
In Section \ref{ssc:Holographic dictionary for scalar SPOs} we discussed the holographic dictionary that relates the expectation value of single-particle operators at dimension one and scalar single-particle operators at dimension two in term of gauge-invariant combination of KK fields. This formulation has the advantage of being general and manifestly gauge invariant: it can be used to probe any microstate solution in any coordinate system. From a practical point of view, however, it is useful to formulate the dictionary in a preferred gauge, as this makes it more manageable to use. In this appendix we study the dictionary in such a gauge.

In this appendix we consider the general class of solutions for which the four-dimensional base metric $ds_4^2$ in~\eqref{metric superstrata},~\eqref{ansatzSummary} is flat $\mathbb{R}^4$. This class includes all solutions discussed in this paper.
For solutions in which the base metric is not flat, one should instead use the gauge-invariant formulation of the dictionary in Eq.~\eqref{gauge invariant old dict}.

We choose coordinates such that the flat $\mathbb{R}^4$ base metric takes the form:
\begin{equation}\label{flat ds4}
 ds^2_4=dr^2+r^2(d\theta^2+\sin^2\theta d\phi^2+\cos^2\theta d\psi^2)\,.
 \end{equation} 
In general, the functions $Z_1,Z_2$ and $Z_4$
obey a Poisson-type equation, where the $\Theta_i$ play the role of sources; see Eq.~\eqref{eq:firstlayer}.
For $1/4$-BPS solutions, one has $\Theta_1=\Theta_2=\Theta_4=0$, and so the $Z_a$ are harmonic. For all the ($1/8$-BPS) superstrata we deal with in the present work, the first term in the large $r$ expansion of the source $\Theta_a \wedge d\beta$ is of order $1/r^k$ where $k\geq 7$.

For convenience we perform the transformation in Eq.~\eqref{transformation standard ads} and reabsorb the overall factor $\sqrt{Q_1Q_5}$ in the metric.
We thus define the following expansion\footnote{Note that this equation is more general than \cite[Eq.\;(3.6)]{Giusto:2019qig}, which applies only to solutions in which the $Z_a$ are harmonic.} \cite{Kanitscheider:2006zf,Kanitscheider:2007wq}:
\begin{equation}\label{eq:geometryexpansion}
\begin{aligned}
\tilde Z_1 &=\frac{Q_1}{\tilde r^2} \left(1+ \sum_{k=1}^2 \sum_{m_k,\bar{m}_k=-k/2}^{k/2} f^{\,(m_k,\bar m_k)}_{1k} \,\frac{Y_k^{m_k,\bar m_k}}{\tilde r^k}+O(\tilde r^{-3})\right)\,,\\
\tilde Z_2&=\frac{Q_5}{\tilde r^2} \left(1+ \sum_{k=1}^2 \sum_{m_k,\bar{m}_k=-k/2}^{k/2} f^{k\,(m_k,\bar{m}_k)}_{5k} \,\frac{Y_k^{m_k,\bar m_k}}{\tilde r^k}+O(\tilde r^{-3})\right)\,,\\
\tilde Z_4&=\frac{\sqrt{Q_1Q_5}}{\tilde r^2} \left(\sum_{k=1}^2 \sum_{m_k,\bar m_k=-k/2}^{k/2} \mathcal{A}_{k}^{(m_k,\bar m_k)} \,\frac{Y_k^{m_k,\bar m_k}}{\tilde r^k} +O(\tilde r^{-3})\right)\,,\\
\tilde  A&= \frac{1}{\tilde  r^2}\, \sum_{a=1}^3 (a_{a+} Y_1^{a +} + a_{a-} Y_1^{a -})+O(\tilde r^{-3})\,,\qquad \mathcal{F} = -\frac{2 Q_p}{a_0 \tilde r^2} + O(r^{-3})\,,\!\!\!\!\!\!\!\!
\end{aligned}
\end{equation}
 where we have denoted with $Y_k^{m_k,\bar m_k}$ the scalar harmonics on $S^3$ of degree $k$ and with $Y_1^{a \pm}$ the vector harmonics of degree 1 (see Appendix~\ref{sec:harm S3 and AdS3}), and where $a_0$ was defined below \eqref{transformation standard ads}. 
 We are still left with the freedom of choosing the origin of the coordinate system on the flat base. We fix this redundancy by requiring
 \begin{equation}\label{center of mass D1D5}
 f_{11}^{(m_1,\bar m_1)}+f^{(m_1,\bar m_1)}_{51}=0\,,
 \end{equation}
which corresponds to placing the origin at the center of mass of the D1-D5 system.
The choices in Eqs.~\eqref{eq:geometryexpansion} and~\eqref{center of mass D1D5} imply that, at order $k=1$, all fields in~\eqref{0611 geom harmonics expansion} that can be set to zero by a gauge transformation vanish.
With these choices, the expansion of the bulk quantities in the dictionary~\eqref{gauge invariant old dict} in terms of \eqref{eq:geometryexpansion} yields the gauge-fixed single-particle fields in supergravity:
\begin{equation}\label{gauge fixed exp of sp}
\begin{aligned}
 \Big[s^{(6)(\alpha,\dot\alpha)}_{k=1}\Big]&=-2\sqrt{2} f_{51}^{(\alpha,\dot\alpha)}\,,\qquad  
 \Big[s^{(6)(a,\dot a)}_{k=2}\Big]=\sqrt{\frac{3}{2}} (f_{12}^{(a,\dot a)}-f_{52}^{(a,\dot a)})\,,\\
 \Big[s^{(7)(\alpha,\dot\alpha)}_{k=1}\Big]&=2\sqrt{2} \mathcal{A}_1^{(\alpha,\dot\alpha)}\,,\qquad~\;  
 \Big[s^{(7)(a,\dot a)}_{k=2}\Big]=\sqrt{6} (\mathcal{A}_2^{(a,\dot a)})\,,\\
\Big[\tilde{\sigma}^{(a,\dot a)}_{k=2}\Big] &=-\frac{1}{\sqrt{2}}  \big(f_{12}^{(a,\dot a)}+f_{52}^{(a,\dot a)}\big)+2\sqrt{2}\big(f_{51}^{(\alpha,\dot\alpha)}f_{51}^{(\beta,\dot\beta)}+\mathcal{A}_{1}^{(\alpha,\dot\alpha)}\mathcal{A}_{1}^{(\beta,\dot\beta)}\big)a^{(a,\dot a)}_{(\alpha,\dot \alpha)(\beta,\dot \beta)}\\
&\,\quad
+4\sqrt{2}a^{c +}a^{d -}f^{1}_{(a,\dot a)cd}\,,\\
\Big[A^{a(\pm)}_{k=1}\Big]&=-2 a^{-a,\pm}\,,
\end{aligned}
\end{equation}
where $a^{(a,\dot a)}_{(\alpha,\dot \alpha)(\beta,\dot \beta)}$ and $f^{1}_{(a,\dot a)cd}$ are the triple overlap coefficients defined in \eqref{tripleoverlap} and \eqref{triple overlap}.

Let us compare the dictionary for operators of dimension two given in Eq.~\eqref{gauge invariant old dict} with the one given in \cite{Giusto:2019qig}. First, we note that the dictionary for the operator $\tilde O_2$ given in~\cite[Eq.s (5.35)--(5.36)]{Giusto:2019qig} is already formulated in the single-particle basis.

Next, we take~\cite[Eqs.~(5.33)--(5.34)]{Giusto:2019qig} and rotate them to express them in terms of the geometric quantities $g_{a,\dot a}$, $\tilde{g}_{a,\dot a}$. We note that the operator dual to $g_{a,\dot a}$ coincides with $\tilde\Sigma_3$, thus in this rotated basis the holographic dictionary for $g_{a,\dot a}$ is given in terms of a single-particle operator.

By contrast, the operator dual to $\tilde{g}_{a,\dot a}$ is not yet $\tilde\Omega$.
 The dictionary for $\tilde{g}_{a,\dot a}$ at this point reads:
\begin{align}
\label{eq:old-dict}
    \tilde g_{a,\dot a}&=\sqrt{2}\Big(-\big(f_{12}^{(a,\dot a)}+f_{52}^{(a,\dot a)}\big)+8a^{c +}a^{d -}f^{1}_{(a,\dot a)cd}\Big)\\
    &=-(-1)^{a+\dot a}\sqrt{6}\bigg[\frac{1}{N^{3/2}}\langle \Sigma_3^{a\dot a}\rangle+\frac{1}{N^{3/2}}\Big(\langle \Omega^{a\dot a}\rangle-\frac{1}{3}\langle (\Sigma_2\cdot \Sigma_2)^{a\dot a}\rangle-\langle (J\cdot \bar J)^{a\dot a}\rangle+\frac{1}{6}\langle (O\cdot O)^{a\dot a}\rangle\Big)\bigg].
    \nonumber
\end{align}
We now observe that the right-hand side of the first line of~\eqref{eq:old-dict} differs from the second-last line of Eq.~\eqref{gauge fixed exp of sp} through the terms that involve $f_{51}$ and $\cA_1$. These are gauge-fixed versions of gauge-invariant quantities.  
The holographic dictionary for $f_{51}$ and $\cA_1$ is obtained by combining Eqs.~\eqref{gauge fixed exp of sp} and~\eqref{gauge invariant old dict}.
 We can thus improve on the dictionary~\eq{eq:old-dict} by adding the terms in $f_{51}$ and $\cA_1$ to both sides of the equation. Doing so results precisely in the dictionary given in Eq.~\eqref{gauge fixed exp of sp}. 
 This demonstrates the consistency of the results of~\cite{Giusto:2019qig} and the present work, and is a non-trivial check of the independent methods used in the two works.


\newpage

\begin{adjustwidth}{-3mm}{-3mm} 
\bibliographystyle{utphys}      
\bibliography{microstates}       

\providecommand{\href}[2]{#2}\begingroup\raggedright\begin{thebibliography}{10}

\bibitem{Lunin:2001jy}
O.~Lunin and S.~D. Mathur, ``{AdS/CFT duality and the black hole information
  paradox},'' \href{http://dx.doi.org/10.1016/S0550-3213(01)00620-4}{{\em Nucl.
  Phys.} {\bfseries B623} (2002) 342--394},
\href{http://arxiv.org/abs/hep-th/0109154}{{\ttfamily arXiv:hep-th/0109154}}.

\bibitem{Lunin:2002iz}
O.~Lunin, J.~M. Maldacena, and L.~Maoz, ``{Gravity solutions for the D1-D5
  system with angular momentum},''
\href{http://arxiv.org/abs/hep-th/0212210}{{\ttfamily arXiv:hep-th/0212210}}.

\bibitem{Lunin:2004uu}
O.~Lunin, ``{Adding momentum to D1-D5 system},''
  \href{http://dx.doi.org/10.1088/1126-6708/2004/04/054}{{\em JHEP} {\bfseries
  04} (2004) 054},
\href{http://arxiv.org/abs/hep-th/0404006}{{\ttfamily arXiv:hep-th/0404006}}.

\bibitem{Giusto:2004id}
S.~Giusto, S.~D. Mathur, and A.~Saxena, ``{Dual geometries for a set of
  3-charge microstates},''
  \href{http://dx.doi.org/10.1016/j.nuclphysb.2004.09.001}{{\em Nucl. Phys.}
  {\bfseries B701} (2004) 357--379},
\href{http://arxiv.org/abs/hep-th/0405017}{{\ttfamily arXiv:hep-th/0405017}}.

\bibitem{Giusto:2004ip}
S.~Giusto, S.~D. Mathur, and A.~Saxena, ``{3-charge geometries and their CFT
  duals},'' \href{http://dx.doi.org/10.1016/j.nuclphysb.2005.01.009}{{\em Nucl.
  Phys.} {\bfseries B710} (2005) 425--463},
\href{http://arxiv.org/abs/hep-th/0406103}{{\ttfamily arXiv:hep-th/0406103}}.

\bibitem{Kanitscheider:2007wq}
I.~Kanitscheider, K.~Skenderis, and M.~Taylor, ``{Fuzzballs with internal
  excitations},'' {\em JHEP} {\bfseries 06} (2007) 056,
\href{http://arxiv.org/abs/0704.0690}{{\ttfamily arXiv:0704.0690 [hep-th]}}.

\bibitem{Mathur:2011gz}
S.~D. Mathur and D.~Turton, ``{Microstates at the boundary of AdS},''
  \href{http://dx.doi.org/10.1007/JHEP05(2012)014}{{\em JHEP} {\bfseries 05}
  (2012) 014},
\href{http://arxiv.org/abs/1112.6413}{{\ttfamily arXiv:1112.6413 [hep-th]}}.

\bibitem{Lunin:2012gp}
O.~Lunin, S.~D. Mathur, and D.~Turton, ``{Adding momentum to supersymmetric
  geometries},'' \href{http://dx.doi.org/10.1016/j.nuclphysb.2012.11.017}{{\em
  Nucl.Phys.} {\bfseries B868} (2013) 383--415},
\href{http://arxiv.org/abs/1208.1770}{{\ttfamily arXiv:1208.1770 [hep-th]}}.

\bibitem{Giusto:2012yz}
S.~Giusto, O.~Lunin, S.~D. Mathur, and D.~Turton, ``{D1-D5-P microstates at the
  cap},'' \href{http://dx.doi.org/10.1007/JHEP02(2013)050}{{\em JHEP}
  {\bfseries 1302} (2013) 050},
\href{http://arxiv.org/abs/1211.0306}{{\ttfamily arXiv:1211.0306 [hep-th]}}.

\bibitem{Strominger:1996sh}
A.~Strominger and C.~Vafa, ``{Microscopic Origin of the Bekenstein-Hawking
  Entropy},'' \href{http://dx.doi.org/10.1016/0370-2693(96)00345-0}{{\em Phys.
  Lett.} {\bfseries B379} (1996) 99--104},
\href{http://arxiv.org/abs/hep-th/9601029}{{\ttfamily arXiv:hep-th/9601029}}.

\bibitem{Breckenridge:1996is}
J.~Breckenridge, R.~C. Myers, A.~Peet, and C.~Vafa, ``{D-branes and spinning
  black holes},'' \href{http://dx.doi.org/10.1016/S0370-2693(96)01460-8}{{\em
  Phys.Lett.} {\bfseries B391} (1997) 93--98},
\href{http://arxiv.org/abs/hep-th/9602065}{{\ttfamily arXiv:hep-th/9602065
  [hep-th]}}.

\bibitem{Martinec:2017ztd}
E.~J. Martinec and S.~Massai, ``{String Theory of Supertubes},''
  \href{http://dx.doi.org/10.1007/JHEP07(2018)163}{{\em JHEP} {\bfseries 07}
  (2018) 163},
\href{http://arxiv.org/abs/1705.10844}{{\ttfamily arXiv:1705.10844 [hep-th]}}.

\bibitem{Martinec:2018nco}
E.~J. Martinec, S.~Massai, and D.~Turton, ``{String dynamics in NS5-F1-P
  geometries},'' \href{http://dx.doi.org/10.1007/JHEP09(2018)031}{{\em JHEP}
  {\bfseries 09} (2018) 031},
\href{http://arxiv.org/abs/1803.08505}{{\ttfamily arXiv:1803.08505 [hep-th]}}.

\bibitem{Martinec:2019wzw}
E.~J. Martinec, S.~Massai, and D.~Turton, ``{Little Strings, Long Strings, and
  Fuzzballs},'' \href{http://dx.doi.org/10.1007/JHEP11(2019)019}{{\em JHEP}
  {\bfseries 11} (2019) 019}, \href{http://arxiv.org/abs/1906.11473}{{\ttfamily
  arXiv:1906.11473 [hep-th]}}.

\bibitem{Martinec:2020gkv}
E.~J. Martinec, S.~Massai, and D.~Turton, ``{Stringy Structure at the BPS
  Bound},'' \href{http://dx.doi.org/10.1007/JHEP12(2020)135}{{\em JHEP}
  {\bfseries 12} (2020) 135}, \href{http://arxiv.org/abs/2005.12344}{{\ttfamily
  arXiv:2005.12344 [hep-th]}}.

\bibitem{Bufalini:2021ndn}
D.~Bufalini, S.~Iguri, N.~Kovensky, and D.~Turton, ``{Black hole microstates
  from the worldsheet},'' \href{http://arxiv.org/abs/2105.02255}{{\ttfamily
  arXiv:2105.02255 [hep-th]}}.

\bibitem{Bena:2015bea}
I.~Bena, S.~Giusto, R.~Russo, M.~Shigemori, and N.~P. Warner, ``{Habemus
  Superstratum! A constructive proof of the existence of superstrata},''
  \href{http://dx.doi.org/10.1007/JHEP05(2015)110}{{\em JHEP} {\bfseries 05}
  (2015) 110},
\href{http://arxiv.org/abs/1503.01463}{{\ttfamily arXiv:1503.01463 [hep-th]}}.

\bibitem{Bena:2016agb}
I.~Bena, E.~Martinec, D.~Turton, and N.~P. Warner, ``{Momentum Fractionation on
  Superstrata},'' \href{http://dx.doi.org/10.1007/JHEP05(2016)064}{{\em JHEP}
  {\bfseries 05} (2016) 064},
\href{http://arxiv.org/abs/1601.05805}{{\ttfamily arXiv:1601.05805 [hep-th]}}.

\bibitem{Bena:2016ypk}
I.~Bena, S.~Giusto, E.~J. Martinec, R.~Russo, M.~Shigemori, D.~Turton, and
  N.~P. Warner, ``{Smooth horizonless geometries deep inside the black-hole
  regime},'' \href{http://dx.doi.org/10.1103/PhysRevLett.117.201601}{{\em Phys.
  Rev. Lett.} {\bfseries 117} no.~20, (2016) 201601},
\href{http://arxiv.org/abs/1607.03908}{{\ttfamily arXiv:1607.03908 [hep-th]}}.

\bibitem{Bena:2017geu}
I.~Bena, E.~Martinec, D.~Turton, and N.~P. Warner, ``{M-theory Superstrata and
  the MSW String},'' \href{http://dx.doi.org/10.1007/JHEP06(2017)137}{{\em
  JHEP} {\bfseries 06} (2017) 137},
\href{http://arxiv.org/abs/1703.10171}{{\ttfamily arXiv:1703.10171 [hep-th]}}.

\bibitem{Bena:2017xbt}
I.~Bena, S.~Giusto, E.~J. Martinec, R.~Russo, M.~Shigemori, D.~Turton, and
  N.~P. Warner, ``{Asymptotically-flat supergravity solutions deep inside the
  black-hole regime},'' \href{http://dx.doi.org/10.1007/JHEP02(2018)014}{{\em
  JHEP} {\bfseries 02} (2018) 014},
\href{http://arxiv.org/abs/1711.10474}{{\ttfamily arXiv:1711.10474 [hep-th]}}.

\bibitem{Bena:2017upb}
I.~Bena, D.~Turton, R.~Walker, and N.~P. Warner, ``{Integrability and
  Black-Hole Microstate Geometries},''
  \href{http://dx.doi.org/10.1007/JHEP11(2017)021}{{\em JHEP} {\bfseries 11}
  (2017) 021},
\href{http://arxiv.org/abs/1709.01107}{{\ttfamily arXiv:1709.01107 [hep-th]}}.

\bibitem{Mathur:2013nja}
S.~D. Mathur and D.~Turton, ``{Oscillating supertubes and neutral rotating
  black hole microstates},''
  \href{http://dx.doi.org/10.1007/JHEP04(2014)072}{{\em JHEP} {\bfseries 1404}
  (2014) 072},
\href{http://arxiv.org/abs/1310.1354}{{\ttfamily arXiv:1310.1354 [hep-th]}}.

\bibitem{Bena:2013ora}
I.~Bena, S.~F. Ross, and N.~P. Warner, ``{On the Oscillation of Species},''
  \href{http://dx.doi.org/10.1007/JHEP09(2014)113}{{\em JHEP} {\bfseries 1409}
  (2014) 113},
\href{http://arxiv.org/abs/1312.3635}{{\ttfamily arXiv:1312.3635 [hep-th]}}.

\bibitem{Bena:2014rea}
I.~Bena, S.~F. Ross, and N.~P. Warner, ``{Coiffured Black Rings},''
  \href{http://dx.doi.org/10.1088/0264-9381/31/16/165015}{{\em
  Class.Quant.Grav.} {\bfseries 31} (2014) 165015},
\href{http://arxiv.org/abs/1405.5217}{{\ttfamily arXiv:1405.5217 [hep-th]}}.

\bibitem{Shigemori:2020yuo}
M.~Shigemori, ``{Superstrata},''
  \href{http://dx.doi.org/10.1007/s10714-020-02698-8}{{\em Gen. Rel. Grav.}
  {\bfseries 52} no.~5, (2020) 51},
  \href{http://arxiv.org/abs/2002.01592}{{\ttfamily arXiv:2002.01592
  [hep-th]}}.

\bibitem{Kanitscheider:2006zf}
I.~Kanitscheider, K.~Skenderis, and M.~Taylor, ``{Holographic anatomy of
  fuzzballs},'' \href{http://dx.doi.org/10.1088/1126-6708/2007/04/023}{{\em
  JHEP} {\bfseries 04} (2007) 023},
\href{http://arxiv.org/abs/hep-th/0611171}{{\ttfamily arXiv:hep-th/0611171}}.

\bibitem{Taylor:2007hs}
M.~Taylor, ``{Matching of correlators in AdS(3) / CFT(2)},''
  \href{http://dx.doi.org/10.1088/1126-6708/2008/06/010}{{\em JHEP} {\bfseries
  06} (2008) 010},
\href{http://arxiv.org/abs/0709.1838}{{\ttfamily arXiv:0709.1838 [hep-th]}}.

\bibitem{Giusto:2015dfa}
S.~Giusto, E.~Moscato, and R.~Russo, ``{AdS$_{3}$ holography for 1/4 and 1/8
  BPS geometries},'' \href{http://dx.doi.org/10.1007/JHEP11(2015)004}{{\em
  JHEP} {\bfseries 11} (2015) 004},
\href{http://arxiv.org/abs/1507.00945}{{\ttfamily arXiv:1507.00945 [hep-th]}}.

\bibitem{Tormo:2019yus}
J.~Garcia~i Tormo and M.~Taylor, ``{One point functions for black hole
  microstates},'' \href{http://dx.doi.org/10.1007/s10714-019-2566-6}{{\em Gen.
  Rel. Grav.} {\bfseries 51} no.~7, (2019) 89},
  \href{http://arxiv.org/abs/1904.10200}{{\ttfamily arXiv:1904.10200
  [hep-th]}}.

\bibitem{Giusto:2019qig}
S.~Giusto, S.~Rawash, and D.~Turton, ``{AdS$_{3}$ holography at dimension
  two},'' \href{http://dx.doi.org/10.1007/JHEP07(2019)171}{{\em JHEP}
  {\bfseries 07} (2019) 171}, \href{http://arxiv.org/abs/1904.12880}{{\ttfamily
  arXiv:1904.12880 [hep-th]}}.

\bibitem{Ceplak:2018pws}
N.~Ceplak, R.~Russo, and M.~Shigemori, ``{Supercharging Superstrata},''
  \href{http://dx.doi.org/10.1007/JHEP03(2019)095}{{\em JHEP} {\bfseries 03}
  (2019) 095},
\href{http://arxiv.org/abs/1812.08761}{{\ttfamily arXiv:1812.08761 [hep-th]}}.

\bibitem{Heidmann:2019zws}
P.~Heidmann and N.~P. Warner, ``{Superstratum Symbiosis},''
  \href{http://dx.doi.org/10.1007/JHEP09(2019)059}{{\em JHEP} {\bfseries 09}
  (2019) 059}, \href{http://arxiv.org/abs/1903.07631}{{\ttfamily
  arXiv:1903.07631 [hep-th]}}.

\bibitem{Heidmann:2019xrd}
P.~Heidmann, D.~R. Mayerson, R.~Walker, and N.~P. Warner, ``{Holomorphic Waves
  of Black Hole Microstructure},''
  \href{http://dx.doi.org/10.1007/JHEP02(2020)192}{{\em JHEP} {\bfseries 02}
  (2020) 192}, \href{http://arxiv.org/abs/1910.10714}{{\ttfamily
  arXiv:1910.10714 [hep-th]}}.

\bibitem{Mayerson:2020tcl}
D.~R. Mayerson, R.~A. Walker, and N.~P. Warner, ``{Microstate Geometries from
  Gauged Supergravity in Three Dimensions},''
  \href{http://arxiv.org/abs/2004.13031}{{\ttfamily arXiv:2004.13031
  [hep-th]}}.

\bibitem{Klebanov:1999tb}
I.~R. Klebanov and E.~Witten, ``{AdS / CFT correspondence and symmetry
  breaking},'' \href{http://dx.doi.org/10.1016/S0550-3213(99)00387-9}{{\em
  Nucl. Phys. B} {\bfseries 556} (1999) 89--114},
  \href{http://arxiv.org/abs/hep-th/9905104}{{\ttfamily arXiv:hep-th/9905104}}.

\bibitem{Skenderis:2006uy}
K.~Skenderis and M.~Taylor, ``{Kaluza-Klein holography},''
  \href{http://dx.doi.org/10.1088/1126-6708/2006/05/057}{{\em JHEP} {\bfseries
  05} (2006) 057},
\href{http://arxiv.org/abs/hep-th/0603016}{{\ttfamily arXiv:hep-th/0603016
  [hep-th]}}.

\bibitem{Mathur:2005zp}
S.~D. Mathur, ``{The fuzzball proposal for black holes: An elementary
  review},'' \href{http://dx.doi.org/10.1002/prop.200410203}{{\em Fortsch.
  Phys.} {\bfseries 53} (2005) 793--827},
\href{http://arxiv.org/abs/hep-th/0502050}{{\ttfamily arXiv:hep-th/0502050}}.

\bibitem{Skenderis:2008qn}
K.~Skenderis and M.~Taylor, ``{The fuzzball proposal for black holes},''
  \href{http://dx.doi.org/10.1016/j.physrep.2008.08.001}{{\em Phys. Rept.}
  {\bfseries 467} (2008) 117--171},
\href{http://arxiv.org/abs/0804.0552}{{\ttfamily arXiv:0804.0552 [hep-th]}}.

\bibitem{Mathur:2012zp}
S.~D. Mathur, ``{Black Holes and Beyond},''
  \href{http://dx.doi.org/10.1016/j.aop.2012.05.001}{{\em Annals Phys.}
  {\bfseries 327} (2012) 2760--2793},
\href{http://arxiv.org/abs/1205.0776}{{\ttfamily arXiv:1205.0776 [hep-th]}}.

\bibitem{Bena:2013dka}
I.~Bena and N.~P. Warner, ``{Resolving the Structure of Black Holes:
  Philosophizing with a Hammer},''
\href{http://arxiv.org/abs/1311.4538}{{\ttfamily arXiv:1311.4538 [hep-th]}}.

\bibitem{Aharony:1999ti}
O.~Aharony, S.~S. Gubser, J.~M. Maldacena, H.~Ooguri, and Y.~Oz, ``{Large N
  field theories, string theory and gravity},''
  \href{http://dx.doi.org/10.1016/S0370-1573(99)00083-6}{{\em Phys. Rept.}
  {\bfseries 323} (2000) 183--386},
  \href{http://arxiv.org/abs/hep-th/9905111}{{\ttfamily arXiv:hep-th/9905111}}.

\bibitem{David:2002wn}
J.~R. David, G.~Mandal, and S.~R. Wadia, ``{Microscopic formulation of black
  holes in string theory},''
  \href{http://dx.doi.org/10.1016/S0370-1573(02)00271-5}{{\em Phys. Rept.}
  {\bfseries 369} (2002) 549--686},
\href{http://arxiv.org/abs/hep-th/0203048}{{\ttfamily arXiv:hep-th/0203048}}.

\bibitem{Arutyunov:1999en}
G.~Arutyunov and S.~Frolov, ``{Some cubic couplings in type IIB supergravity on
  AdS(5) x S**5 and three point functions in SYM(4) at large N},''
  \href{http://dx.doi.org/10.1103/PhysRevD.61.064009}{{\em Phys. Rev. D}
  {\bfseries 61} (2000) 064009},
  \href{http://arxiv.org/abs/hep-th/9907085}{{\ttfamily arXiv:hep-th/9907085}}.

\bibitem{DHoker:1999jke}
E.~D'Hoker, D.~Z. Freedman, S.~D. Mathur, A.~Matusis, and L.~Rastelli,
  ``{Extremal correlators in the AdS / CFT correspondence},''
\href{http://arxiv.org/abs/hep-th/9908160}{{\ttfamily arXiv:hep-th/9908160
  [hep-th]}}.

\bibitem{Arutyunov:2000ima}
G.~Arutyunov and S.~Frolov, ``{On the correspondence between gravity fields and
  CFT operators},'' \href{http://dx.doi.org/10.1088/1126-6708/2000/04/017}{{\em
  JHEP} {\bfseries 04} (2000) 017},
\href{http://arxiv.org/abs/hep-th/0003038}{{\ttfamily arXiv:hep-th/0003038
  [hep-th]}}.

\bibitem{DHoker:2000xhf}
E.~D'Hoker, J.~Erdmenger, D.~Z. Freedman, and M.~Perez-Victoria, ``{Near
  extremal correlators and vanishing supergravity couplings in AdS / CFT},''
  \href{http://dx.doi.org/10.1016/S0550-3213(00)00534-4}{{\em Nucl. Phys. B}
  {\bfseries 589} (2000) 3--37},
  \href{http://arxiv.org/abs/hep-th/0003218}{{\ttfamily arXiv:hep-th/0003218}}.

\bibitem{Corley:2001zk}
S.~Corley, A.~Jevicki, and S.~Ramgoolam, ``{Exact correlators of giant
  gravitons from dual N = 4 SYM theory},'' {\em Adv. Theor. Math. Phys.}
  {\bfseries 5} (2002) 809--839,
\href{http://arxiv.org/abs/hep-th/0111222}{{\ttfamily arXiv:hep-th/0111222}}.

\bibitem{Uruchurtu:2011wh}
L.~I. Uruchurtu, ``{Next-next-to-extremal Four Point Functions of N=4 1/2 BPS
  Operators in the AdS/CFT Correspondence},''
  \href{http://dx.doi.org/10.1007/JHEP08(2011)133}{{\em JHEP} {\bfseries 08}
  (2011) 133}, \href{http://arxiv.org/abs/1106.0630}{{\ttfamily arXiv:1106.0630
  [hep-th]}}.

\bibitem{Rastelli:2017udc}
L.~Rastelli and X.~Zhou, ``{How to Succeed at Holographic Correlators Without
  Really Trying},'' \href{http://dx.doi.org/10.1007/JHEP04(2018)014}{{\em JHEP}
  {\bfseries 04} (2018) 014},
\href{http://arxiv.org/abs/1710.05923}{{\ttfamily arXiv:1710.05923 [hep-th]}}.

\bibitem{Aprile:2018efk}
F.~Aprile, J.~Drummond, P.~Heslop, and H.~Paul, ``{Double-trace spectrum of
  $N=4$ supersymmetric Yang-Mills theory at strong coupling},''
  \href{http://dx.doi.org/10.1103/PhysRevD.98.126008}{{\em Phys. Rev. D}
  {\bfseries 98} no.~12, (2018) 126008},
  \href{http://arxiv.org/abs/1802.06889}{{\ttfamily arXiv:1802.06889
  [hep-th]}}.

\bibitem{Aprile:2020uxk}
F.~Aprile, J.~Drummond, P.~Heslop, H.~Paul, F.~Sanfilippo, M.~Santagata, and
  A.~Stewart, ``{Single Particle Operators and their Correlators in Free
  $\mathcal{N}=4$ SYM},'' \href{http://arxiv.org/abs/2007.09395}{{\ttfamily
  arXiv:2007.09395 [hep-th]}}.

\bibitem{Maldacena:1997re}
J.~M. Maldacena, ``{The large N limit of superconformal field theories and
  supergravity},'' {\em Adv. Theor. Math. Phys.} {\bfseries 2} (1998) 231--252,
\href{http://arxiv.org/abs/hep-th/9711200}{{\ttfamily arXiv:hep-th/9711200}}.

\bibitem{Vafa:1995zh}
C.~Vafa, ``{Gas of D-branes and Hagedorn density of BPS states},''
  \href{http://dx.doi.org/10.1016/0550-3213(96)00025-9}{{\em Nucl. Phys.}
  {\bfseries B463} (1996) 415--419},
\href{http://arxiv.org/abs/hep-th/9511088}{{\ttfamily arXiv:hep-th/9511088
  [hep-th]}}.

\bibitem{Deger:1998nm}
S.~Deger, A.~Kaya, E.~Sezgin, and P.~Sundell, ``{Spectrum of D = 6, N = 4b
  supergravity on AdS(3) x S(3)},''
  \href{http://dx.doi.org/10.1016/S0550-3213(98)00555-0}{{\em Nucl. Phys.}
  {\bfseries B536} (1998) 110--140},
\href{http://arxiv.org/abs/hep-th/9804166}{{\ttfamily arXiv:hep-th/9804166
  [hep-th]}}.

\bibitem{Larsen:1998xm}
F.~Larsen, ``{The Perturbation spectrum of black holes in N=8 supergravity},''
  \href{http://dx.doi.org/10.1016/S0550-3213(98)00564-1}{{\em Nucl. Phys.}
  {\bfseries B536} (1998) 258--278},
\href{http://arxiv.org/abs/hep-th/9805208}{{\ttfamily arXiv:hep-th/9805208
  [hep-th]}}.

\bibitem{deBoer:1998ip}
J.~de~Boer, ``{Six-dimensional supergravity on S**3 x AdS(3) and 2-D conformal
  field theory},'' \href{http://dx.doi.org/10.1016/S0550-3213(99)00160-1}{{\em
  Nucl. Phys.} {\bfseries B548} (1999) 139--166},
\href{http://arxiv.org/abs/hep-th/9806104}{{\ttfamily arXiv:hep-th/9806104
  [hep-th]}}.

\bibitem{Larsen:1999uk}
F.~Larsen and E.~J. Martinec, ``{U(1) charges and moduli in the D1-D5
  system},'' {\em JHEP} {\bfseries 06} (1999) 019,
\href{http://arxiv.org/abs/hep-th/9905064}{{\ttfamily arXiv:hep-th/9905064}}.

\bibitem{Eberhardt:2019ywk}
L.~Eberhardt, M.~R. Gaberdiel, and R.~Gopakumar, ``{Deriving the
  AdS$_{3}$/CFT$_{2}$ correspondence},''
  \href{http://dx.doi.org/10.1007/JHEP02(2020)136}{{\em JHEP} {\bfseries 02}
  (2020) 136}, \href{http://arxiv.org/abs/1911.00378}{{\ttfamily
  arXiv:1911.00378 [hep-th]}}.

\bibitem{Gaberdiel:2020ycd}
M.~R. Gaberdiel, R.~Gopakumar, B.~Knighton, and P.~Maity, ``{From Symmetric
  Product CFTs to ${\rm AdS}_3$},''
  \href{http://arxiv.org/abs/2011.10038}{{\ttfamily arXiv:2011.10038
  [hep-th]}}.

\bibitem{Shigemori_2019}
M.~Shigemori, ``Counting superstrata,''
  \href{http://dx.doi.org/10.1007/jhep10(2019)017}{{\em Journal of High Energy
  Physics} {\bfseries 2019} no.~10, (Oct, 2019) }.
  \url{http://dx.doi.org/10.1007/JHEP10(2019)017}.

\bibitem{Avery:2010qw}
S.~G. Avery, ``{Using the D1D5 CFT to Understand Black Holes},''
\href{http://arxiv.org/abs/1012.0072}{{\ttfamily arXiv:1012.0072 [hep-th]}}.

\bibitem{Baggio:2012rr}
M.~Baggio, J.~de~Boer, and K.~Papadodimas, ``{A non-renormalization theorem for
  chiral primary 3-point functions},''
  \href{http://dx.doi.org/10.1007/JHEP07(2012)137}{{\em JHEP} {\bfseries 07}
  (2012) 137},
\href{http://arxiv.org/abs/1203.1036}{{\ttfamily arXiv:1203.1036 [hep-th]}}.

\bibitem{Chakrabarty:2015foa}
B.~Chakrabarty, D.~Turton, and A.~Virmani, ``{Holographic description of
  non-supersymmetric orbifolded D1-D5-P solutions},''
  \href{http://dx.doi.org/10.1007/JHEP11(2015)063}{{\em JHEP} {\bfseries 11}
  (2015) 063},
\href{http://arxiv.org/abs/1508.01231}{{\ttfamily arXiv:1508.01231 [hep-th]}}.

\bibitem{Shigemori:2021pir}
M.~Shigemori, ``{Interpolating between multi-center microstate geometries},''
  \href{http://arxiv.org/abs/2105.11639}{{\ttfamily arXiv:2105.11639
  [hep-th]}}.

\bibitem{Romans:1986er}
L.~Romans, ``{Selfduality for Interacting Fields: Covariant Field Equations for
  Six-dimensional Chiral Supergravities},''
  \href{http://dx.doi.org/10.1016/0550-3213(86)90016-7}{{\em Nucl. Phys. B}
  {\bfseries 276} (1986) 71}.

\bibitem{Giusto:2013rxa}
S.~Giusto, L.~Martucci, M.~Petrini, and R.~Russo, ``{6D microstate geometries
  from 10D structures},''
  \href{http://dx.doi.org/10.1016/j.nuclphysb.2013.08.018}{{\em Nucl.Phys.}
  {\bfseries B876} (2013) 509--555},
\href{http://arxiv.org/abs/1306.1745}{{\ttfamily arXiv:1306.1745 [hep-th]}}.

\bibitem{Balasubramanian:2000rt}
V.~Balasubramanian, J.~de~Boer, E.~Keski-Vakkuri, and S.~F. Ross,
  ``{Supersymmetric conical defects: Towards a string theoretic description of
  black hole formation},''
  \href{http://dx.doi.org/10.1103/PhysRevD.64.064011}{{\em Phys. Rev.}
  {\bfseries D64} (2001) 064011},
\href{http://arxiv.org/abs/hep-th/0011217}{{\ttfamily arXiv:hep-th/0011217}}.

\bibitem{Maldacena:2000dr}
J.~M. Maldacena and L.~Maoz, ``{De-singularization by rotation},'' {\em JHEP}
  {\bfseries 12} (2002) 055,
\href{http://arxiv.org/abs/hep-th/0012025}{{\ttfamily arXiv:hep-th/0012025}}.

\bibitem{Jevicki:1998bm}
A.~Jevicki, M.~Mihailescu, and S.~Ramgoolam, ``{Gravity from CFT on S**N(X):
  Symmetries and interactions},''
  \href{http://dx.doi.org/10.1016/S0550-3213(00)00147-4}{{\em Nucl.Phys.}
  {\bfseries B577} (2000) 47--72},
\href{http://arxiv.org/abs/hep-th/9907144}{{\ttfamily arXiv:hep-th/9907144
  [hep-th]}}.

\bibitem{Lunin:2001pw}
O.~Lunin and S.~D. Mathur, ``{Three-point functions for M(N)/S(N) orbifolds
  with N = 4 supersymmetry},''
  \href{http://dx.doi.org/10.1007/s002200200638}{{\em Commun. Math. Phys.}
  {\bfseries 227} (2002) 385--419},
\href{http://arxiv.org/abs/hep-th/0103169}{{\ttfamily arXiv:hep-th/0103169}}.

\bibitem{Mihailescu_2000}
M.~Mihailescu, ``Correlation functions for chiral primaries in d = 6
  supergravity on ads3 × s3,''
  \href{http://dx.doi.org/10.1088/1126-6708/2000/02/007}{{\em Journal of High
  Energy Physics} {\bfseries 2000} no.~02, (Jan, 2000) 007–007}.
  \url{http://dx.doi.org/10.1088/1126-6708/2000/02/007}.

\bibitem{Arutyunov:2000by}
G.~Arutyunov, A.~Pankiewicz, and S.~Theisen, ``{Cubic couplings in D = 6 N=4b
  supergravity on AdS(3) x S**3},''
  \href{http://dx.doi.org/10.1103/PhysRevD.63.044024}{{\em Phys. Rev. D}
  {\bfseries 63} (2001) 044024},
  \href{http://arxiv.org/abs/hep-th/0007061}{{\ttfamily arXiv:hep-th/0007061}}.

\bibitem{Yang:2021kot}
P.~Yang, Y.~Jiang, S.~Komatsu, and J.-B. Wu, ``{D-branes and Orbit Average},''
  \href{http://arxiv.org/abs/2103.16580}{{\ttfamily arXiv:2103.16580
  [hep-th]}}.

\bibitem{McFadden_2011}
P.~McFadden and K.~Skenderis, ``Holographic non-gaussianity,''
  \href{http://dx.doi.org/10.1088/1475-7516/2011/05/013}{{\em Journal of
  Cosmology and Astroparticle Physics} {\bfseries 2011} no.~05, (May, 2011)
  013–013}. \url{http://dx.doi.org/10.1088/1475-7516/2011/05/013}.

\bibitem{Bruni_1997}
M.~Bruni, S.~Matarrese, S.~Mollerach, and S.~Sonego, ``Perturbations of
  spacetime: gauge transformations and gauge invariance at second order and
  beyond,'' \href{http://dx.doi.org/10.1088/0264-9381/14/9/014}{{\em Classical
  and Quantum Gravity} {\bfseries 14} no.~9, (Sep, 1997) 2585–2606}.
  \url{http://dx.doi.org/10.1088/0264-9381/14/9/014}.

\bibitem{Hansen:2006wu}
J.~Hansen and P.~Kraus, ``{Generating charge from diffeomorphisms},''
  \href{http://dx.doi.org/10.1088/1126-6708/2006/12/009}{{\em JHEP} {\bfseries
  0612} (2006) 009},
\href{http://arxiv.org/abs/hep-th/0606230}{{\ttfamily arXiv:hep-th/0606230
  [hep-th]}}.

\bibitem{Lunin:2000yv}
O.~Lunin and S.~D. Mathur, ``{Correlation functions for M(N)/S(N) orbifolds},''
  \href{http://dx.doi.org/10.1007/s002200100431}{{\em Commun. Math. Phys.}
  {\bfseries 219} (2001) 399--442},
\href{http://arxiv.org/abs/hep-th/0006196}{{\ttfamily arXiv:hep-th/0006196}}.

\bibitem{Galliani_2016}
A.~Galliani, S.~Giusto, E.~Moscato, and R.~Russo, ``Correlators at large c
  without information loss,''
  \href{http://dx.doi.org/10.1007/jhep09(2016)065}{{\em Journal of High Energy
  Physics} {\bfseries 2016} no.~9, (Sep, 2016) }.
  \url{http://dx.doi.org/10.1007/JHEP09(2016)065}.

\bibitem{Galliani:2017jlg}
A.~Galliani, S.~Giusto, and R.~Russo, ``{Holographic 4-point correlators with
  heavy states},'' \href{http://dx.doi.org/10.1007/JHEP10(2017)040}{{\em JHEP}
  {\bfseries 10} (2017) 040},
\href{http://arxiv.org/abs/1705.09250}{{\ttfamily arXiv:1705.09250 [hep-th]}}.

\bibitem{Bombini:2017sge}
A.~Bombini, A.~Galliani, S.~Giusto, E.~Moscato, and R.~Russo, ``{Unitary
  4-point correlators from classical geometries},''
  \href{http://dx.doi.org/10.1140/epjc/s10052-017-5492-3}{{\em Eur. Phys. J.}
  {\bfseries C78} no.~1, (2018) 8},
\href{http://arxiv.org/abs/1710.06820}{{\ttfamily arXiv:1710.06820 [hep-th]}}.

\bibitem{Bombini:2019vnc}
A.~Bombini and A.~Galliani, ``{AdS$_3$ four-point functions from
  $\frac{1}{8}$-BPS states},''
\href{http://arxiv.org/abs/1904.02656}{{\ttfamily arXiv:1904.02656 [hep-th]}}.

\bibitem{Giusto:2018ovt}
S.~Giusto, R.~Russo, and C.~Wen, ``{Holographic correlators in AdS$_{3}$},''
  \href{http://dx.doi.org/10.1007/JHEP03(2019)096}{{\em JHEP} {\bfseries 03}
  (2019) 096}, \href{http://arxiv.org/abs/1812.06479}{{\ttfamily
  arXiv:1812.06479 [hep-th]}}.

\bibitem{Rastelli_2019}
L.~Rastelli, K.~Roumpedakis, and X.~Zhou, ``Ads3× s3 tree-level correlators:
  hidden six-dimensional conformal symmetry,''
  \href{http://dx.doi.org/10.1007/jhep10(2019)140}{{\em Journal of High Energy
  Physics} {\bfseries 2019} no.~10, (Oct, 2019) }.
  \url{http://dx.doi.org/10.1007/JHEP10(2019)140}.

\bibitem{Giusto:2019pxc}
S.~Giusto, R.~Russo, A.~Tyukov, and C.~Wen, ``{Holographic correlators in
  AdS$_3$ without Witten diagrams},''
  \href{http://dx.doi.org/10.1007/JHEP09(2019)030}{{\em JHEP} {\bfseries 09}
  (2019) 030}, \href{http://arxiv.org/abs/1905.12314}{{\ttfamily
  arXiv:1905.12314 [hep-th]}}.

\bibitem{Giusto:2020neo}
S.~Giusto, R.~Russo, A.~Tyukov, and C.~Wen, ``{The CFT$_6$ origin of all
  tree-level 4-point correlators in AdS$_3 \times S^3$},''
  \href{http://dx.doi.org/10.1140/epjc/s10052-020-8300-4}{{\em Eur. Phys. J. C}
  {\bfseries 80} no.~8, (2020) 736},
  \href{http://arxiv.org/abs/2005.08560}{{\ttfamily arXiv:2005.08560
  [hep-th]}}.

\bibitem{Aprile:2021mvq}
F.~Aprile and M.~Santagata, ``{Two-particle spectrum of tensor multiplets
  coupled to $AdS_3\times S^3$ gravity},''
  \href{http://arxiv.org/abs/2104.00036}{{\ttfamily arXiv:2104.00036
  [hep-th]}}.

\bibitem{Giusto:2020mup}
S.~Giusto, M.~R.~R. Hughes, and R.~Russo, ``{The Regge limit of AdS$_{3}$
  holographic correlators},''
  \href{http://dx.doi.org/10.1007/JHEP11(2020)018}{{\em JHEP} {\bfseries 11}
  (2020) 018}, \href{http://arxiv.org/abs/2007.12118}{{\ttfamily
  arXiv:2007.12118 [hep-th]}}.

\bibitem{Ceplak:2021wak}
N.~Ceplak and M.~R.~R. Hughes, ``{The Regge limit of AdS$_3$ holographic
  correlators with heavy states: towards the black hole regime},''
  \href{http://arxiv.org/abs/2102.09549}{{\ttfamily arXiv:2102.09549
  [hep-th]}}.

\bibitem{Ceplak:2021wzz}
N.~Ceplak, S.~Giusto, M.~R.~R. Hughes, and R.~Russo, ``{Holographic correlators
  with multi-particle states},''
  \href{http://arxiv.org/abs/2105.04670}{{\ttfamily arXiv:2105.04670
  [hep-th]}}.

\bibitem{Ribault:2014hia}
S.~Ribault, ``{Conformal field theory on the plane},''
\href{http://arxiv.org/abs/1406.4290}{{\ttfamily arXiv:1406.4290 [hep-th]}}.

\bibitem{Black:2010uq}
W.~Black, R.~Russo, and D.~Turton, ``{The Supergravity fields for a D-brane
  with a travelling wave from string amplitudes},''
  \href{http://dx.doi.org/10.1016/j.physletb.2010.09.059}{{\em Phys.Lett.}
  {\bfseries B694} (2010) 246--251},
\href{http://arxiv.org/abs/1007.2856}{{\ttfamily arXiv:1007.2856 [hep-th]}}.

\bibitem{Giusto:2011fy}
S.~Giusto, R.~Russo, and D.~Turton, ``{New D1-D5-P geometries from string
  amplitudes},'' \href{http://dx.doi.org/10.1007/JHEP11(2011)062}{{\em JHEP}
  {\bfseries 11} (2011) 062},
\href{http://arxiv.org/abs/1108.6331}{{\ttfamily arXiv:1108.6331 [hep-th]}}.

\bibitem{Bena:2011dd}
I.~Bena, S.~Giusto, M.~Shigemori, and N.~P. Warner, ``{Supersymmetric Solutions
  in Six Dimensions: A Linear Structure},''
  \href{http://dx.doi.org/10.1007/JHEP03(2012)084}{{\em JHEP} {\bfseries 1203}
  (2012) 084},
\href{http://arxiv.org/abs/1110.2781}{{\ttfamily arXiv:1110.2781 [hep-th]}}.

\bibitem{Gutowski:2003rg}
J.~B. Gutowski, D.~Martelli, and H.~S. Reall, ``{All supersymmetric solutions
  of minimal supergravity in six dimensions},''
  \href{http://dx.doi.org/10.1088/0264-9381/20/23/008}{{\em Class. Quant.
  Grav.} {\bfseries 20} (2003) 5049--5078},
\href{http://arxiv.org/abs/hep-th/0306235}{{\ttfamily arXiv:hep-th/0306235}}.

\end{thebibliography}\endgroup

\end{adjustwidth}


\end{document}